\documentclass[utf8]{FrontiersinHarvard} % for articles in journals using the Harvard Referencing Style (Author-Date), for Frontiers Reference Styles by Journal: https://zendesk.frontiersin.org/hc/en-us/articles/360017860337-Frontiers-Reference-Styles-by-Journal
\usepackage{url,hyperref,lineno,microtype,subcaption}
\usepackage[onehalfspacing]{setspace}
\usepackage{graphicx}

\usepackage{bm}
\usepackage{booktabs}

\graphicspath{{./frontier_compatible_figures/}}

\def\keyFont{\fontsize{8}{11}\helveticabold }
\def\firstAuthorLast{Liaudat {et~al.}} %use et al only if is more than 1 author
\def\Authors{Tobías I. Liaudat\,$^{1,2,3,*}$, Jean-Luc Starck\,$^{1,4}$ and Martin Kilbinger\,$^{1}$}
\def\Address{$^{1}$Université Paris-Saclay, Université Paris Cité, CEA, CNRS, AIM, 91191, Gif-sur-Yvette, France;\\
$^{2}$Department of Computer Science, University College London (UCL), London WC1E 6BT, UK;\\
$^{3}$ Mullard Space Science Laboratory, University College London (UCL), Dorking, RH5 6NT, UK;\\
$^{4}$ Institute of Astrophysics, Foundation for Research and Technology-Hellas (FORTH), Heraklion, 70013, Greece;}
\def\corrAuthor{Tobías I. Liaudat}

\def\corrEmail{tobiasliaudat@gmail.com}

\begin{document}
\onecolumn
\firstpage{1}

\title[PSF modelling for astronomical telescopes: a review focused on weak lensing]{Point spread function modelling for astronomical telescopes: a review focused on weak gravitational lensing studies} 

\author[\firstAuthorLast ]{\Authors} %This field will be automatically populated
\address{} %This field will be automatically populated
\correspondance{} %This field will be automatically populated

\extraAuth{}% If there are more than 1 corresponding author, comment this line and uncomment the next one.
%\extraAuth{corresponding Author2 \\ Laboratory X2, Institute X2, Department X2, Organization X2, Street X2, City X2 , State XX2 (only USA, Canada and Australia), Zip Code2, X2 Country X2, email2@uni2.edu}

\maketitle

\begin{abstract}

%%% Leave the Abstract empty if your article does not require one, please see the Summary Table for full details.
% For full guidelines regarding your manuscript please refer to \href{http://www.frontiersin.org/about/AuthorGuidelines}{Author Guidelines}.
% 
\section{}
The accurate modelling of the Point Spread Function (PSF) is of paramount importance in astronomical observations, as it allows for the correction of distortions and blurring caused by the telescope and atmosphere. PSF modelling is crucial for accurately measuring celestial objects' properties. The last decades brought us a steady increase in the power and complexity of astronomical telescopes and instruments. Upcoming galaxy surveys like \textit{Euclid} and LSST will observe an unprecedented amount and quality of data. Modelling the PSF for these new facilities and surveys requires novel modelling techniques that can cope with the ever-tightening error requirements. The purpose of this review is three-fold. First, we introduce the optical background required for a more physically-motivated PSF modelling and propose an observational model that can be reused for future developments. Second, we provide an overview of the different physical contributors of the PSF, including the optic- and detector-level contributors and the atmosphere. We expect that the overview will help better understand the modelled effects. Third, we discuss the different methods for PSF modelling from the parametric and non-parametric families for ground- and space-based telescopes, with their advantages and limitations. Validation methods for PSF models are then addressed, with several metrics related to weak lensing studies discussed in detail. Finally, we explore current challenges and future directions in PSF modelling for astronomical telescopes.

\tiny
 \keyFont{\section{Keywords:} point spread function, inverse problems, weak gravitational lensing, image processing, super-resolution.} 
%All article types: you may provide up to 8 keywords; at least 5 are mandatory.
\end{abstract}

\section{Introduction}
Any astronomical image is observed through an optical system that introduces deformations and distortions. Even the most powerful imaging system introduces distortions to the observed object. How to characterise these distortions is a subject of study known as PSF modelling. Specific science applications, like weak gravitational lensing (WL) in cosmology (see \citet{kilbinger2015,mandelbaum2018_bis3} for reviews), require very accurate and precise measurements of galaxy shapes. A crucial step of any weak lensing mission is to estimate the PSF at any position of the observed images. If the PSF is not considered when measuring galaxy shapes, the measurement will be biased, resulting in unacceptably biased WL studies. Furthermore, the PSF can be the predominant source of systematic errors and biases in WL studies. This fact makes PSF modelling a vital task. Forthcoming astronomical telescopes, such as the \textit{Euclid} space telescope \citep{laureijs2011}, the \textit{Roman} space telescope \citep{wfirst, akeson2019}, and the Vera C. Rubin Observatory \citep{LSST2009, ivezic2019}, raise many challenges for PSF models as the instruments are getting more complex and the imposed scientific requirements tighter. These factors have triggered and continue to trigger developments in the PSF modelling literature.

\begin{figure}
    \centering
    \includegraphics[width=0.75\textwidth,trim={0cm 0.25cm 0 0},clip,]{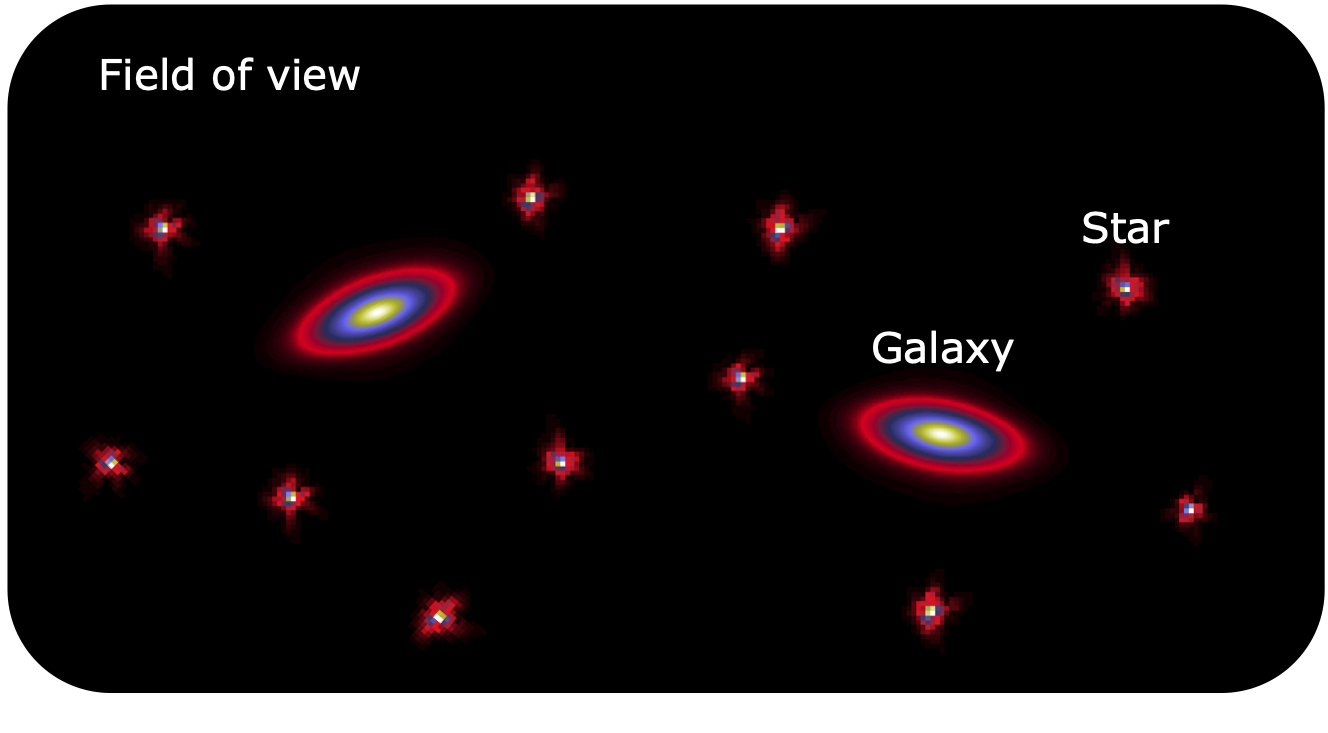}
    \caption{An illustration of a field of view showing the PSF modelling problem. First, the PSF model should be estimated from the stars. The model should then be used to estimate the PSF at the target positions, e.g., galaxy positions.}
    \label{fi:psf_modelling_problem}
\end{figure}

PSF modelling is an interdisciplinary problem which requires knowledge of optics, inverse problems, and the target science application, in our case, weak gravitational lensing studies. The objective is to estimate the PSF at target positions, e.g., galaxy positions, from degraded star observations and complementary sources of information. Figure \ref{fi:psf_modelling_problem} shows an illustration of the problem. The PSF modelling problem is challenging as the model should account for the different variations of the PSF in the field of view, i.e., spatial, spectral and temporal. This review is related to these three scientific fields, it discusses in detail the PSF and aims to help understand the different PSF modelling choices. We start by introducing optical concepts required to analyse optical imaging systems that are required to understand the more physically-based PSF models in Section \ref{sc_02:intro_optics}. Then, motivated by the optical introduction, we describe the adopted general observational model in Section \ref{sc_02:general_observational_forward_model}. Section \ref{sc_02:contributors_PSF} introduces the different contributors to the PSF at the optical and detector level. Section \ref{sc_02:psf_modelling} gives an overview of state-of-the-art PSF modelling techniques and leads to Section \ref{sc_02:PSF_modelling_comments}, which includes comments on the desirable properties of a PSF model. We end the review by describing different techniques for validating PSF models in Section \ref{sc_02:validation_psf} and concluding in Section \ref{sc:conclusions}. In addition, we include Table \ref{tb:variable}, which summarizes the notation and the different coordinates used throughout this article.

\begin{table}
    \begin{center}       
    \begin{tabular}{cl} 
    \toprule
    Variable & \multicolumn{1}{c}{Description}   \\
    \midrule
    \multicolumn{2}{c}{\textit{Coordinates}} \\
    \midrule
    $(x,y)$                 & Pupil plane or output aperture plane coordinates \\
    $(u,v)$                 & Image or focal plane coordinates \\
    $(\xi, \eta)$           & Object plane coordinates \\
    $(\bar{u}, \bar{v})$    & Pixel coordinates, the discrete counterpart of the image plane  \\
    $\mathbf{p}_i$          & 3D spatial coordinate \\
    $\lambda$               & Wavelength  \\
    $t$                     & Time  \\
    \midrule
    \multicolumn{2}{c}{\textit{Notation}} \\
    \midrule
    $\mathcal{I}, \mathcal{H}, \ldots$  & Calligraphic uppercase variables are continuous functions \\
    $I, H, \ldots$                      & Uppercase variables are matrices \\
    $c_m, b_{1}^{k}, \ldots$            & Lowercase variables are scalars \\
    $I_\mathrm{img}(\bar{u},\bar{v};t|u_i,v_i) \in \mathbb{R}$ & Pixel value at position $(\bar{u},\bar{v})$ for the image $I_\mathrm{img}$ with its \\
    & centroid at position $(u_i,v_i)$ observed at time $t$. \\
    $I_{\mathrm{img}, (\cdot|u_i,v_i)} \in \mathbb{R}^{p \times p}$ & Observed image  with its centroid at position $(u_i,v_i)$  \\
    \bottomrule
    \end{tabular}
    \caption{Coordinates and notation used throughout this article.}
    \label{tb:variable}
    \end{center}
\end{table}

\section{Gentle introduction to optics}
\label{sc_02:intro_optics}

A rigorous treatment of the optics involved in the formation of the PSF on complex optical systems could be the sole topic of a review article. In this section we introduce simplified optical concepts to motivate a more physical understanding of the PSF, how to model it, and certain implicit assumptions usually adopted. This review follows the optic formalism of \citet{goodman2005}. For a profound and rigorous description of optical theory, we refer the reader to the seminal book of \citet{born1999_7th_ed} or more concise works \citep{gross2005, hecht2017, gaskill1978}. We refer to \citet{schmidt2010} for more information on practical wave propagation. If the reader is familiar with the Fourier optics literature, we recommend continuing to Section \ref{sc_02:general_observational_forward_model}.

This introduction is based on the scalar formulation of diffraction. It starts by presenting diffraction equations from a general perspective with the Huygens-Fresnel principle to the more simplified formulations of Fresnel and Fraunhofer. The introduction continues with the diffraction analysis of the effects of a thin single-lens optical system. The results motivate the analysis of more general optical systems that are treated with the black box concept from \citet{goodman2005}. The section proceeds by introducing the modelling of aberrations in the optical system, and then extending the monochromatic to the polychromatic analysis briefly studying the coherent and incoherent cases. The optical introduction ends by mentioning several assumptions usually adopted in the PSF modelling literature.

\subsection{Scalar diffraction theory}

\setcounter{subfigure}{0}
\begin{subfigure}
\setcounter{subfigure}{0}
    \centering
    \begin{minipage}[b]{0.98\textwidth}
        \centering
        \includegraphics[width=0.75\linewidth,trim={0cm 0cm 0 0},clip,]{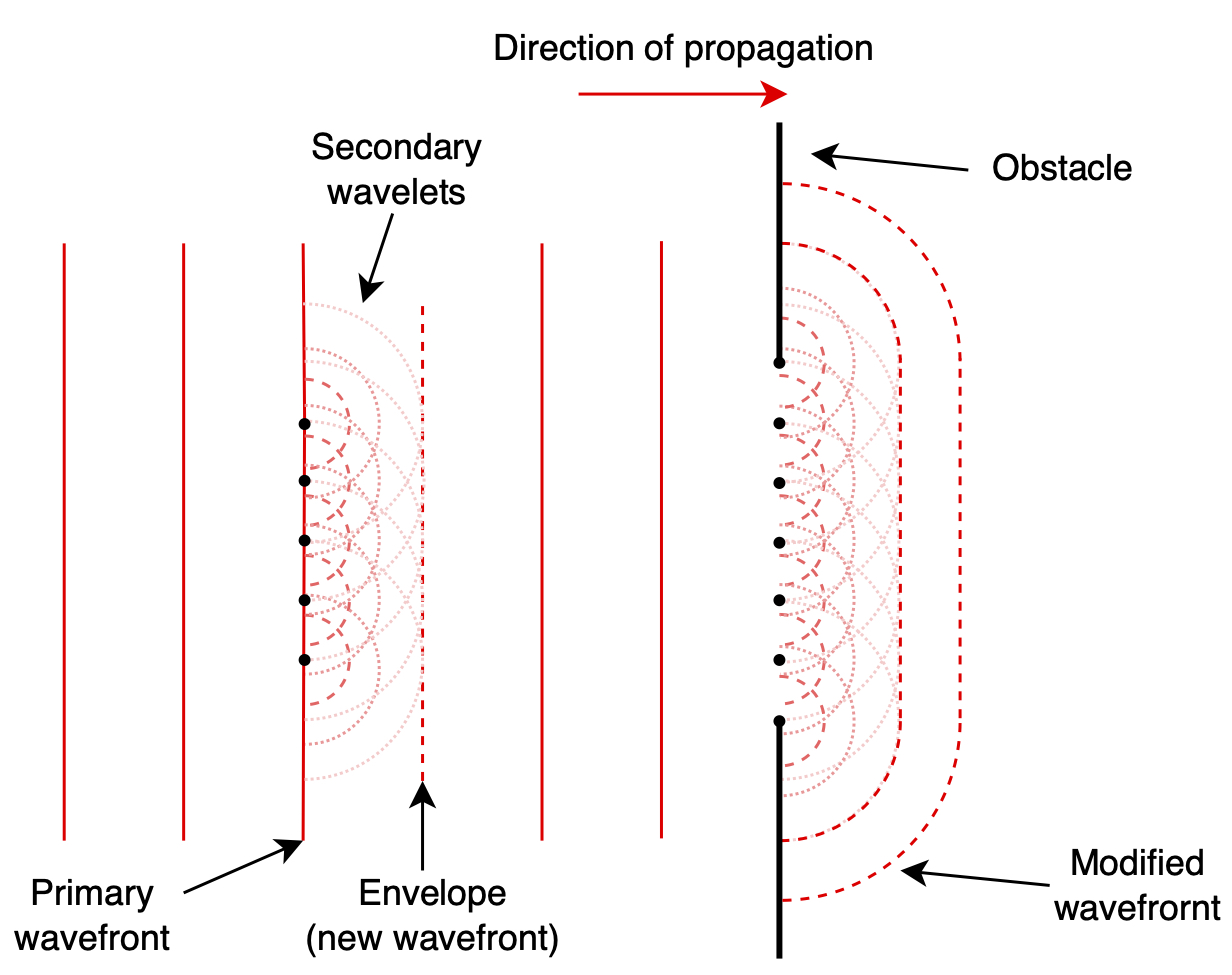}
        \caption{}
        \label{fi_02:huygens_principle}
    \end{minipage}  
\setcounter{subfigure}{1}
    \begin{minipage}[b]{0.98\textwidth}
        \centering
        \includegraphics[width=\linewidth,trim={0cm 0cm 0 0},clip,]{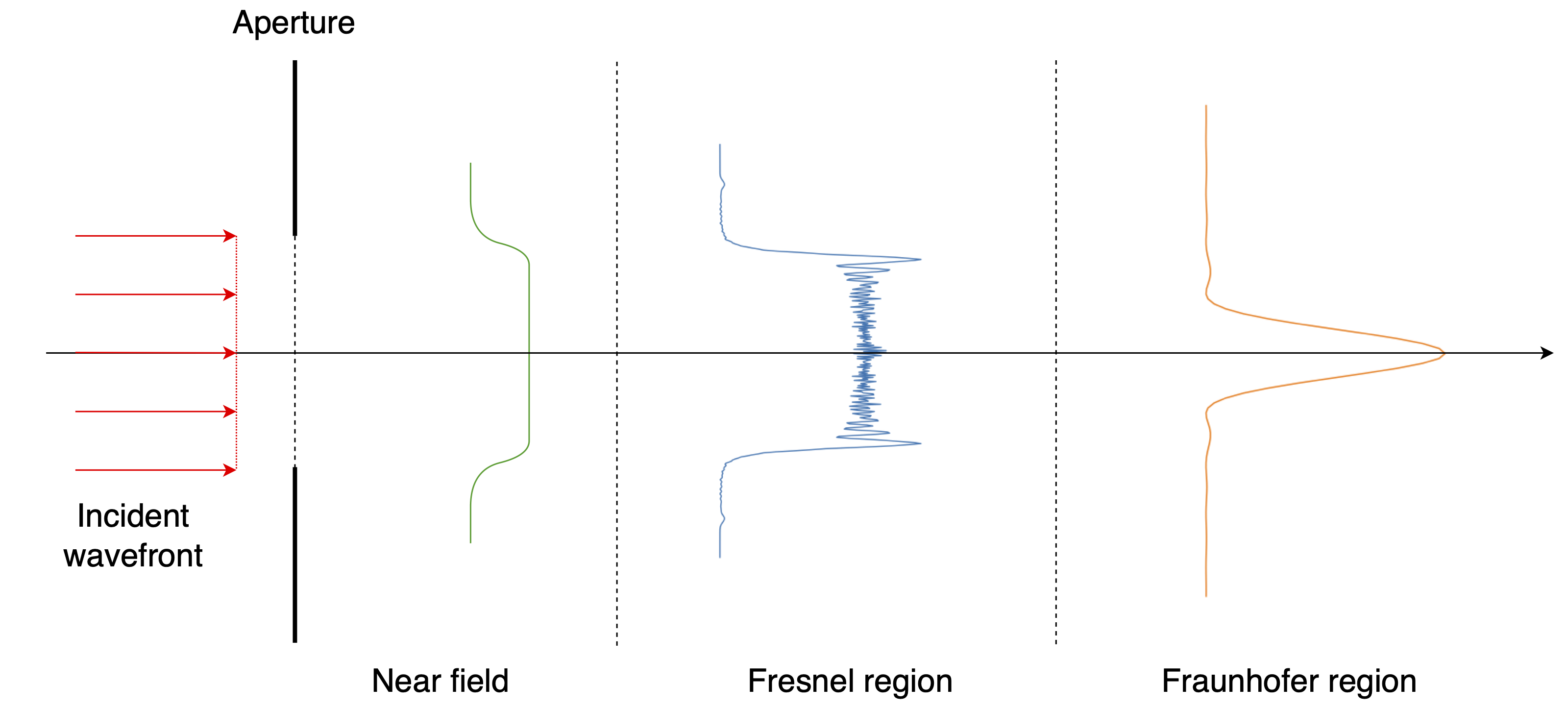}
        \caption{}
        \label{fi_02:diffraction_approximations}
    \end{minipage}
\setcounter{subfigure}{-1}
    \caption{\textbf{(a)} Illustration of the Huygens-Fresnel principle and the modification of a wavefront due to an obstacle. Reproduced from \citet{liaudat2022_thesis}. \textbf{(b)} Illustration of the different diffraction regions behind an aperture.} 
    \label{fi:diffraction_figures}
\end{subfigure}

%%%
\subsubsection{The Huygens-Fresnel principle}

When studying the PSF, we are examining how an optical system with a specific instrument contributes to and modifies our observations. To understand how the optical system interacts with the propagation of light, we need to dig into the nature of light, an electromagnetic (EM) wave. To make a fundamental analysis, one would need to use Maxwell's equations, solve them with the optical system under study, and obtain the electric and magnetic fields. Solving a set of coupled partial differential equations is an arduous task. Several approximations can be made, if some conditions are met, to alleviate the mathematical burden of solving Maxwell's equations without introducing much error into the analysis.

Diffraction theory provides a fundamental framework for analysing light propagation through an optical system. This is especially the case when working with EM waves in the optical range when the optical image is close to the focus region. The \textit{Huygens-Fresnel principle} \citep{huygens1690, fresnel1819, crew1900} states that every point of a wavefront may be considered a secondary disturbance giving rise to spherical wavelets. At any later instant, the wavefront may be regarded as the envelope of all the disturbances. Fresnel's contribution to the principle is that the secondary wavelets mutually interfere. This principle provides a powerful method of analysis of luminous wave propagation. In Figure \ref{fi_02:huygens_principle}, the propagation of an incident plane wavefront through an obstacle, a single slit, is shown. The secondary wavelets constitute the plane wavefront before the obstacle. Then, the wavefront shape is modified due to the obstacle, following the Huygens-Fresnel principle.

The secondary waves mutually interfere constructively or destructively, according to their phases. The analysis of the light propagation in a homogeneous medium is simple as the spherical wavelets interfere without obstacles, and the total wavefront propagates spherically in the medium. However, suppose the wave encounters an obstacle. In that case, the secondary waves in the vicinity of the boundaries of the obstacle will interfere in ways that are not obvious from the incident wavefront. 

\begin{figure}
    \centering
    \includegraphics[width=\textwidth]{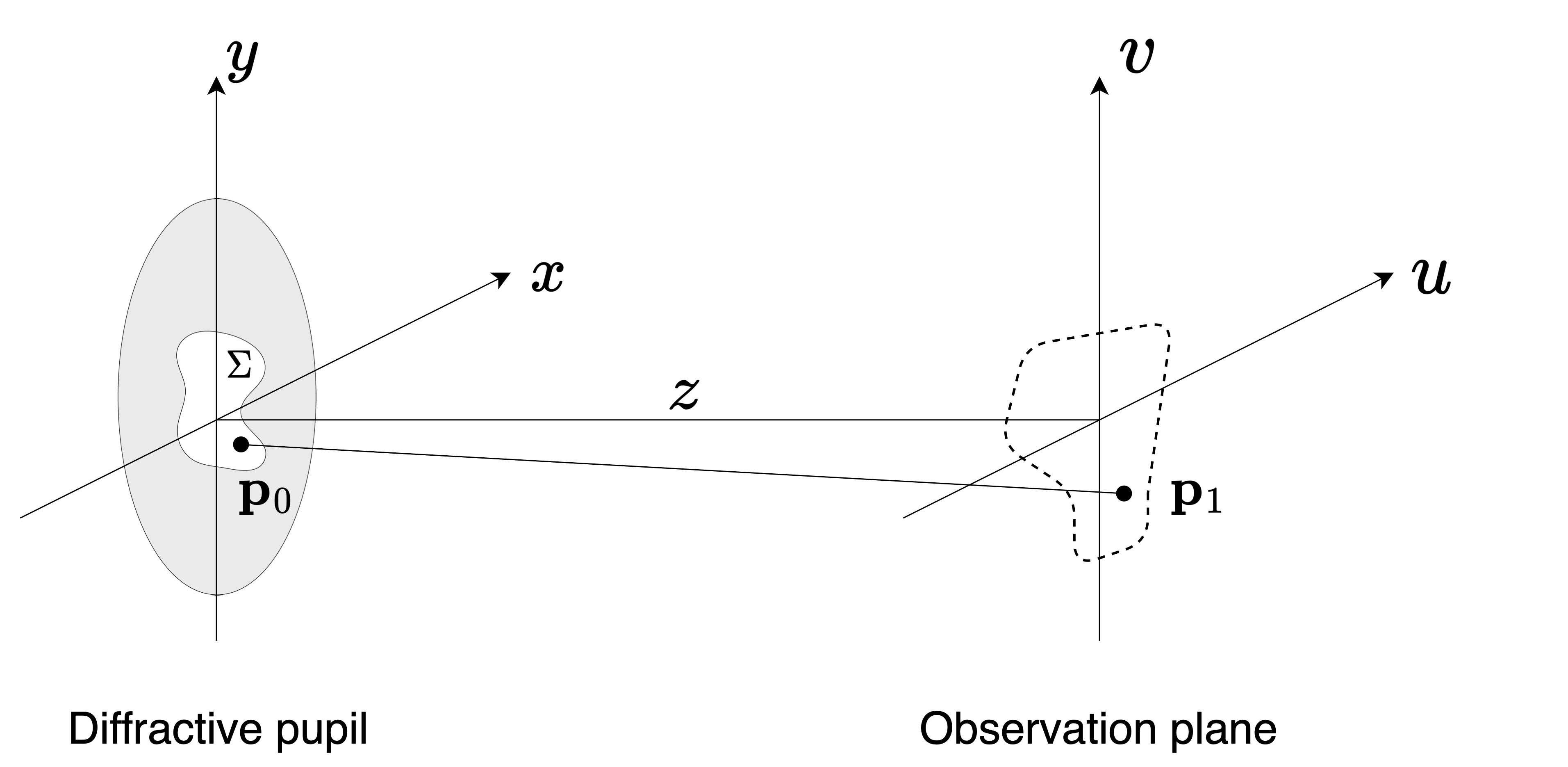}
    \caption{Illustration of the coordinate system for the diffraction equations. Figure adapted from \citet{liaudat2022_thesis}.}
    \label{fi_02:coordinate_system_diffraction}
\end{figure}

Let us study the Huygens-Fresnel principle and consider a diffractive aperture in a plane $(x,y)$ illuminated in the positive $z$ direction. We analyse the diffracted wave in a parallel plane $(u,v)$ at a normal distance $z$ from the first plane. The $z$-axis is orthogonal to both planes and intersects them at their origins. Figure \ref{fi_02:coordinate_system_diffraction} illustrates the coordinate system described above. The diffracted wave, which can be intuitively understood as the superposition of spherical waves, is written as
\begin{equation}
    \mathcal{U}(\mathbf{p}_1) = \frac{z}{\text{j} \lambda} \iint_{\Sigma} \mathcal{U}(x,y;0) \frac{\exp\left[\,\text{j}\, k\, r_{01}\right]}{r_{01}^{2}} \,\mathrm{d} x \mathrm{d} y \,,
    \label{eq_02:r_s_diffraction}
\end{equation}
where $\text{j}$ denotes the imaginary unit, $\lambda$ is the wavelength, $k = 2\pi / \lambda$, $\mathbf{p}_0 = (x_0,y_0;0)$, $\mathbf{p}_1 = (u_1,v_1;z)$, $r_{01} = \| \mathbf{p}_1 - \mathbf{p}_0 \|_{2}$, $\Sigma$ is the aperture in the $(x,y)$ plane, and $\mathcal{U}$ is the electric field. The incident wave is $\mathcal{U}(\mathbf{p}_0)$ and the diffracted wave is $\mathcal{U}(\mathbf{p}_1)$.

There are two main approximations in the derivation of Equation \ref{eq_02:r_s_diffraction}. The first approximation is that we are considering a \textit{scalar theory of diffraction}, a scalar electric and magnetic field, and not the fields in their complete vectorial form. The scalar theory provides a full description of the EM fields in a dielectric medium that is linear, isotropic, homogeneous, and non-dispersive. However, even if the medium verifies these properties, if some boundary conditions are imposed on a wave, e.g., an aperture, some coupling is introduced between the EM field components and the scalar theory is no longer exact. Nevertheless, the EM fields are modified only at the edges of the aperture, and the effects extend over only a few wavelengths into the aperture. Therefore, if the aperture is large compared to the wavelength, the error introduced by the scalar theory is negligible. Refractive optical elements can also induce polarisation of the EM field. The level of accuracy desired will determine if the bias introduced can be neglected or needs to be taken into account.

Although the current formulation is powerful in representing the diffraction phenomena, it is still challenging to work with the integral from Equation \ref{eq_02:r_s_diffraction}. As a consequence, we will explore further approximations that will give origin to the \textit{Fresnel diffraction} and \textit{Fraunhofer diffraction}.

\subsubsection{Fresnel diffraction}
\label{sc:fresnel_approx}
The Fresnel approximation is based on the binomial expansion of the square root in the expression $\sqrt{1 + b}$ for some $b$\footnote{The binomial expansion is given by $\sqrt{1 + b} = 1 + \frac{1}{2} b - \frac{1}{8} b^{2} + \cdots$.}. The distance $r_{01}$ can be expressed as
\begin{equation}
    r_{01} = z \sqrt{1 + \left(\frac{u_1 - x_0}{z}\right)^{2} + \left(\frac{v_1 - y_0}{z}\right)^{2}} \,,
    \label{eq_02:fresnel_full}
\end{equation}
which can be approximated, using the first two terms of the binomial expansion, as
\begin{equation}
    r_{01} \approx z \left( 1 + \frac{1}{2} \left(\frac{u_1 - x_0}{z}\right)^{2} + \frac{1}{2} \left(\frac{v_1 - y_0}{z}\right)^{2} \right) \, .
    \label{eq_02:fresnel_approx}
\end{equation}
The $r_{01}$ appearing in the exponential of Equation \ref{eq_02:r_s_diffraction} has much more influence in the result than the $r_{01}^{2}$ in the divisor. Therefore, we use Equation \ref{eq_02:fresnel_approx} to approximate $r_{01}$ in the exponential, and for the divisor we approximate $r_{01}^{2} \approx z^{2}$. Then, we can express the diffracted field as
\begin{equation}
    \mathcal{U}(u_1,v_1;z) = \frac{\exp\left[\,\text{j} k z\right]}{\text{j} \lambda z} \iint_{\Sigma} \mathcal{U}(x,y;0) \exp\left[\text{j}\, \frac{k}{2 z} \left[ \left( u_1 - x \right)^{2} + \left(v_1 - y \right)^{2}  \right] \right] \,\mathrm{d} x \mathrm{d} y \,,
    \label{eq_02:fresnel_diffraction_first}
\end{equation}
and if we expand the terms in the exponential, we get
\begin{align}
    \label{eq_02:fresnel_diffraction_full}
    \mathcal{U}(u_1,v_1;z) =  \frac{\exp\left[\,\text{j} k z\right]}{\text{j} \lambda z} & \exp\left[\text{j}\, \frac{k}{2 z} \left( u_1^{2} + v_1^{2} \right)\right] \\
    \times \iint_{\Sigma} & \left\{ \mathcal{U}(x,y;0) \exp\left[\text{j}\, \frac{k}{2 z} \left( x^{2} + y^{2} \right)\right] \right\} \exp\left[-\text{j}\, \frac{2 \pi}{\lambda z} \left( u_1 x + v_1 y \right)\right] \mathrm{d} x \mathrm{d} y \,.\nonumber
\end{align}
The Fourier transform (FT) expression can be recognised with some multiplicative factors in Equation \ref{eq_02:fresnel_diffraction_full}. The diffracted wave is the FT of the product of the incident wave and a quadratic phase exponential. In this case, we have approximated the spherical secondary waves of the Huygens-Fresnel principle by parabolic wavefronts. The approximation in the Fresnel diffraction formula is equivalent to the \textit{paraxial approximation} \citet[\S 4.2.3]{goodman2005}. This last approximation consists of a \textit{small-angle approximation} as it restricts the rays to be close to the optical axis. This restriction also allows us to approximate Equation \ref{eq_02:fresnel_full} with Equation \ref{eq_02:fresnel_approx}. The region where the approximation is valid is known as the \textit{region of Fresnel diffraction}. In this region, the major contributions to the integral come from points $(x, y)$ for which $x \approx u$ and $y \approx v$, meaning that the higher-order terms in the expansion that we are not considering are unimportant. The \textit{region of Fresnel diffraction} can be seen as the coordinates $(u,v,z)$ that verify
\begin{equation}
    z^{3} \gg \frac{\pi}{4 \lambda} \left( \left( u - x \right)^{2} +  \left( v - y \right)^{2} \right)^{2} \,, \quad \forall (x, y) \in \Sigma.
\end{equation}
A more practical and widely-used condition is the \textit{Fresnel number} \citep[\S 10.3.3]{hecht2017} which can be written as follows
\begin{equation}
    N_{F} = \frac{r^{2}}{\lambda z}\,,
\end{equation}
where $r$ is the radius of a circular aperture, and $z$ is de distance from the aperture. If $N_{F}$ is close to unity, the Fresnel diffraction is a good approximation. However, if $N_{F} \ll 1$, then Fraunhofer's approximation, which we will introduce in the following section, is valid. For more information on the validity of the Fresnel approximation, we refer the reader to \citet{southwell81,rees1987}.

\subsubsection{Fraunhofer diffraction}
We continue to present a further approximation that, if valid, can significantly simplify the calculations. The Fraunhofer approximation assumes that the exponential term with a quadratic dependence of $(x, y)$ is approximately unity over the aperture. The region where the approximation is valid is the \textit{far field} or \textit{Fraunhofer region}. Figure \ref{fi_02:diffraction_approximations} illustrates the different diffraction approximations as a function of the aperture's distance. The required condition to be in this region reads
\begin{equation}
    z \gg \frac{k \left( x^{2} + y^{2} \right)}{2} \;, \quad \forall (x, y) \in \Sigma.
\end{equation}
The Fraunhofer diffraction formula is given by
\begin{equation}
    \mathcal{U}(u_1,v_1;z) = \frac{\exp\left[\,\text{j} k z\right]}{\text{j} \lambda z} \exp\left[\text{j}\, \frac{k}{2 z} \left( u_1^{2} + v_1^{2} \right)\right] \iint_{\Sigma} \left\{ \mathcal{U}(x,y;0) \right\} \exp\left[-\text{j}\, \frac{2 \pi}{\lambda z} \left( u_1 x + y v_1 \right)\right] \,\mathrm{d} x \mathrm{d} y \,,
    \label{eq_02:fraunhofer_diffraction_full}
\end{equation}
where we can reformulate the previous equation using the FT as follows 
\begin{equation}
    \mathcal{U}(u_1,v_1;z) = \frac{\exp\left[\,\text{j} k z\right]}{\text{j} \lambda z} \exp\left[\,\text{j}\, \frac{k}{2 z} \left( u_1^{2} + v_1^{2} \right)\right] \text{FT}\left\{ \iota_{\Sigma}(x,y;0) \mathcal{U}(x,y;0) \right\}\left(\frac{u_1}{\lambda z}, \frac{v_1}{\lambda z}\right) \,,
    \label{eq_02:fraunhofer_diffraction_FT}
\end{equation}
where $\iota_{\Sigma}$ is an indicator function over the aperture taking values in $\{0, 1 \}$. It is also possible to consider image vignetting and multiply the indicator with a weight function so that the resulting function takes values in $[0, 1]$. Cameras are sensitive to the light's intensity reaching their detectors. The instantaneous intensity of an EM wave is equal to its squared absolute value. Therefore, we can write the intensity of the diffracted wave as
\begin{equation}
    \mathcal{I}(u_1,v_1;z) = \left| \mathcal{U}(u_1,v_1;z) \right|^{2} = \frac{1}{\lambda^{2} z^{2}} \left| \text{FT}\left\{ \iota_{\Sigma}(x,y;0) \mathcal{U}(x,y;0) \right\} \left(\frac{u_1}{\lambda z}, \frac{v_1}{\lambda z}\right)  \right|^{2} \,,
    \label{eq_02:fraunhofer_diffraction_intensity}
\end{equation}
which is significantly simpler than the original Rayleigh-Sommerfeld expression from Equation \ref{eq_02:r_s_diffraction}.

\subsection{Modelling diffraction in a simple optical system}
\label{sc_02:effect_of_lenses}

\begin{figure}
    \centering
    \includegraphics[width=\textwidth]{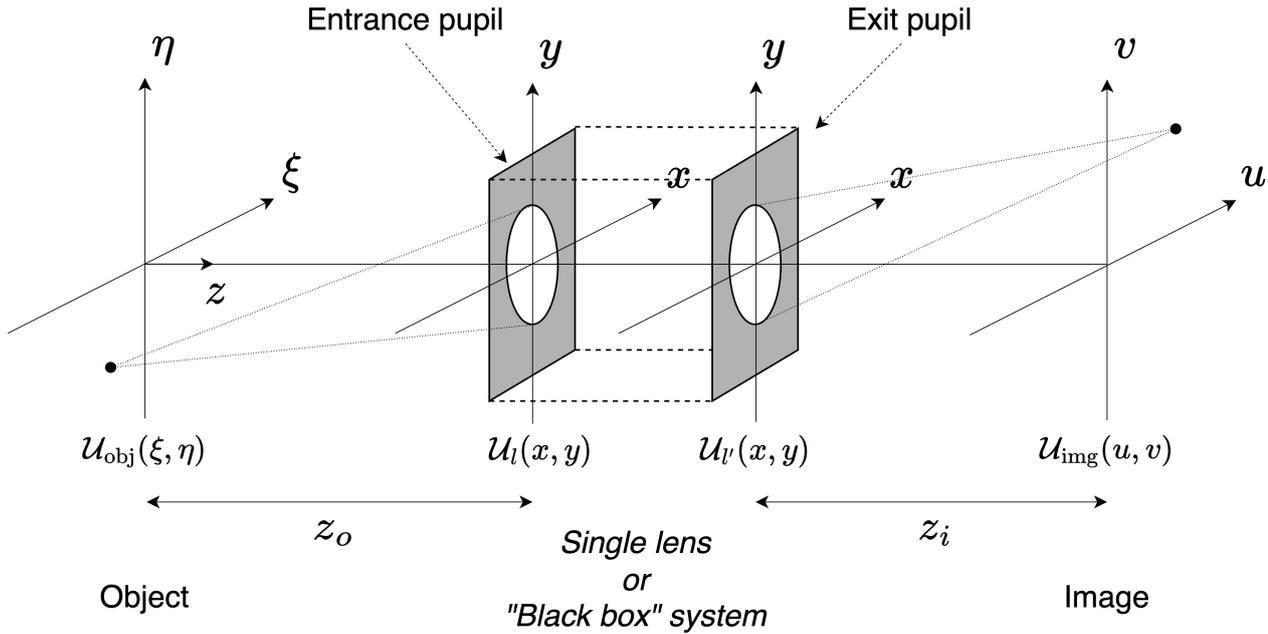}
    \caption{Illustration of the coordinate system of the imaging systems we are studying. The central imaging system can be a single positive lens or the generalised \textit{black box} concept of an imaging system. The image plane coordinates are $(u,v)$, the input and output aperture plane coordinates are $(x,y)$, and the object plane coordinates are $(\xi, \eta)$. This figure has been adapted from \citet{goodman2005}.}
    \label{fi_02:black_box_imaging_system}
\end{figure}

The study of the diffraction phenomena is necessary but not sufficient to describe the effects of an optical system. Optical imaging systems are generally based on lenses or mirrors, which have the ability to form images. To simplify the analysis we study the effect of a single positive (converging) thin lens illuminated with monochromatic illumination and compute the impulse response of such a system. The coordinate system used for this analysis is shown in Figure \ref{fi_02:black_box_imaging_system}.

Let us write the output wave as a function of the input wave using the superposition integral as follows
\begin{equation}
    \mathcal{U}_{\rm img}(u,v) = \iint_{- \infty}^{+ \infty} \mathcal{H}(u,v;\xi,\eta) \, \mathcal{U}_{\rm obj}(\xi,\eta) \mathrm{d}\xi \mathrm{d}\eta \,,
    \label{eq_02:superposition_int_1}
\end{equation}
where $\mathcal{H}$ is the field's value at image coordinates $(u,v)$ due to a unitary point-source object at position $(\xi, \eta)$. 
We describe a monochromatic input wave reaching the entrance pupil of the lens coming from a point source located at $(\xi,\eta)$ at the object plane, which is located at a distance $z_o$ from the lens. Following the \textit{paraxial approximation} we can write the waves at the entrance pupil as follows
\begin{equation}
    \mathcal{U}_l (x,y) = \frac{1}{\text{j} \lambda z_{o}} \exp\left[\, \text{j} \frac{k}{2\, z_o} \left( \left( x - \xi \right)^{2} + \left( y - \eta \right)^{2} \right)\right]\,,
\end{equation}
and the wave at the output pupil is as follows
\begin{equation}
    \mathcal{U}_{l'} (x,y) = \mathcal{U}_l (x,y) \mathcal{P}(x,y) \exp\left[-\text{j} \frac{k}{2 f} \left( x^{2} + y^{2} \right) \right]\,,
\end{equation}
where $f$ is the focal length of the lens and $\mathcal{P}$ is the pupil function of the lens which accounts for the finite dimension of the lens, i.e., the obscured and unobscured areas. We have implicitly assumed that the pupil function is constant for any $(u,v)$ position considered. This assumption does not hold for wide-field imagers where there are obscurations involved in the pupil function. We continue by using the Fresnel diffraction formula from Section \ref{sc:fresnel_approx} to compute the diffraction effect from the lens' exit pupil to the image plane. Replacing $\mathcal{U}$ in Equation \ref{eq_02:fresnel_diffraction_full} with the output lens wave $\mathcal{U}_{l'}$ to compute the impulse response, we obtain
\begin{align}
    \label{eq:impulse_response_full}
    \mathcal{H}(u,v;\xi,\eta) = \frac{1}{\lambda^{2} z_o z_i} & \overbrace{\exp\left[\,\text{j} \frac{k}{2\, z_i} \left( u^{2} + v^{2} \right)\right]}^{\text{\normalsize (II)}} \overbrace{\exp\left[\,\text{j} \frac{k}{2\, z_o} \left( \xi^{2} + \eta^{2} \right)\right]}^{\text{\normalsize (III)}} \\
    & \times \iint_{- \infty}^{ \infty} \mathcal{P}(x,y) \underbrace{\exp\left[\, \text{j} \frac{k}{2} \left( \frac{1}{z_o} + \frac{1}{z_i} - \frac{1}{f} \right) \left( x^{2} + y^{2} \right)\right]}_{\text{\normalsize (I)}} \nonumber \\ 
    & \times \exp\left[-\text{j}k \left( \left( \frac{\xi}{z_o} + \frac{u}{z_i} \right)x + \left( \frac{\eta}{z_o} + \frac{v}{z_i}  \right)y \right)\right] \mathrm{d}x \mathrm{d}y .\nonumber
\end{align}
The previous formula of the impulse response of a positive lens is hard to exploit in a practical sense due to the quadratic phase terms. However, several approximations can be exploited to remove them:

\begin{itemize}
    \item We start studying the term (I) inside the integrand. We consider the image plane to coincide with the focal plane, i.e., $z_i = f$, and the imaged object to be very far away from the entrance pupil. Consequently, the term (I) is approximately one. The part of the exponent which is close to zero is the following one
    \begin{equation}
        \frac{1}{z_o} + \frac{1}{z_i} - \frac{1}{f} \approx 0,
    \end{equation}
    which, in the case of equality, is known as the \textit{lens law} of geometrical optics.
    \item The term (II) only depends on the image coordinates $(u,v)$. The term can be ignored as we are interested in the intensity distribution of the image and it is not being integrated in Equation \ref{eq_02:superposition_int_1}.
    \item The term (III) depends on the object coordinates, is integrated in the convolution operation in Equation \ref{eq_02:superposition_int_1} and therefore, might change significantly the imaged object. We can neglect the influence of this term if its phase changes by a small amount, i.e., a small fraction of a radian, within the region of the object that mostly contributes to the image position $(u,v)$. A deeper discussion about the validity of the term (III) approximation can be found in \citet[\S 5.3.2]{goodman2005} and references therein.
\end{itemize}

We can now apply the previous approximations to the calculation of the impulse response of an optical system with a positive lens. Under Fresnel diffraction, we simplify Equation \ref{eq:impulse_response_full} to obtain
\begin{equation}
    \mathcal{H}(u,v;\tilde{\xi},\tilde{\eta}) \approx \frac{1}{\lambda^{2} z_o f} \iint_{- \infty}^{+ \infty} \mathcal{P}(x,y) \exp\left[-\text{j} \frac{2 \pi}{\lambda f} \left( ( u - \tilde{\xi} ) x + ( v - \tilde{\eta} ) y \right)\right] \mathrm{d}x \mathrm{d}y \,,
    \label{eq:impulse_response}
\end{equation}
where $m = - f / z_o$ is the magnification of the system, which could be positive or negative depending if the image is inverted or not, and the normalized (or reduced) object-plane coordinates are $\tilde{\xi} = m\,\xi$ and $\tilde{\eta} = m\,\eta$. The diffraction pattern is centred on the image coordinates, $u=m\,\xi$ and $v= m\,\eta$, which are the transformed coordinates of the impulse response's position $(\xi,\eta)$. 

The impulse response obtained in Equation \ref{eq:impulse_response} is Fraunhofer's diffraction pattern centred in $(u=\tilde{\xi}, v=\tilde{\eta})$ and up to a scaling factor of $1/\lambda z_o$. This result is a consequence of the choice of $z_i$, such that it verifies the lens law, allowing us to drop out quadratic phase terms in the integral. We have obtained a simple formulation for the impulse response, but the optical system we studied is not used in practice to carry out galaxy imaging surveys. We need to extend the analysis to more general optical systems.

\subsection{Analysis of a general optical imaging system}
\label{sc_02:optical_imaging_systems}

Let us now analyse a general optical imaging system composed of one or many lenses or mirrors of possibly different characteristics. We treat the optical system as a \textit{black box} characterised by the transformations applied to an incident \textit{object} scalar wave, $\mathcal{U}_{\rm obj}$, into an output \textit{image} wave, $\mathcal{U}_{\rm img}$. Figure \ref{fi_02:black_box_imaging_system} illustrates the new interpretation of the optical system, where we have replaced the previous single lens system with a \textit{black box}. In this general model, we assume that the effect of the optical system between the entrance and exit pupils is well described by geometrical optics, which is an affine transformation. We also assume that all the diffraction effects can be associated with one of the two pupils, input or output (see \citet[S. 6.1]{goodman2005} for more discussion on both assumptions). We choose the latter one and consider the diffraction of the output wave between the output pupil and the image plane. For the moment, our analysis continues to assume an ideal monochromatic illumination.

The \textit{ideal image}, $\mathcal{U}_{\rm g}$, is defined as the input image when applied the effect of the geometrical optics inside the \textit{black box} and writes
\begin{equation}
    \mathcal{U}_{\rm g}\left(\tilde{\xi},\tilde{\eta}\right) = \frac{1}{|m|} \mathcal{U}_{\rm obj}\left(\frac{\tilde{\xi}}{m},\frac{\tilde{\eta}}{m}\right)\;, \quad \text{and} \quad \tilde{\xi} = m \xi , \quad \tilde{\eta} = m \eta \,,
\end{equation}  
where $m$ is the magnification factor of the optical system, and we expressed the images in \textit{reduced coordinates}.

Our analysis is based on the impulse response developed in the previous section. The approximations applied and the use of reduced object coordinates have made the system spatially invariant. This fact translates to having $\mathcal{H}(u,v;\tilde{\xi},\tilde{\eta}) = \mathcal{H}(u -\tilde{\xi}, v - \tilde{\eta})$, as the approximated impulse response from Equation \ref{eq:impulse_response} depends only in the difference of the image coordinates and the reduced object coordinates. The impulse response writes
\begin{equation}
    \mathcal{H}(u-\tilde{\xi},v-\tilde{\eta}) = \frac{a}{\lambda f} \iint_{- \infty}^{+ \infty} \mathcal{P}(x,y) \exp\left[-\text{j} \frac{2 \pi}{\lambda f} \left( ( u - \tilde{\xi} ) x + ( v - \tilde{\eta} ) y \right)\right] \mathrm{d}x \mathrm{d}y \,,
    \label{eq:impulse_response_invariant}
\end{equation}
where $a$ is a constant amplitude that does not depend on the optical system under study.
The superposition integral in Equation \ref{eq_02:superposition_int_1} relates the waves at the object an image position with the impulse response in a \textit{spatially variant} system. However, if the system is \textit{spatially invariant}, the equation can be reformulated as the convolution equation, which writes
\begin{equation}
    \mathcal{U}_{\rm img}(u,v) = \iint_{- \infty}^{+ \infty} \mathcal{H}(u -\tilde{\xi}, v - \tilde{\eta}) \, \mathcal{U}_{\rm g}(\tilde{\xi},\tilde{\eta}) \; \mathrm{d}\tilde{\xi} \, \mathrm{d}\tilde{\eta} \,.
    \label{eq_02:superposition_int_approx_2}
\end{equation}
The previous equation can be rewritten with the usual convolution notation as
\begin{equation}
    \mathcal{U}_{\rm img}(u,v) = \left( \mathcal{U}_{\rm g} \star \mathcal{H} \right) (u,v)\,.
    \label{eq:psf_convolution}
\end{equation}
In this general case of a system without aberrations and under the aforementioned approximations, we see that the output image is formed by a geometrical-optics transformation followed by a convolution with an impulse response from the Fresnel diffraction of the exit aperture.

\subsubsection{Introducing optical aberrations}
\begin{figure}
    \centering
    \includegraphics[width=0.8\textwidth]{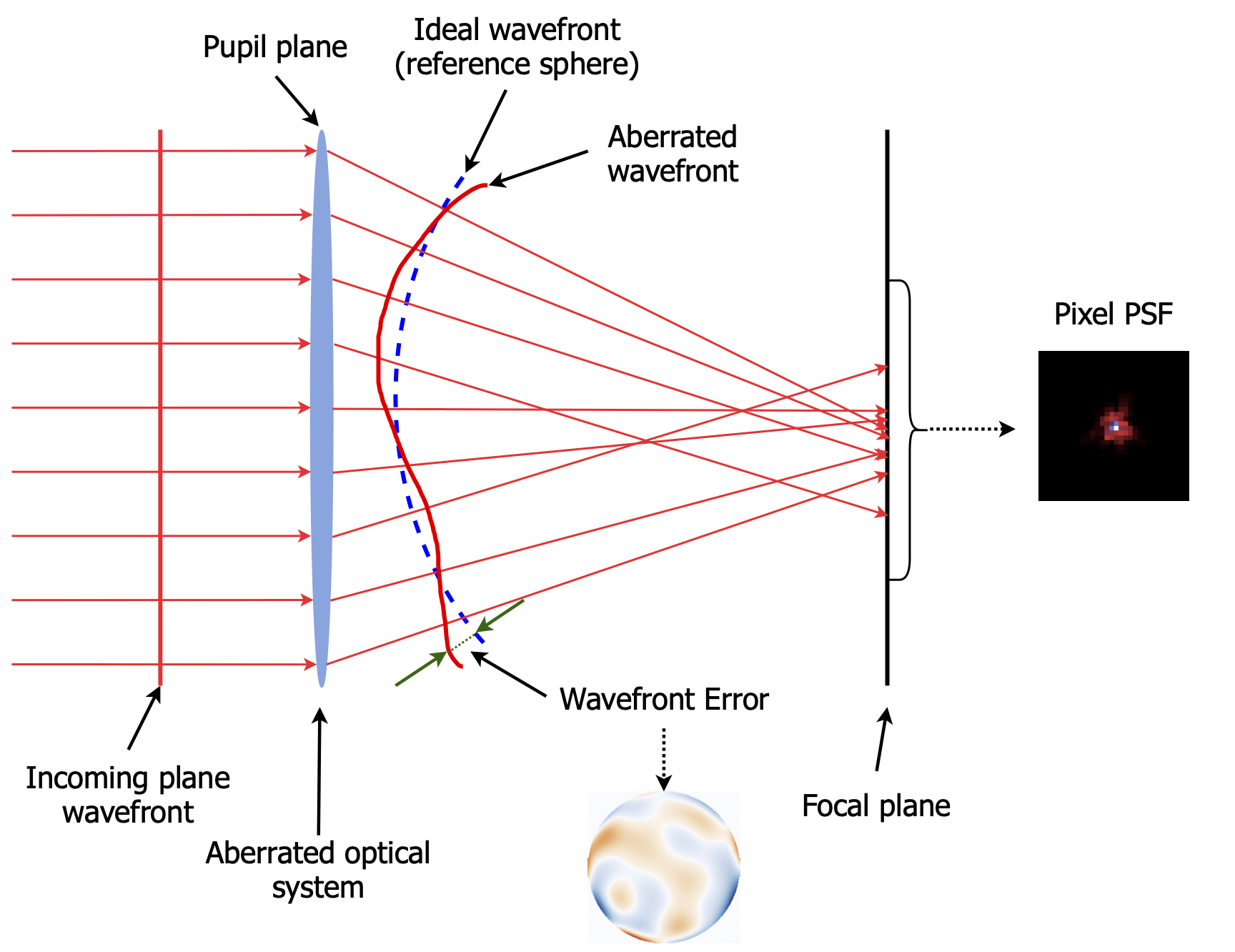}
    \caption{Illustration of the wavefront error in a one-dimensional projection of an ideal setting where the optical system is represented as a single lens. Figure reproduced from \citet{liaudat2023}.}
    \label{fi_02:wfe_illustration}
\end{figure}

In the previous development, we considered an ideal optical system without any aberrations, known as \textit{diffraction-limited}. An aberrated optical system produces the imperfect convergence of rays, which can be expressed equivalently in wavefront space by deviations from the ideal reference sphere. The aberrations produce leads and lags in the wavefront with respect to the ideal sphere, see Figure \ref{fi_02:wfe_illustration}. A complementary interpretation, from \citet{goodman2005}, is that we start with the previous diffraction-limited system producing converging spherical wavefronts. Then, we add a phase-shifting plate representing the system's aberrations. The plate is located in the aperture after the exit pupil and affects the output wave's phase. To characterise the aberrations, we will use the \textit{generalised pupil function} that generalises the pupil function $\mathcal{P}$ from Equation \ref{eq:impulse_response_invariant} and writes
\begin{equation}
    \mathcal{P}_{\rm gen}(x,y;u,v) = \mathcal{P}(x,y;u,v) \exp\left[\text{j} \frac{2 \pi}{\lambda} \mathcal{W}(x,y;u,v)\right] \,,
    \label{eq_02:generalized_pupil_function}
\end{equation}
where $\lambda$ is the central wavelength of the incident wave, $\mathcal{P}$ is the pupil function including the telescope's obscurations, and $\mathcal{W}$ represents the optical path differences (OPD) between a perfectly spherical and the aberrated wavefront. We also refer to the OPD as the wavefront error (WFE). Figure \ref{fi_02:wfe_illustration} illustrates the concept of WFE. It is common to represent the WFE using a Zernike polynomial decomposition \citep{noll1976} as they are orthogonal in the unit disk and we generally use circular apertures in telescopes and optical systems in general. Figure \ref{fi:zernike_example} shows the first Zernike polynomials. 

\setcounter{subfigure}{0}
\begin{subfigure}
\setcounter{subfigure}{0}
    \centering
    \begin{minipage}[b]{0.28\textwidth}
        \centering
        \includegraphics[width=0.85\linewidth,trim={0cm 0cm 0 0},clip,]{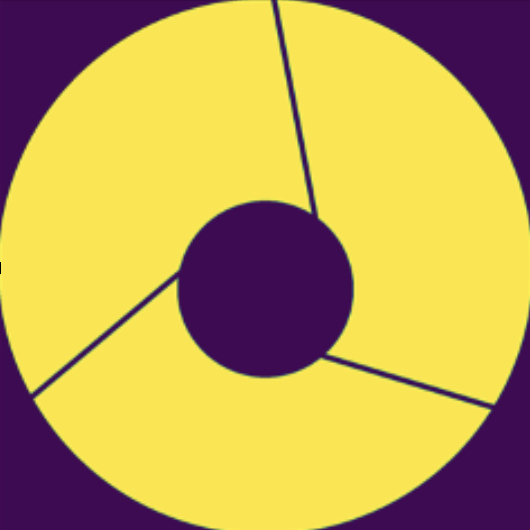}
        \caption{}
        \label{fi:obscuration_example}
    \end{minipage}  
\setcounter{subfigure}{1}
    \begin{minipage}[b]{0.7\textwidth}
        \centering
        \includegraphics[width=0.90\linewidth,trim={0cm 0cm 0 0},clip,]{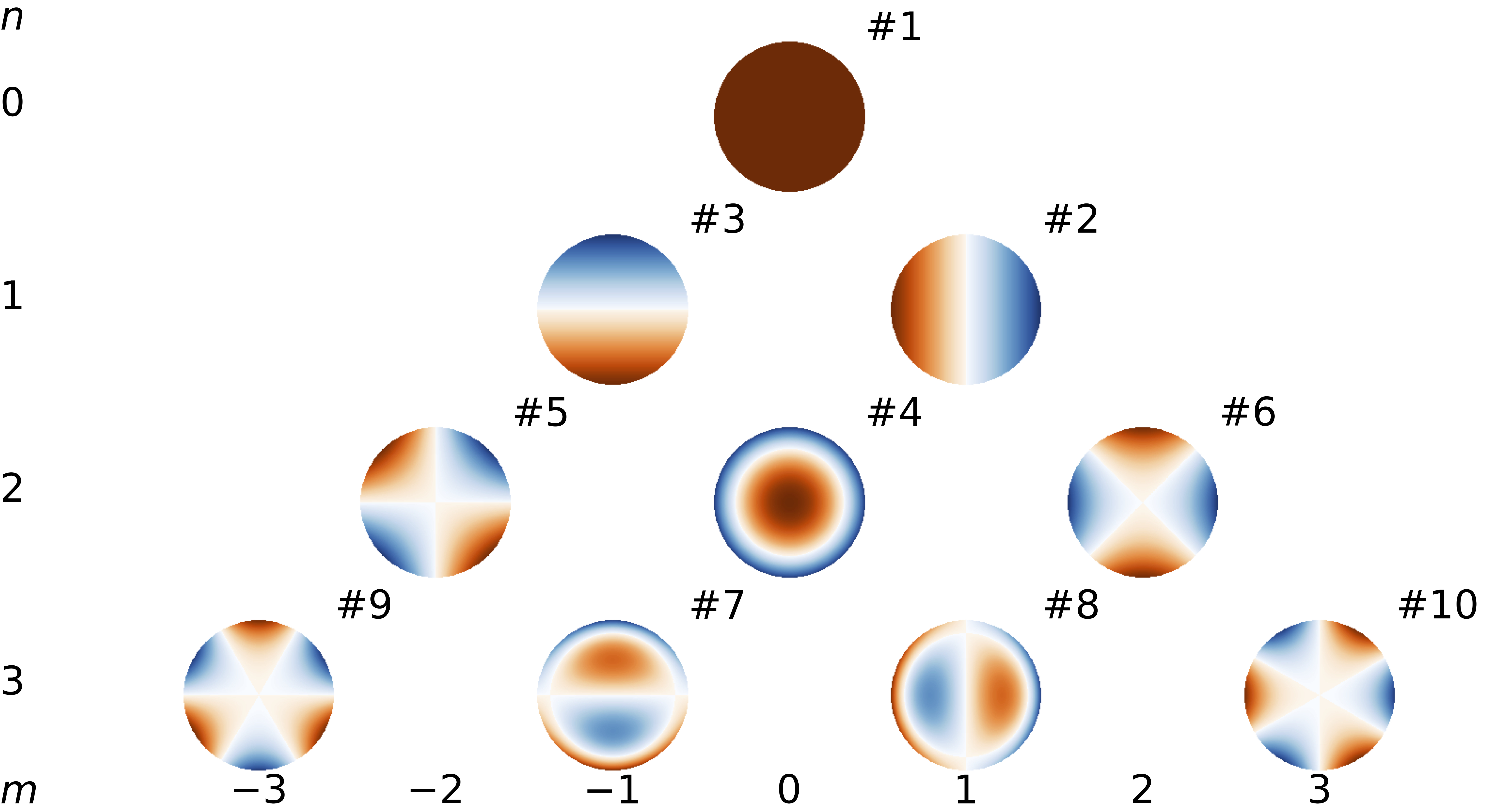}
        \caption{}
        \label{fi:zernike_example}
    \end{minipage}
\setcounter{subfigure}{-1}
    \caption{\textbf{(a)} An illustration of \textit{Euclid}'s pupil function, which can be seen in \citet{venancio2020}, in the $(x,y)$ plane for a given position in the $(\xi,\eta)$. \textbf{(b)} Example of the first Zernike polynomial maps.} 
    \label{fi:zernike_and_obscurations}
\end{subfigure}

The aberrations, $\mathcal{W}$, and the pupil function, $\mathcal{P}$, depend on the object's position in the focal plane as is seen in the $(u,v)$ coordinate dependence in Equation \ref{eq_02:generalized_pupil_function}. Large telescopes with wide focal planes have spatially varying aberrations. The path travelled by the light rays changes considerably between distant points in the focal plane, changing the aberrations, $\mathcal{W}$, too. The obscurations and the aperture, represented by the pupil function $\mathcal{P}$, also change with the focal plane position. For example, Figure \ref{fi:obscuration_example} illustrates the obscurations from the \textit{Euclid} telescope. One can notice a circular aperture with several obscurations in it, a secondary mirror and three arms supporting the mirror. What we observe in Figure \ref{fi:obscuration_example} is a $2$D projection of a $3$D structure. This projection changes as a function of the focal plane position we are analysing, making the function $\mathcal{P}$ dependent on the $(u,v)$ coordinates.

In the impulse response of the optical system without aberrations from Equation \ref{eq:impulse_response_invariant}, we had a spatially invariant system. This invariance allowed us to use the convolution rather than the superposition integral, which is a computationally practical formulation. If we now consider aberrations, we must inject the generalized pupil function appearing in Equation \ref{eq_02:generalized_pupil_function} into the impulse response formula from Equation \ref{eq:impulse_response_invariant}. The addition of the $(u,v)$ dependency in $\mathcal{P}_{\text{gen}}$ makes the impulse response $\mathcal{H}$ spatially variant again.

\begin{figure}
    \centering
    \includegraphics[width=0.85\textwidth]{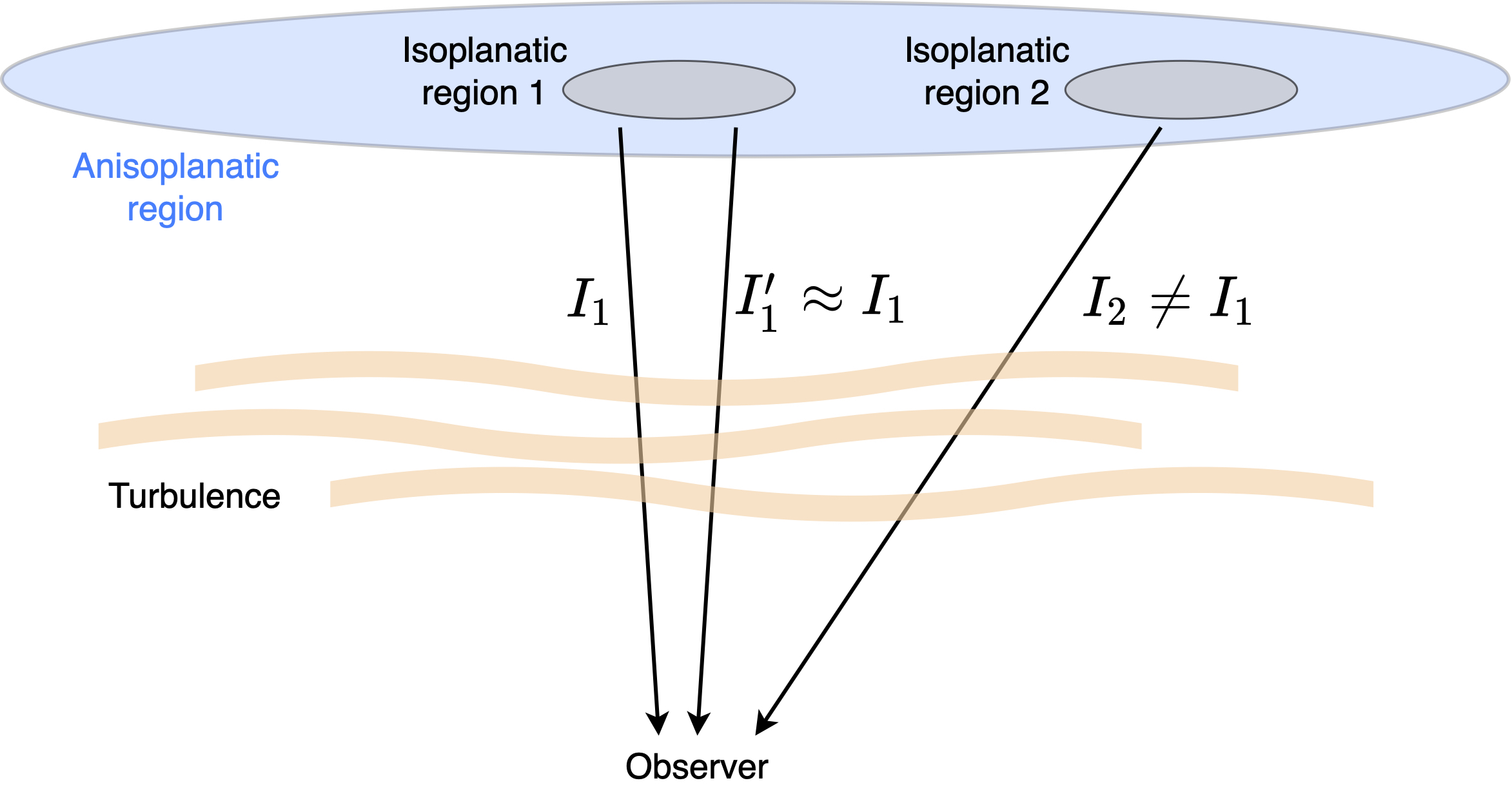}
    \caption{Illustration of isoplanatic regions. Two rays from the same isoplanatic region travel through almost the same turbulence and suffer almost the same distortions. Figure reproduced from \citet{liaudat2022_thesis}.}
    \label{fi_02:isoplanatic_region}
\end{figure}

The study of $\mathcal{H}$, the impulse response and the main topic of this review, is strongly spatially variant in systems with a large focal plane. Nevertheless, we can consider $\mathcal{H}$ spatially invariant in its \textit{isoplanatic region}. This region consists of close-by points in the focal plane, where the light has travelled similar paths giving small deviations of $\mathcal{H}$. We are assuming a certain regularity in $\mathcal{H}$ due to the optical system under study that allows the deviations to be small. In other words, we consider $\mathcal{H}$ to be locally spatially invariant or spatially invariant in patches. Figure \ref{fi_02:isoplanatic_region} illustrates the idea of an \textit{isoplanatic region}. This local invariance assumption limits the size of the imaged objects under study, as they should have a certain size range with respect to the support of $\mathcal{H}$ so that all the objects being imaged lie within the aforementioned region. We consider the generalized pupil function evaluated at the centroid of the imaged object, $(u_i, v_i)$, and note the locally spatially invariant generalized pupil function as follows
\begin{equation}
    \mathcal{P}_{\rm gen}^{*}(x,y|u_i,v_i) = \mathcal{P}^{*}(x,y|u_i,v_i) \exp\left[\text{j} \frac{2 \pi}{\lambda} \mathcal{W}^{*}(x,y|u_i,v_i)\right] \,.
    \label{eq_02:generalized_pupil_function_invariant}
\end{equation}
Injecting Equation \ref{eq_02:generalized_pupil_function_invariant}, instead of Equation \ref{eq_02:generalized_pupil_function}, to the impulse response in Equation \ref{eq:impulse_response_invariant} reads as follows
\begin{align}
    \mathcal{H}(u-\tilde{\xi},v-\tilde{\eta}|u_i,v_i) = \frac{a}{\lambda f} \iint_{- \infty}^{+ \infty} \mathcal{P}^{*}(x,y|u_i,&v_i) \exp\left[\text{j} \frac{2 \pi}{\lambda} \mathcal{W}^{*}(x,y|u_i,v_i)\right] \nonumber \\
    & \times \exp\left[-\text{j} \frac{2 \pi}{\lambda f} \left( ( u - \tilde{\xi} ) x + ( v - \tilde{\eta} ) y \right)\right] \mathrm{d}x \mathrm{d}y \,,
    \label{eq:impulse_response_aberrated_invariant}
\end{align}
where we have made the system spatially invariant again allowing us to exploit the convolution formula in Equation \ref{eq_02:superposition_int_approx_2}.

We have considered aberrations that only depend on the object's position in the focal plane, also known as achromatic aberrations. However, depending on the optical system under study, there might be wavelength-dependent aberrations. For example, some refractive components, or some components implementing complex thin film coatings, may introduce spurious spectral dependences to the optical system's response. If this is the case, we can add a wavelength dependence to the WFE function $\mathcal{W}$ to account for these effects.

\subsubsection{Polychromatic illumination: the coherent and the incoherent case}
\label{sc:poly_illumination}

We studied until now a system with ideal monochromatic light. It is time to shift to polychromatic light as telescopes have filters with finite bandwidths and hence allow multiple frequencies of light. For a more rigorous analysis of polychromatic illumination, we refer the reader to the theory of partial coherence \citet[\S 10]{beran1964, goodman1985, born1999_7th_ed}. Even if we study the system's behaviour to light with a particular wavelength, this is practically never the case, as real illumination is never perfectly chromatic, even for lasers. Therefore, we consider a narrowband polychromatic illumination centred at a given wavelength $\lambda$. The \textit{narrowband assumption} states that the bandwidth occupied is small with respect to the central wavelength. For polychromatic light, we follow \citet{goodman2005} and consider a time-varying phasor of the field, $\mathcal{U}_{\rm img}(u,v;t)$, where its intensity is given by the time integration of its instantaneous intensity
\begin{equation}
    \mathcal{I}_{\rm img}(u,v) = \left\langle \left| \mathcal{U}_{\rm img}(u,v;t) \right|^{2} \right\rangle_{t} = \frac{1}{T} \int_{-T/2}^{T/2} \left| \mathcal{U}_{\rm img}(u,v;t) \right|^{2} \mathrm{d}t \,,
    \label{eq_02:intensity_poly}
\end{equation}
where $T$ is the detector integration time that is considered much greater than the optical wave period. We can generalise the field expression from Equation \ref{eq_02:superposition_int_approx_2} by considering that light is polychromatic and that the impulse response $\mathcal{H}$ is wavelength independent due to the narrowband assumption. The field then writes
\begin{equation}
    \mathcal{U}_{\rm img}(u,v;t) = \iint_{- \infty}^{+ \infty} \mathcal{H}\left(u -\tilde{\xi}, v - \tilde{\eta}\right) \mathcal{U}_{\rm g}\left(\tilde{\xi}, \tilde{\eta}; t - \tau \right) \; \mathrm{d}\tilde{\xi} \mathrm{d}\tilde{\eta} \,,
    \label{eq_02:superposition_int_approx_poly}
\end{equation}
where $\tau$ represents the delay of the wave propagation from $(\tilde{\xi}, \tilde{\eta})$ to $(u,v)$. Continuing with the polychromatic analysis, we rewrite the intensity from Equation \ref{eq_02:intensity_poly} as
\begin{equation}
    \mathcal{I}_{\rm img}(u,v) = \iint\limits_{- \infty}^{+ \infty} \mathrm{d}\tilde{\xi}_{1} \mathrm{d}\tilde{\eta}_{1} \iint\limits_{- \infty}^{+ \infty}  \mathrm{d}\tilde{\xi}_{2} \mathrm{d}\tilde{\eta}_{2} \mathcal{H}\left(u -\tilde{\xi}_{1}, v - \tilde{\eta}_{1}\right) \mathcal{H}^{*}\left(u -\tilde{\xi}_{2}, v - \tilde{\eta}_{2}\right) \mathcal{J}_{\rm g}\left( \tilde{\xi}_{1}, \tilde{\eta}_{1}; \tilde{\xi}_{2}, \tilde{\eta}_{2}\right),
    \label{eq_02:poly_intensity}
\end{equation}
where $\mathcal{H}^{*}$ is the conjugate of $\mathcal{H}$, $\mathcal{J}_{\rm g}$ is known as the \textit{mutual intensity} which describes the spatial coherence of $\mathcal{U}_{\rm g}$ at two points and writes
\begin{equation}
    \mathcal{J}_{\rm g}\left( \tilde{\xi}_{1}, \tilde{\eta}_{1}; \tilde{\xi}_{2}, \tilde{\eta}_{2}\right) = \left\langle \mathcal{U}_{\rm g}\left(\tilde{\xi}_{1}, \tilde{\eta}_{1}; t \right) \mathcal{U}_{\rm g}^{*}\left(\tilde{\xi}_{2}, \tilde{\eta}_{2}; t \right) \right\rangle \,.
\end{equation}
We can distinguish two types of illuminations, \textit{coherent} and \textit{incoherent}. \textit{Coherent} illumination refers to waves whose phases vary in a perfectly correlated way. This illumination is approximately the case of a laser. In \textit{incoherent} illumination, the wave's phases vary in an uncorrelated fashion. Most natural light sources can be considered incoherent sources. The \textit{mutual intensity} is helpful to represent both types of illumination. In the case of coherent light, we obtain,
\begin{equation}
    \mathcal{J}_{\rm g}^{\text{co}}\left( \tilde{\xi}_{1}, \tilde{\eta}_{1}; \tilde{\xi}_{2}, \tilde{\eta}_{2}\right) = \mathcal{U}_{\rm g}\left(\tilde{\xi}_{1}, \tilde{\eta}_{1} \right) \mathcal{U}_{\rm g}^{*}\left(\tilde{\xi}_{2}, \tilde{\eta}_{2} \right) \,,
    \label{eq_02:coherent_mutual_intensity}
\end{equation}
where $\mathcal{U}_{\rm g}\left(\tilde{\xi}_{1}, \tilde{\eta}_{1} \right)$ and $\mathcal{U}^{*}_{\rm g}\left(\tilde{\xi}_{2}, \tilde{\eta}_{2} \right)$ are time-independent phasor amplitudes relative to their time-varying counterpart. As both time-varying phasors are synchronized, we have taken a reference phasor and normalized them against their amplitude with respect to a reference point that can be the origin $(0,0)$. For example, 
\begin{equation}
    \mathcal{U}_{\rm g}\left(\tilde{\xi}_{1}, \tilde{\eta}_{1}; t \right) = \mathcal{U}_{\rm g}\left(\tilde{\xi}_{1}, \tilde{\eta}_{1} \right) \frac{ \mathcal{U}_{\rm g}\left(0, 0; t \right)}{ \left\langle \left|\mathcal{U}_{\rm g}\left(0, 0; t \right) \right|^{2}  \right\rangle^{\frac{1}{2}}}  \,.
\end{equation}
Substituting Equation \ref{eq_02:coherent_mutual_intensity} into Equation \ref{eq_02:poly_intensity} we obtain
\begin{align}
    \label{eq_02:output_intensity_coherent}
    \mathcal{I}_{\rm img}^{\text{co}}(u,v) =& \left| \mathcal{U}_{\rm img}^{\text{co}}(u, v) \right|^{2} = \left| \iint_{- \infty}^{+ \infty} \mathcal{H}\left(u -\tilde{\xi}, v - \tilde{\eta}\right) \mathcal{U}_{\rm g}\left(\tilde{\xi}, \tilde{\eta} \right) \mathrm{d}\tilde{\xi} \mathrm{d}\tilde{\eta} \right|^{2} \,, \\
    \mathcal{I}_{\rm img}^{\text{co}}(u,v) =& \left| \left( \mathcal{U}_{\rm g} \star \mathcal{H} \right)(u,v) \right|^{2} \,, \nonumber
\end{align}
where we observe the \textit{coherent illumination gives a linear system in the complex amplitude of the field} $\mathcal{U}_{\rm g}$. The previous result is related to the interference of coherent waves. If we now consider incoherent illumination, the mutual intensity writes
\begin{equation}
    \mathcal{J}_{\rm g}^{\text{in}}\left( \tilde{\xi}_{1}, \tilde{\eta}_{1}; \tilde{\xi}_{2}, \tilde{\eta}_{2}\right) = \kappa \, \mathcal{I}_{g}\left(\tilde{\xi}_{1}, \tilde{\eta}_{1} \right) \delta\left(\tilde{\xi}_{1} - \tilde{\xi}_{2} , \tilde{\eta}_{1} - \tilde{\eta}_{2} \right) \,,
    \label{eq_02:incoherent_mutual_intensity}
\end{equation}
where $\kappa$ is a real constant, $\delta$ is Dirac delta distribution, and $\mathcal{I}_{\rm g}$ is the intensity of the $U_{\rm g}$ field. The constant $\kappa$ is a result of a simplification from statistical optics giving origin to Equation \ref{eq_02:incoherent_mutual_intensity}. The constant depends on the degree of the extension of coherence when the evanescent-wave phenomenon \citep{beran1964} is taken fully into account. If the coherence extends over a wavelength, $\kappa$ is equal to $\bar{\lambda}^{2} / \pi$, where $\bar{\lambda}$ is the mean wavelength. See \citet[\S 5.5.2]{goodman1985} for a deeper discussion on incoherent illumination and the $\kappa$ constant. Replacing Equation \ref{eq_02:incoherent_mutual_intensity} in Equation \ref{eq_02:poly_intensity} the output (image) intensity writes
\begin{align}
    \label{eq_02:output_intensity_incoherent}
    \mathcal{I}_{\rm img}^{\text{in}}(u,v) &= \left| \mathcal{U}_{\rm img}^{\text{in}}(u, v) \right|^{2} = \kappa \iint_{- \infty}^{+ \infty} \left| \mathcal{H}\left(u -\tilde{\xi}, v - \tilde{\eta}\right) \right|^{2} \mathcal{I}_{\rm g}\left(\tilde{\xi}, \tilde{\eta} \right) \mathrm{d}\tilde{\xi} \mathrm{d}\tilde{\eta} \,, \\
    \mathcal{I}_{\rm img}^{\text{in}}(u,v) &= \kappa \left( \mathcal{I}_{\rm g} \star \left| \mathcal{H} \right|^{2} \right)(u,v) = \kappa \left( \mathcal{I}_{\rm g} \star \mathcal{H}_{\text{int}} \right)(u,v) \,, \nonumber
\end{align}
where $\mathcal{H}_{\text{int}} = \left| \mathcal{H} \right|^{2}$ is the \textit{intensity impulse response}, also known as the PSF. In this case, \textit{an optical system illuminated with incoherent light is linear in intensity}. Equation \ref{eq_02:output_intensity_incoherent} shows a commonly exploited fact; the output intensity is the convolution of the intensity PSF with ideal image intensity $\mathcal{I}_{\rm g}$.

\subsection{Usual assumptions adopted in PSF modelling}
PSF modelling articles generally implicitly assume specific hypotheses. We provide some of them on the following list:
\begin{itemize}
    \item The scalar diffraction theory is valid.
    \item The lens law is verified, the paraxial approximation is valid, and the approximations discussed in Section \ref{sc_02:effect_of_lenses} hold. These approximations allow us to discard quadratic phase terms from Fresnel's diffraction and exploit the simpler Fraunhofer diffraction formula.
    \item The incoming light from natural sources is assumed to be ideally incoherent. Then, the optical system is linear in intensity, as seen in Equation \ref{eq_02:output_intensity_incoherent}.
    \item The PSF is considered to be spatially invariant in its isoplanatic region. In other words, PSF is assumed not to change on objects' typical length scales. This assumption allows us to use the convolution equation, i.e. Equation \ref{eq_02:output_intensity_incoherent}, rather than the superposition integral, i.e., Equation \ref{eq_02:superposition_int_1}.
\end{itemize}
Although the previous assumptions are standard, certain precision levels require dropping simplifications. For example, the \textit{Euclid} mission requirements on the PSF model accuracy as read in \citet{laureijs2011} and \citet{racca2016} is of $2 \times 10^{-4}$ for the root mean square (RMS) error on each ellipticity component ($\delta e_{i}^{\text{PSF}}$), and $1 \times 10^{-3}$ for the relative RMS error on the size ($\delta R_{\text{PSF}}^{2} / R_{\text{PSF}}^{2}$). The PSF model might need to include light polarisation to fulfil these extremely tight PSF requirements. Other assumptions might also be dropped for the precise imaging of widespread objects. This case might require discarding the spatially invariant assumption of the PSF or reducing the size of the isoplanatic region.

To conclude, the usual formulation of the PSF, i.e., the intensity of the impulse response, convolving an image seen in many articles, comes from the previous assumptions using the results from Equation \ref{eq_02:generalized_pupil_function_invariant}, Equation \ref{eq:impulse_response_aberrated_invariant} and Equation \ref{eq_02:output_intensity_incoherent}. We rewrite this formula as follows 
\begin{equation}
    \mathcal{I}_{\rm img}(u,v) = \left( \mathcal{H}_{\text{int}} \star \mathcal{I}_{g} \right)(u,v) \,,
    \label{eq_02:basic_psf_convolution}
\end{equation}
where we remind the reader that $(u,v)$ is the image plane, we have dropped the $\kappa$ term from Equation \ref{eq_02:output_intensity_incoherent}, and $\mathcal{H}_{\text{int}}$ is the intensity impulse response or PSF that writes
\begin{align}
    \mathcal{H}_{\text{int}}(u,v|u_i,v_i) = \frac{{a'}^{2}}{\lambda^{2} \, f^{2}} \bigg| \iint_{- \infty}^{+ \infty} \mathcal{P}^{*}(x,y|u_i,v_i)  \exp\bigg[\text{j} \frac{2 \pi}{\lambda} \mathcal{W}^{*}&(x,y|u_i,v_i)\bigg]  \nonumber \\
    & \times \exp\bigg[-\text{j} \frac{2 \pi}{\lambda f}  \left( u x + v y \right)\bigg] \mathrm{d}x \mathrm{d}y \bigg|^{2},
    \label{eq_02:intensity_psf}
\end{align}
where we are studying the PSF for a specific wavelength and focal plane position.

%%% 
\section{General observational model}
\label{sc_02:general_observational_forward_model}

We consider the PSF as the intensity impulse response, $\mathcal{H}_{\text{int}}$, of the imaging system under study to a point source. The concept of PSF \citep{born1999_7th_ed} is used throughout many imaging applications, including astronomical imaging \citep{liaudat2023,schmitz2019}, medical imaging \citep{dougherty2001,joyce2018}, or microscopy \citep{soulez2012,debarnot2021,debarnot2021_b}. The central idea behind a PSF is that it represents transformations done to the imaged object by the imaging system. The PSF is, in a certain way, a characterisation of the imaging system. Considering incoherent illumination and that the hypotheses from the previous section hold, we can affirm that the optical system behaves linearly as in Equation \ref{eq_02:basic_psf_convolution}. Consequently, the PSF is considered the impulse response of the optical system and affects the ground truth image through a convolution operation. Focusing on astronomical imaging, the definition of the imaging system can vary between the different use cases and telescopes. For example, in a ground-based telescope, we will consider that the atmosphere belongs to the imaging system we are modelling. However, naturally, the atmosphere will not be considered in a space-based telescope. This article will focus on optical systems, which work with electromagnetic radiation with a wavelength close to the visible spectrum. For example, \textit{Euclid} VIS instrument's theoretical wavelength range is from $550$nm to $900$nm.

The PSF describes the effects of the imaging system in the imaging process of the object of interest. The PSF is a convolutional kernel, as we have seen in Section \ref{sc_02:optical_imaging_systems}. However, this convolutional kernel varies spatially, spectrally, and temporally. We give a non-exhaustive list that motivates each of these variations:

\begin{itemize}
    \item \textit{Spatial variations:} The optical system presents a certain \textit{optical axis}, which is an imaginary line where the system has some degree of rotational symmetry. In simpler words, it can be considered as the direction of the light ray that produces a PSF in the centre of the focal plane for an unaberrated optical system. The angle of incidence is defined as the angle between an incoming light ray and the optical axis. The main objective of the optical systems we study is to make the incoming light rays converge in the focal plane, where there will be some measurement instruments, e.g., a camera. Depending on the angle of incidence, the image will form in different positions in the focal plane. The path of the incoming light will be different for each angle of incidence, and therefore the system's response will be different too. In other words, the PSF will change depending on the angle of incidence or spatial position in the focal plane where the image is forming. Optical systems with wide focal planes, generally associated with wide field-of-views (FOVs), present significant PSF spatial variations.
    \item \textit{Spectral variations:} Principally due to the diffraction phenomena and its well-known wavelength dependence covered in Section \ref{sc_02:intro_optics}. Refractive\footnote{Refraction refers to the change of direction in the propagation of a wave passing from one medium to another. Most of the wave energy is transmitted to the new medium. Reflection refers to the abrupt change of direction of the wave propagation due to a boundary between mediums. In this last case, most of the oncoming wave energy remains in the same medium.} components of the optical system under study are also a source of spectral variations \citep{baron2022}. Other sources of spectral variations are detector electronic components \citep{meyers2015_b} and atmospheric chromatic effects \citep{meyers2015}.
    \item \textit{Temporal variations:} The state of the telescope changes with respect to time; therefore, the imaged object's transformation also changes. In space-based telescopes, high-temperature gradients cause mechanical dilations and contractions that affect the optical system. In ground-based telescopes, the atmosphere composition changes with time. Consequently, it temporally affects the response of the optical system, i.e., the PSF.
\end{itemize}

The PSF convolutional kernel varies with space, time and wavelength. Once we have set up a specific wavelength and time to analyse our system, we will have a different convolutional kernel for each position in the field of view.
Let us refer to the PSF field $\mathcal{H}_{\text{int}}$ as all the PSFs representing an optical system. Then $\mathcal{H}_{\text{int}}(u,v;\lambda;t|u_i,v_i)$ is a specific PSF where $(u_i,v_i)$ represents its centroid, i.e., the first order moments. The same notation is maintained from Figure \ref{fi_02:black_box_imaging_system}, where the $(u,v)$ variables represent the image plane. We can define the PSF field as a varying convolutional kernel $\mathcal{H}_{\text{int}}: \mathbb{R}^{2} \times \mathbb{R}_{+} \times \mathbb{R}_{+} \times \mathbb{R}^{2} \to \mathbb{R}$. This definition would accurately describe how the PSF affects the images considering the assumptions from Section \ref{sc_02:intro_optics} are valid. We recall that we adopted the approximation that considers the PSF locally invariant in its \textit{isoplanatic region} \citep[\S~9.5.1]{born1999_7th_ed}, see Figure \ref{fi_02:isoplanatic_region} for an illustration. This approximation means that in the vicinity of an observed object, we will consider that the PSF only varies with time and wavelength, thus facilitating the computation of the convolution. The close vicinity, or the isoplanatic region, will be defined as the postage stamp to image the object of interest. The typical galaxies observed for weak lensing have a comparable size with respect to the PSF size (see \citet[Figure 7]{mandelbaum2018_bis2} for a distribution of relative galaxy to PSF size in the HSC survey). Consequently, the approximation error is kept low as it is only done for small patches of the focal plane.

Let us define our object of interest with the subscript ground truth (GT), $\mathcal{I}_\mathrm{GT}(u,v;\lambda;t|u_i,v_i)$, that is the $I_{g}$ object from Section \ref{sc_02:intro_optics}, as a continuous light distribution $\mathcal{I}_\mathrm{GT}: \mathbb{R}^{2} \times \mathbb{R}_{+} \times \mathbb{R}_{+} \times \mathbb{R}^{2} \to \mathbb{R}$. In this review, we are not considering transient objects, i.e., the time dependence scale of the object is comparable with the exposure time used to image it. Therefore, we can ignore the temporal dependency of the GT object, $\mathcal{I}_\mathrm{GT}(u,v;\lambda;t) \neq f(t)$. Let us write our general observational model that relates our GT object of interest, our PSF and our observed image as follows
\begin{equation}
    I_\mathrm{img}(\bar{u},\bar{v};t|u_i,v_i) = \mathcal{F}_{p} \left\{ \int_{0}^{+\infty} \mathcal{T}(\lambda) \; (\mathcal{I}_\mathrm{GT} \star \mathcal{H}_{\text{int}})(u,v;\lambda;t|u_i,v_i) \; \mathrm{d} \lambda \right\} \circ N(\bar{u},\bar{v};t|u_i,v_i) \,,
    \label{eq_02:general_forward_obs_model}
\end{equation}
where $\mathcal{F}_{p}$ is a degradation operator discretising the image to $\mathbb{R}^{p \times p}$ which includes the image sampling from the instrument. The variables $(\bar{u},\bar{v})$ denote the discrete (pixelised) version of the $(u,v)$ variables. Then, $I_\mathrm{img}(\bar{u},\bar{v};t|u_i,v_i) \in \mathbb{R}$ corresponds to the instrument's measurement at a single pixel $(\bar{u},\bar{v})$, and $I_{\mathrm{img}, (\cdot;t|u_i,v_i)} \in \mathbb{R}^{p \times p}$ to the entire image. The variables $(u_i, v_i)$ correspond to the centre location of the target object $i$. The instrument's transmission is represented by $\mathcal{T}: \mathbb{R}_{+} \to \mathbb{R}_{+}$, a function with finite support, and $N_{(\cdot;t|u_i,v_i)} \in \mathbb{R}^{p \times p}$ corresponds to the noise affecting our observation and possibly a modelling error, where $\circ$ is some composition operator. We have carried out the spectral integration \citep{hopkins1957, eriksen2018} on the instrument's passband defined in $\tau$. Although Equation \ref{eq_02:general_forward_obs_model} provides a general observational model, it can be unpractical. The continuous functions $\mathcal{H}_{\text{int}}$, $\mathcal{T}$, and $\mathcal{I}_{\text GT}$ are practically inaccessible. We make several assumptions to simplify the problem: 
\begin{itemize}
    \item[(a)] the continuous functions $\mathcal{H}_{\text{int}}$ and $\mathcal{I}_{\text GT}$ are well approximated by piece-wise constant functions over a regular grid in $\mathbb{R}^{2}$. We assume $\mathcal{H}_{\text{int}} \approx H$ and $\mathcal{I}_{\text GT} \approx I_{\text GT}$, where $H, I_{\text GT} \in \mathbb{R}^{P \times P}$ with $P \geq p$. The resolution of these two variables has to be greater or equal to the observation resolution,
    \item[(b)] the noise is additive, i.e. $\circ \equiv +$, although the formulation could be adapted to consider other types of noise, e.g., Poisson,
    \item[(c)] the degradation operator is approximated by its discrete counterpart, $\mathcal{F}_p \approx F_p$, where $F_p: \mathbb{R}^{P \times P} \to \mathbb{R}^{p \times p}$, that has been discretized in a regular grid. We assume that the degradation operator is \textit{linear}, and that includes pixellation, possibly downsampling, intra-pixel shifts and linear detector effects,
    \item[(d)] we keep the approximation that the PSF is locally constant within the postage stamp of $P \times P$ values of the target image,
    \item[(e)] the integral can be well approximated by a discretised version using $n_{\lambda}$ bins.
\end{itemize}
Taking into account the aforementioned assumptions, we can define our practical observational model as follows
\begin{equation}
    I_\mathrm{img}(\bar{u},\bar{v};t|u_i,v_i) = {F}_{p} \left\{ \sum_{k=1}^{n_{\lambda}} T(\lambda_k) \; ({I}_\mathrm{GT} \star {H})(\bar{u},\bar{v};\lambda_k;t|u_i,v_i) \; \Delta \lambda_k \right\} + N(\bar{u},\bar{v};t|u_i,v_i) \,,
    \label{eq_02:approx_forward_obs_model}
\end{equation}
where $I_{\mathrm{img},(\cdot;t|u_i,v_i)} \in \mathbb{R}^{p \times p}$, $T$ is a discretized version of $\mathcal{T}$, and $b^{k} = [b^{k}_{0}, b^{k}_{1}]$ is the $k$-th wavelength bin centred in $\lambda_k$, with a width of $\Delta \lambda_k = b^{k}_{1} - b^{k}_{0}$.

\subsection*{Particular case: a star observation}
\begin{figure}
    \centering
    \includegraphics[width=0.9\textwidth,trim={0cm 0.75cm 0 0},clip,]{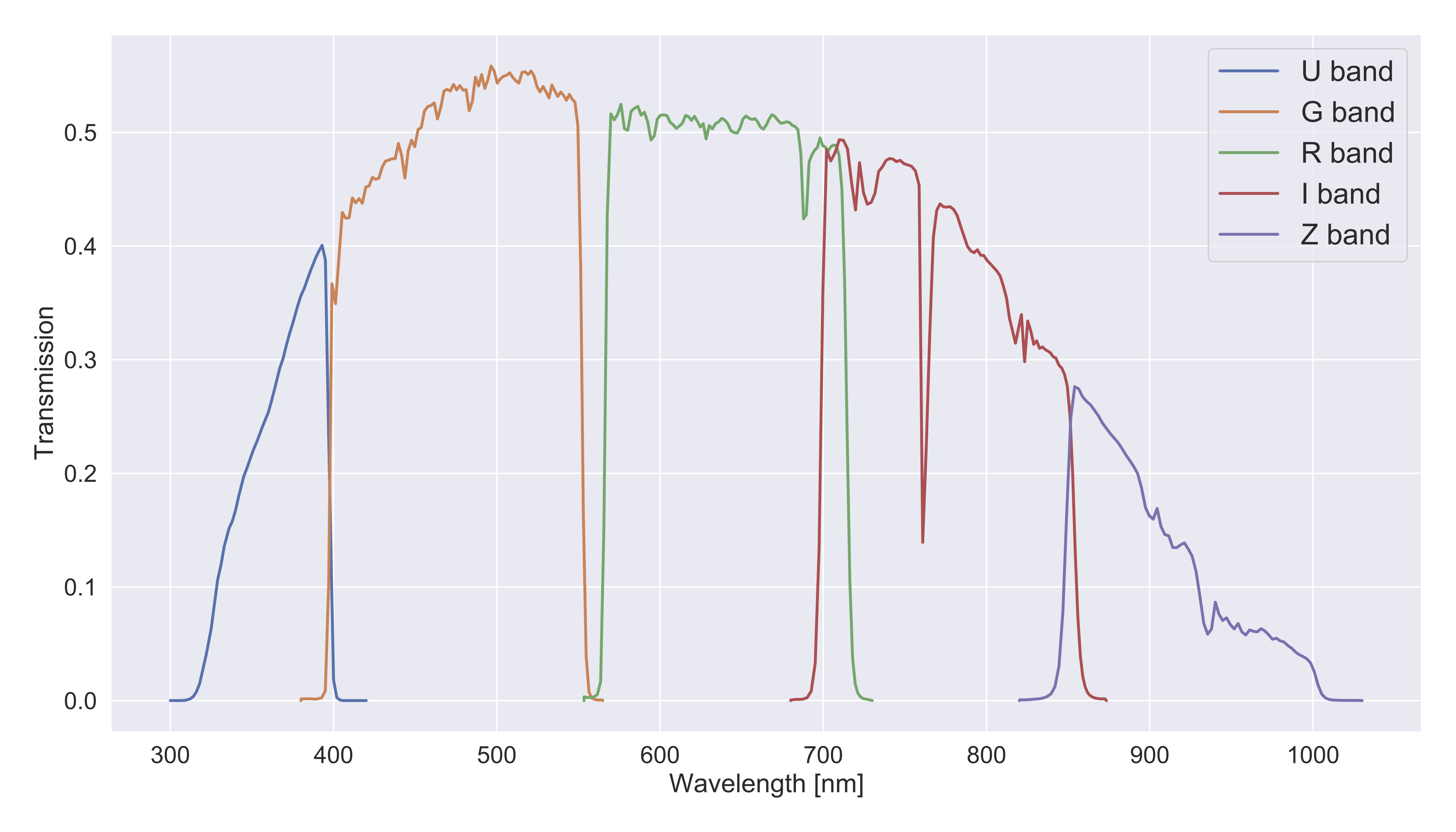}
    \caption{The $3$rd generation set of filters of the MegaCam instrument at Canada-France-Hawaii Telescope that is currently being used for the Canada-France Imaging Survey. The transmission filter response includes the full telescope and $1.25$ airmasses of atmospheric attenuation. The full telescope includes mirrors, optics, and CCDs.}
    \label{fi_02:megacam_bands}
\end{figure}
The case of star observations is of particular interest, as some stars in the FOV can be considered as a spatial impulse, i.e., $\mathcal{I}_\mathrm{star}(u,v;\lambda|u_i,v_i) = \delta(u_i,v_i;\lambda) = f_{(u_i,v_i)}(\lambda)$. Therefore, if we plug the impulse in Equation \ref{eq_02:approx_forward_obs_model}, we obtain a degraded observation of the PSF field. These observations will be crucial to constrain the PSF models. Unluckily, we do not always have access to the star's spectral variation, $f_{(u_i,v_i)}(\lambda)$. However, we dispose of complementary photometric observations that can be useful to characterise the spectral variations. These observations provide us with the star's spectral energy distribution (SED), which can be defined as the calibrated flux density as a function of wavelength, usually at low spectral resolution. The photometric observations are done in several spectral bands. Figure \ref{fi_02:megacam_bands} shows the bands from the MegaCam instrument at the Canada-France-Hawaii Telescope (CFHT)\footnote{The filter curves can be downloaded from the Spanish Virtual Observatory (SVO) webpage, \href{http://svo2.cab.inta-csic.es/svo/theory/fps/index.php}{http://svo2.cab.inta-csic.es/svo/theory/fps/index.php} .}. We refer the reader to \citet{hogg2022} for more information about SEDs and stellar photometry. The SED is a normalised low-resolution sampling of the star's spectral variations. We can write the SED definition we will use as
\begin{equation}
    \text{SED}_{b^{k}}(\lambda_k)= \frac{1}{z_{n_{\lambda}(b)}} \int_{b^{k}_{0}}^{b^{k}_{1}} f_{(u_i,v_i)}(\lambda) \;\mathrm{d} \lambda \,,
    \label{eq_02:sed_approx_model}
\end{equation}
where we continued to use the $b^k$ bin definition from Equation \ref{eq_02:approx_forward_obs_model}, and $z_{n_{\lambda}(b)}$ is a constant used so that the SED is normalised to unity. We have that $\sum_{k=1}^{n_{\lambda}} \text{SED}_{b^{k}}(\lambda_k) = 1$. We continue by considering that the GT image in Equation \ref{eq_02:approx_forward_obs_model} is a star, and we use the spectral bins from the SED definition to discretise the spectral integration. Finally, we write the practical star observation model as
\begin{equation}
    I_{\text{star}}(\bar{u},\bar{v};t|u_i,v_i) = {F}_{p} \left\{ \sum_{k=1}^{n_{\lambda}} T(\lambda_k) \text{SED}_{b^{k}}(\lambda_k)  \; {H}(\bar{u},\bar{v};\lambda_k;t|u_i,v_i) \; \Delta \lambda_k \right\} + N(\bar{u},\bar{v};t|u_i,v_i) \,,
    \label{eq_02:approx_star_obs_model}
\end{equation}
where we consider the star observation $I_{\text{star}, (\cdot;t|u_i,v_i)} \in \mathbb{R}^{p \times p}$ as a degraded version of the PSF field $\tilde{H}_{(\cdot;t|u_i,v_i)} \in \mathbb{R}^{p \times p}$.

%%% 
\section{PSF field contributors and related degradations}
\label{sc_02:contributors_PSF}
So far, we have described how the PSF interacts with the images we observe and how we can model an observation. However, we have not given much information about the different PSF field contributors and the different degradations represented by $F_p$ in Equation \ref{eq_02:approx_forward_obs_model} that can occur when modelling observations. We provide a non-exhaustive list of contributors to the PSF field, sources of known degradations, and the atmosphere's effect on our PSF modelling problem. 

\subsection{Image coaddition}
\label{sc_02:img_coaddition}
A fundamental contributor to the PSF is the choice of image coaddition scheme. A \textit{coadded image} is a composite image created by combining multiple individual exposures of the same region of the sky in some way. This process can help increase the signal-to-noise ratio of the observation. Motivated by the analysis of the LSST data, \citet{mandelbaum2022} explores different coaddition schemes and studies how they affect the PSF of the resulting coadded image. In particular, \citet{mandelbaum2022} define under which schemes the coadded image accepts a well-defined PSF, i.e., the observation can be described by the convolution of an extended object and a uniquely defined coadded PSF. \citet{bosch2017} describes the strategy for image and PSF coaddition in the HSC survey.  

\subsection{Dithering and super-resolution}
\label{sc_02:dithering}
Dithering consists of taking a series of camera exposures shifted by a fractional or a few pixel amount. There are several advantages of using a dithering strategy, which include the removal of cosmic rays and malfunctioning pixels, the improvement of photometric accuracy, the filling of gaps between the detectors, and the improvement of the observed scene's sampling. Dithering allows estimating a sampling density of the images that is denser than the original pixel grid, in other words, to super-resolve the image. Regarding PSFs, it allows recovering Nyquist sampled PSF from undersampled observations. \citet{bernstein2002} studied the effect of dithering and the choice of pixel sizes in imaging strategies. Naturally, as we will later see, the dithering strategy is helpful for space-based telescopes thanks to their stability. In ground-based telescopes, the atmosphere constantly changes the PSF, making a dithering strategy less effective. However, a dithering strategy can be helpful if the telescope is equipped with adaptive optics technology, which will be described in Section \ref{sc_02:adaptive_optics}. An example is the Spectro-Polarimetic High contrast imager for Exoplanets REsearch (SPHERE) instrument \citep{beuzit2019} built for the European Southern Observatory's (ESO) Very Large Telescope (VLT) in Chile.

\citet{lauer1999_b} discusses the limiting accuracy effect of undersampled PSFs in stellar photometry and proposes ways to correct it with dithered data \citep{lauer1999_a}. \citet{fruchter2002} presents the widely-used \textsc{Drizzle} algorithm that consists of shifting and adding the dithered images onto a finer grid.
\citet{rowe2011} proposed a linear coaddition method coined \textsc{Imcom} to obtain a super-resolved image from several undersampled images. \citet{hirata2023} later studied the use of \textsc{Imcom} on simulations \citep{troxel2022} from the \textit{Roman} space telescope, while a companion paper, \citet{yamamoto2023}, explored its implications for weak lensing analyses. \citet{ngole2015} proposed a super-resolution method coined \textsc{SPRITE} targeting the \textit{Euclid} mission based on a sparse regularisation technique. More recent PSF models handle the undersampling of the observations directly in their algorithm for estimating a well-sampled PSF field, as we will later see.

\subsection{Optic-level contributors}
These contributors affect the PSF by modifying the wave propagation in the optical system. In other words, they affect the wavefront's amplitude and phase.

\setcounter{subfigure}{0}
\begin{subfigure}
    \setcounter{subfigure}{0}
        \centering
        \begin{minipage}[b]{0.32\textwidth}
            \includegraphics[width=\linewidth,trim={1cm 1cm 0 0},clip,]{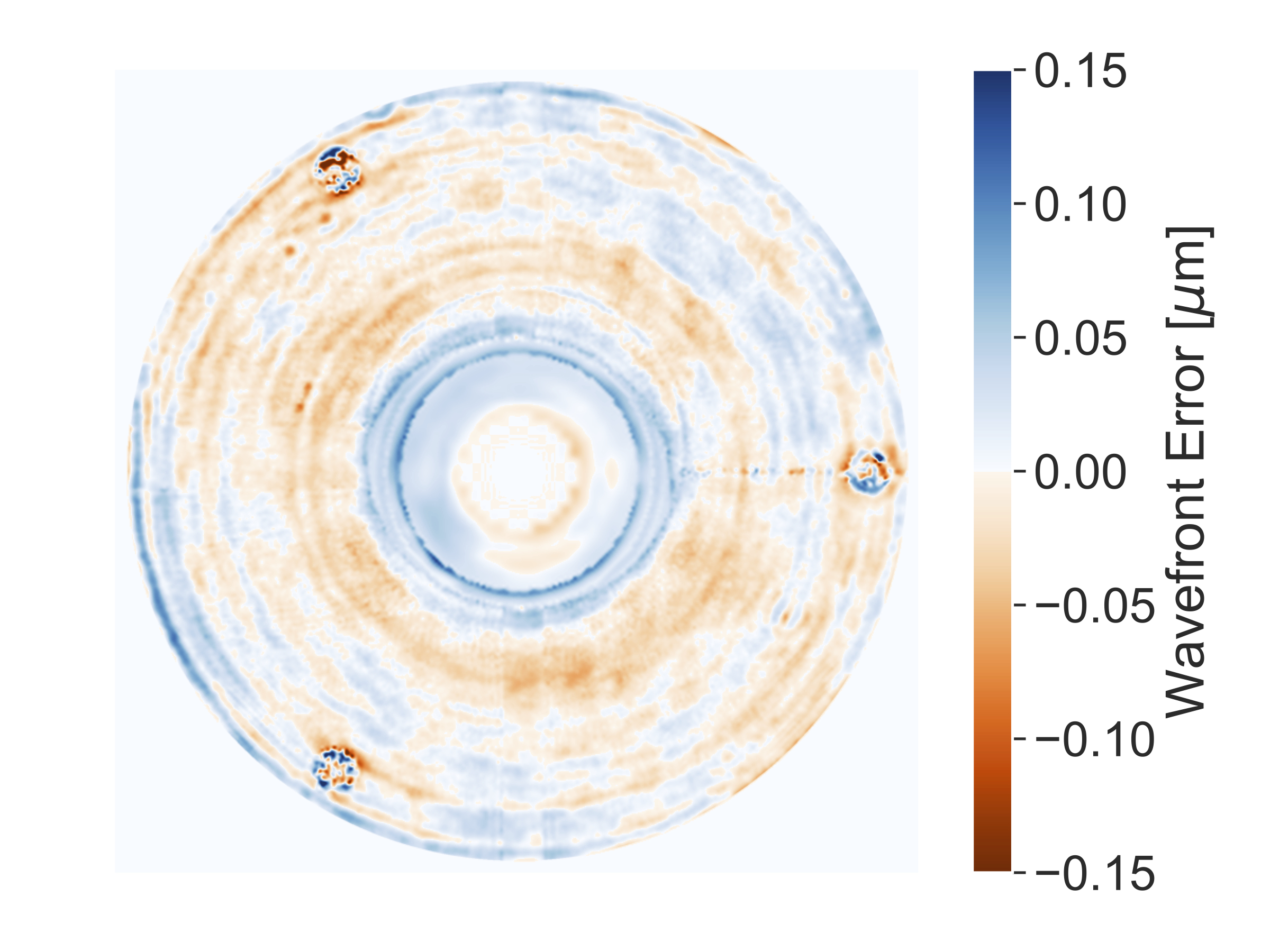}
            \caption{}
            \label{fi_02:surface_errors_HST}
        \end{minipage}
    \setcounter{subfigure}{1}
        \begin{minipage}[b]{0.32\textwidth}
            \includegraphics[width=\linewidth,trim={1cm 1cm 0 0},clip,]{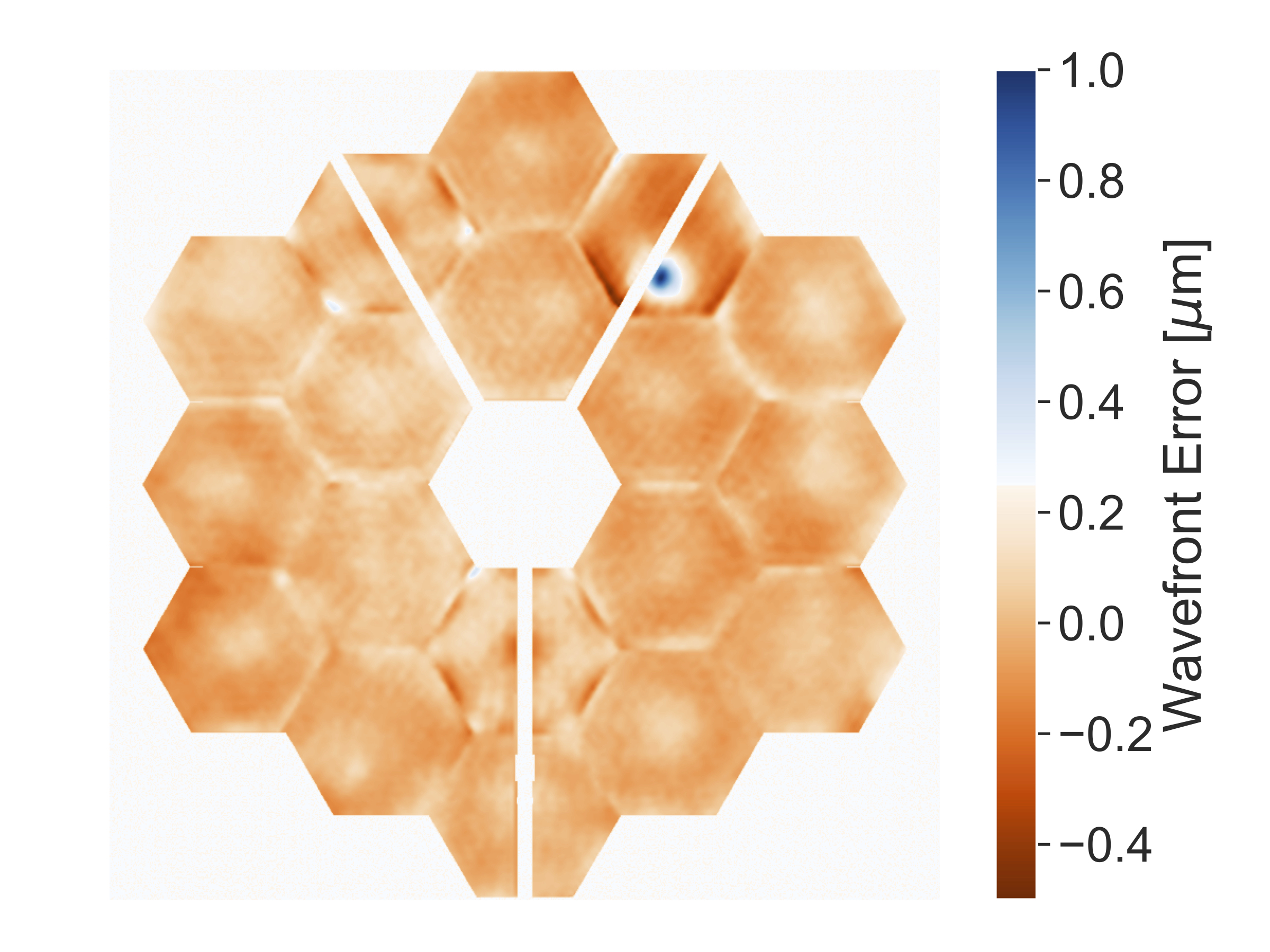}
            \caption{}
            \label{fi:real_JWST_OPD_a}
        \end{minipage}
    \setcounter{subfigure}{2}
        \begin{minipage}[b]{0.32\textwidth}
            \includegraphics[width=\linewidth,trim={1cm 1cm 0 0},clip,]{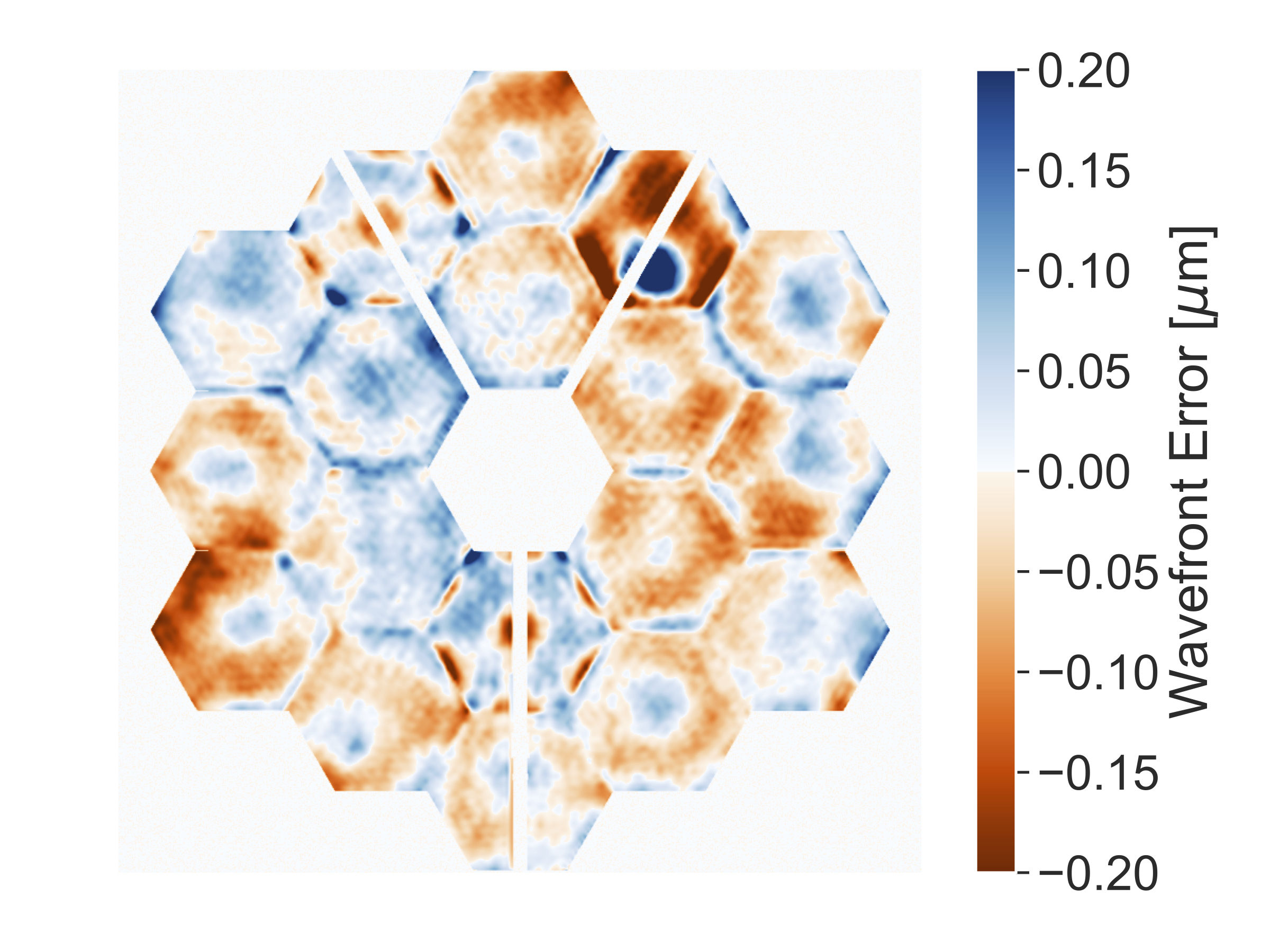}
            \caption{}
            \label{fi:real_JWST_OPD_b}
        \end{minipage}  
    \setcounter{subfigure}{-1}
    \caption{\textbf{(a)} shows a reduced range of variations to show the surface errors measured for the Hubble Space Telescope, where the scale has been reduced from $\pm 22$nm to better show details. \textbf{(b)} and \textbf{(c)} show the wavefront error measurements from the JWST cycle $1$ science operation on the $30$th of July of $2022$ with the total and the reduced range of variation, respectively.}
    \label{fi:real_JWST_OPD}
\end{subfigure}

\begin{itemize}
    \item \textit{Diffraction phenomena and the aperture size:} As we have seen in Section \ref{sc_02:intro_optics}, the diffraction phenomena happening in the optical system play an essential role in the formation of the PSF. The size of the optical system aperture and the wavelength of the light being studied are of particular interest. Equation \ref{eq_02:intensity_psf} shows us that under some approximations, the PSF is the Fourier transform of the aperture. Therefore, the size of the aperture and the PSF are closely related. For example, if we consider an ideal circular aperture, its diffraction pattern is the well-known \textit{Airy disk}. The relation between the width of the PSF and the diameter of the aperture is given by
    \begin{equation}
        \theta_{\text{FWHM}} = 1.025 \frac{\lambda}{d}\,,
        \label{eq_02:airy_psf}
    \end{equation} 
    where $\theta_{\text{FWHM}}$ is the full width at half maximum (FWHM) expressed in radians, $\lambda$ is the wavelength of the light being studied, and $d$ is the diameter of the aperture. The width of the PSF is a fundamental property of an optical system as it defines the resolution of the system. In other words, the PSF size defines the optical system's ability to distinguish small details in the image.
    \item \textit{Optical aberrations:} These aberrations are due to imperfections in the optical elements, for example, a not ideally spherical mirror or a not perfectly aligning of the optical components. The optical aberrations play a significant role in the morphology of the PSF and can be modelled using the WFE introduced in the generalised pupil function from Equation \ref{eq_02:generalized_pupil_function}. Some aberrations have a distinctive name, e.g., coma, astigmatism, and defocus, and they represent a specific Zernike polynomial \citep{noll1976}.
    \item \textit{Surface errors or polishing effects:} One would ideally like perfectly smooth surfaces in mirrors and lenses. However, imperfections arise in the optical surfaces due to imperfect surface polishing. \citet{krist2011} shows the measurement of surface errors (SFE) in the Hubble Space Telescope (HST). \citet[Section 35.2]{gross2006_v3} gives a more in-depth analysis of surface errors focusing on the tolerancing of SFE. Figure \ref{fi_02:surface_errors_HST} shows the surface errors measured for the Hubble space telescope (HST). \citet{krist1995b} studied HST's SFE before and after its iconic repair in $1993$ with parametric and non-parametric \citep{gerchberg1972} phase retrieval algorithms.
    \item \textit{Obscurations:} Complex optical systems have telescope designs where some elements can obscure some part of the pupil. Obscurations are an essential contributor to PSF morphology and result from projecting a $3$D structure onto the $2$D focal plane. The resulting projection depends on the considered position of the focal plane. Accurate modelling of telescopes with wide-field imagers, e.g., \textit{Euclid}, requires the computation of the obscuration's position dependence arising from the $3$D projection. The \textit{Euclid}'s obscurations are presented in Figure \ref{fi:obscuration_example}. \citet{fienup1993} and \citet{fienup1994} studied HST's obscuration from phase retrieval algorithms and noticed a misalignment that caused a pupil shift.
    \item \textit{Stray and scattered light:} Optical elements and instruments give rise to light reaching the detectors. \citet{krist1995c} studied the problem for the HST. \citet{storkey2004} developed methods to clean observations with scattered light from the SuperCOSMOS Sky Survey \cite{hambly2001}. \citet{sandin2014} studied the effect of scattered light on the outer parts of the PSF.
    \item \textit{Material outgassing and ice contamination:}
    Material outgassing leads to molecular contamination that alters different properties of the imaging system. Water is the most common contaminant in cryogenic spacecraft, which then turns into thin ice films. A notable example is the \textit{Gaia} mission that suffered from ice contamination (see \citet[Section 4.2.1]{gaia}) and required several decontamination procedures to slowly remove the ice from the optical system. \citet{schirmer2023} studied the ice formation and contamination for \textit{Euclid}. The article also reviews the lessons learned from other spacecraft on the topic of material outgassing. A companion paper of \citet{schirmer2023} is expected to be published soon addressing the quantification of iced optics impact on \textit{Euclid}'s data. 
    \item \textit{Chromatic optical components:} These components have a particular wavelength dependence, excluding the natural chromaticity due to diffraction. They are usually spectral filters and depend on the optical system design. A particular example is a dichroic filter which ideally serves as an ideal band-pass filter. The \textit{Euclid} optical system includes a dichroic filter which allows the use of both instruments, VIS and NISP, simultaneously as their passbands are disjoint. A dichroic filter is made of a stack of thin coatings of specific materials and thicknesses. Even if these components have a high-quality manufacturing process, they can induce significant chromatic variations in reflection affecting the PSF morphology. \citet{baron2022} proposed a test bench to characterise \textit{Euclid}'s dichroic filter and a numerical model of its chromatic dependence.
    \item \textit{Light polarisation:} In Section \ref{sc_02:intro_optics}, we studied the scalar diffraction theory, thus neglecting light polarisation. First, the optical system can induce polarisation even when the incoming light is not polarised. \citet{breckinridge2015} studied the effect of polarisation aberrations on the PSF of astronomical telescopes. The study of polarisation is carried out using Jones matrices \citep{jones1941}. These matrices describe a ray's polarisation change when going through an optical system. See \citet{mcguire1990,mcguire1991,yun2011} for more information on polarisation aberrations. Second, there are some regions in space where the incoming light has been polarised by different sources, e.g., Galactic foreground dust. \citet{lin2020} studied the impact of light polarisation on weak lensing systematics for the Roman space telescope \citep{wfirst}. The study found that the systematics introduced by light polarisation are comparable to Roman's requirements.
    \item \textit{Thermal variations:} The thermal variations in a telescope introduce mechanical variations in its structure that affect the performance of the optical system. The origin of the thermal variations is strong temperature gradients due to the sun's illumination. It is sometimes referred to as the \textit{telescope's breathing} \citep{bely1993} for the periodical pattern consequence of its orbit. Thermal variations can introduce a small defocusing of the system that will change the PSF morphology. This phenomenon was first identified in the HST \citep{hasan1993}. \citet{nino2007} studied HST focus variations with temperature and \citet{lallo2006} studied HST temporal optical behaviour, where temperature variations play a principal role. Later works \citep{makidon2006, sahu2007, suchkov1997} studied the impact of the thermal variations, and consequently PSF variations, on different science applications. A Structural-Thermal-Optical Performance (STOP) test helps predict thermal variations' impact on the optical system. This effect is naturally more significant in space-based telescopes as the temperature gradients in space are considerably more prominent than the ones found on the ground. Space-based telescopes located at the stable ${\rm L}_{2}$ Lagrange point, e.g., \textit{Euclid} and JWST, are less prone to thermal variations compared to telescopes orbiting the Earth, e.g., HST.  
\end{itemize}

As an example, Figures \ref{fi:real_JWST_OPD_a} and \ref{fi:real_JWST_OPD_b} show the measured optical contribution for the James Webb Space Telescope (JWST) PSF. \citet{rigby2022} presents a detailed analysis of JWST's state, from its commissioning, including its PSF. 

\subsection{Detector-level degradations}
\label{sc_02:detector_degradations}
The detector-level degradations are related to the detectors being used and, therefore, to the intensity of the PSF. They affect the observed images through the degradation operator $F_p$ from Equation \ref{eq_02:approx_forward_obs_model}, and as we will use star images, or eventually other observations, to constrain PSF models, it is necessary to consider their effects. Some of these degradations are non-convolutional and will not be well-modelled by a convolutional kernel. Nevertheless, we expect that image preprocessing steps will mainly correct these effects. However, the correction will not be perfect, and some modelling errors can propagate to the observations.

\begin{figure}
    \centering
    \includegraphics[width=0.9\textwidth]{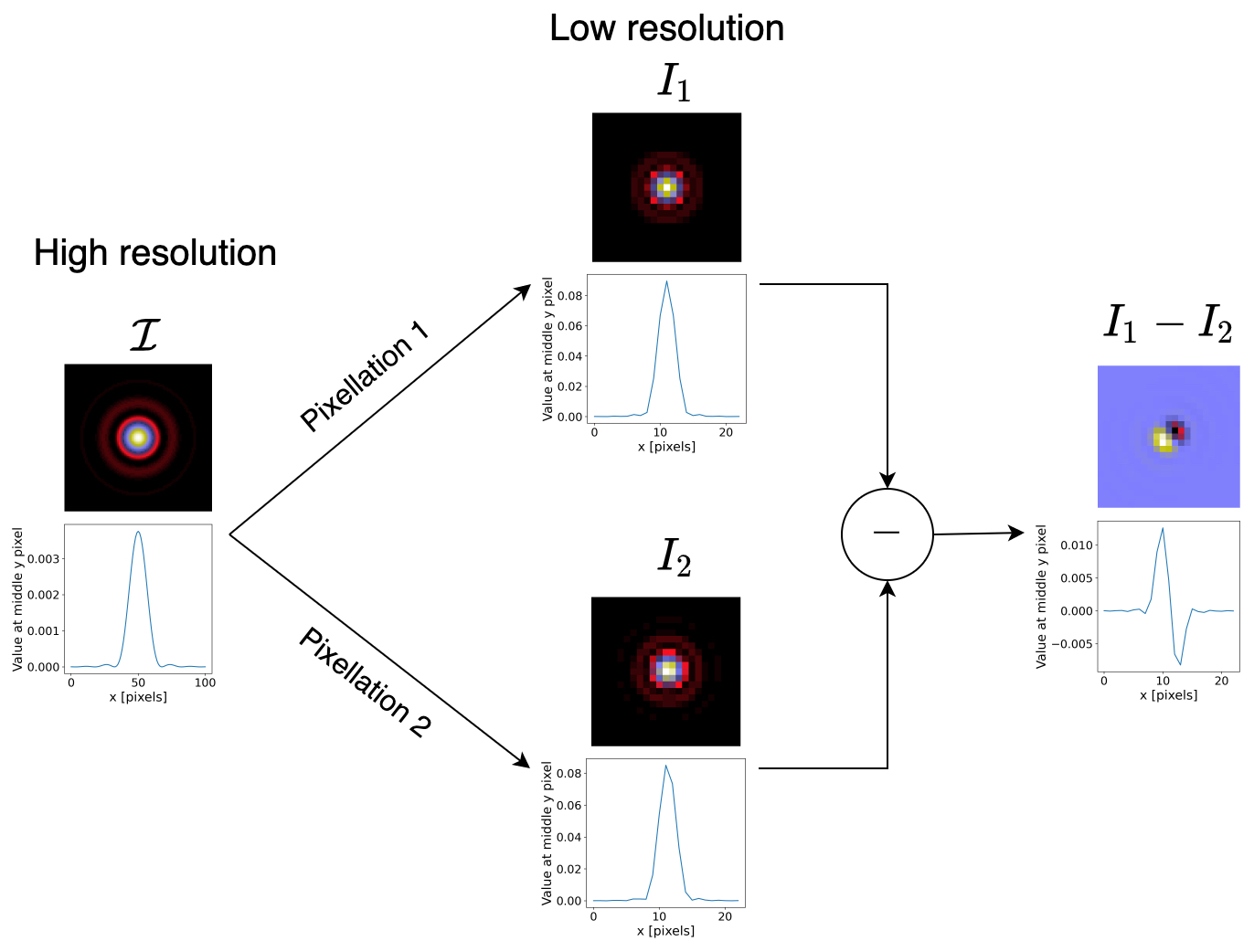}
    \caption{Example of two different pixellations on the same high-resolution image representing an Airy PSF. The difference between the two pixellations is an intra-pixel shift of $(\Delta x, \Delta y) = (0.35, 0.15)$ between them. Figure reproduced from \citet{liaudat2022_thesis}.}
    \label{fi_02:psf_pixellation_example}
\end{figure}

\begin{itemize}
    \item \textit{Undersampling and pixellation:} The EM wave that arrives at the detectors is a continuous function. The discrete pixels in the detectors integrate the functions and measure the intensity of the wave in their respective area. We name this process \textit{pixellation}, also known as \textit{sampling}. Some authors, e.g., \citet{anderson2000, bernstein2002,kannawadi2016}, define an \textit{effective} PSF as the convolution of the optical PSF, i.e., the flux distribution at the focal plane from a point source, with the pixel response of the instrument, e.g., a $2$D top-hat function. \citet{high2007} performed an early study on the effects of pixellation in WL and the choice of pixel scale for a WL space-based mission. \citet[Section 3]{krist2011} gives some insight on pixellation effects for HST. Two aspects of pixellation play a crucial role in PSF modelling. First, the sampling is done with the same grid, but it is indispensable to consider that the continuous function is not necessarily centred on the grid. This difference means that intra-pixel shifts between the different pixellations will be found. Figure \ref{fi_02:psf_pixellation_example} shows how two pixel representations of the same light profile change due to two different pixellations. When optimising a PSF model to reproduce some observed stars, the centroids of both images must be the same. Suppose the image centroids are the same, and the underlying model represents the observations satisfactorily. In that case, the residual image between the two pixellated images will be close to zero. If the centroids are not the same, the residual can be far from zero even though the model is a good representation of the observation, as illustrated in the residual image in Figure \ref{fi_02:psf_pixellation_example}. The second aspect is related to the Nyquist-Shannon sampling theorem. The theorem states the required number of samples to perfectly determine a signal of a given bandwidth. In the telescopes we study, the bandwidth and number of samples are related to the aperture's diameter and pixel size. Depending on the telescope's design, the sampling may not verify the Nyquist-Shannon theorem. If the images are undersampled, meaning that the theorem is not verified, a super-resolution step is required in the PSF modelling, which is the case of \textit{Euclid}. Using an observation strategy with dithering, as described in Section \ref{sc_02:dithering}, can significantly mitigate the undesired effects of undersampling and pixellation. \citet{kannawadi2021} studies ways of mitigating the effects of undersampling in WL shear estimations using \textsc{metacalibration} \citep{huff2017, sheldon2017, sheldon2020}, which is a method for measuring WL shear from well-sampled galaxy images. \citet{finner2023} studies near-IR weak-lensing (NIRWL) measurements in the CANDELS fields from HST images. The authors find that the most significant contributing systematic effect to WL measurements is caused by undersampling.
    \item \textit{Optical throughput and CCD Quantum Efficiency (QE):} The optical throughput of the system is the combined effect of the different elements composing the optical system, such as mirrors and optical elements like coatings \citep{venancio2016}. The filter being used in the telescope forms part of the optical throughput, as can be seen in Figure \ref{fi_02:megacam_bands} for the MegaCam set of filters. Figure \ref{fi_02:megacam_bands} also includes the CCD QE, which describes the sensibility of the CCD to detect photos of different wavelengths. Commonly, CCDs do not have a uniform response to the different wavelengths. Therefore, we must multiply the CCD QE with the telescope's optical throughput to compute the total transmission. 
    \item \textit{CCD misalignments:} Ideally, we expect that all the CCDs in the detector lie in a single plane that happens to be the focal plane of the optical system. However, this is not the case in practice, as there might be small misalignments between the CCDs introducing small defocuses that change from CCD to CCD. See \citet[Figure 8]{jee2011} for a study of this effect for the Vera C. Rubin Observatory. 
    \item \textit{Guiding errors:} Even if space telescopes are expected to be very stable when doing observations thanks to the attitude and orbit control systems (AOCS), there will exist a small residual motion that is called pointing jitter. The effect on the observation is introducing a small blur that can be modelled by a specific convolutional kernel that depends on the pointing time series. \citet[Section 4.8.3]{fenechconti2017} proposes to model the effect for \textit{Euclid} with a Gaussian kernel.
    \item \textit{Charge Transfer Inefficiency (CTI):} CCD detectors are in charge of converting incoming photons to electrons and collecting them in a potential well in the pixel during an exposure. The charge on each pixel is read when the exposure finishes. The collected electrons are transferred through a chain of pixels to the edge of the CCD, amplified and then read. High energy radiation above the Earth's atmosphere gradually damages CCD detector \citep{prodhomme2014_b, prodhomme2014_a}. The silicon damage in the detectors creates traps for the electrons that are delayed during the reading procedure. This effect is known as CTI, producing a trailing of bright objects and blurring the image. This effect is noticeably significant for space telescopes, given the harsh environment. CTI effects are expected to be corrected in the VIS image preprocessing. \citet{rhodes2010} carried out a study on the impact of CTI on WL studies. \citet{massey2009} developed a model to correct for CTI for the HST and later improved it in \citet{massey2014}.
    \item \textit{Brighter-fatter effect (BFE):} The assumption that each pixel photon count is independent of its neighbours does not hold in practice. There is a photoelectron redistribution in the pixels as a function of the number of photoelectrons in each pixel. The BFE is due to the accumulation of charge in the pixels' potential wells and the build-up of a transverse electric field. The effect is stronger for bright sources. \citet{antilogus2014} studied the effect and observed that the images from the CCDs do not scale linearly with flux, so bright star sizes appear larger than fainter stars. \citet{guyonnet2015} and \citet{coulton2018} proposed methods to model and correct this effect. The preprocessing of VIS images is supposed to correct for the BFE, but there might be some residuals.
    \item \textit{Wavelength dependent sub-pixel response:} There exists a charge diffusion between neighbouring pixels in the CCD. \citet{niemi2015_a} studied this effect for an \textit{Euclid}'s VIS CCD and modelled the response of the CCD. \citet{niemi2015_a} proposed to model the effect as a Gaussian convolutional kernel where the standard deviations of the 2D kernel are wavelength dependent, $\sigma_{x}(\lambda)$ and $\sigma_{y}(\lambda)$. \citeauthor{niemi2015_a} measured the proposed model with a reference VIS CCD. \citet{krist2003} studied the charge diffusion in HST and proposed spatially varying blur kernels to model the effect. 
    \item \textit{Noise:} There are several noise sources in the measurements. \textit{Thermal noise} \citep{nyquist1928} refers to the signal measured in the detector due to the random thermal motion of electrons which is usually modelled as Gaussian. \textit{Readout noise} \citep{basden2004} refers to the uncertainty in the photoelectron count due to imperfect electronics in the CCD. \textit{Dark-current shot noise} \citep{baer2006} refers to the random generation of electrons in the CCD, and even though it is related to the temperature, it is not Gaussian. There are also unresolved and undetected background sources that contribute to the observation noise. Which are the statistics of the predominant noise will depend on the imaging setting of the instrument and its properties.
    \item \textit{Tree rings and edge distortions:} There exist electric fields in the detector that are transverse to the surface of the CCD. The origin of these fields includes doping gradients or physical stresses on the silicon lattice. This electric field displaces charge, modifying the effective pixel area. Consequently, it changes the expected astrometric and photometric measurements. This electric field also generates concentric rings, \textit{tree rings}, and bright stripes near the boundaries of the CCD, \textit{edge distortions}. Given the close relationship between this effect and the detector, its importance depends strongly on the instrument being used. This effect is unnoticed in the MegaCam used in the Canada-France Imaging Survey (CFIS) as it depends on the CCD design. However, it was a major concern in the Dark Energy Camera used in the Dark Energy Survey (DES), as shown by \citet{plazas2014}. \citet[Figure 9]{jarvis2020} illustrates the consequence of tree rings in the PSF modelling.
    \item \textit{Other effects:} These effects include detector nonlinearity \citep{plazas2016, stubbs2014}, \textit{sensor interpixel correlation} \citep{lindstrand2019_b}, \textit{interpixel capacitance} \citep{mccullough2008, kannawadi2016, donlon2018}, \textit{charge-induced pixel shifts} \citep{gruen2015}, \textit{persistence} \citep{smith2008_a, smith2008_b}, \textit{reciprocity failure (flux-dependent nonlinearity)} \citep{biesiadzinski2011, bohlin2005}, and \textit{detector analogue-to-digital non-linearity}. 
\end{itemize}

\subsection{The atmosphere}
\label{sc_02:atmosphere}
The atmosphere plays a central role in ground-based telescopes' PSFs. See \citet{roddier1981} for an in-depth study of the subject. How the atmosphere affects our images will strongly depend on the exposure time used to image an object. The PSF induced by the atmosphere for a very short exposure will look like a speckle, while a long exposure will produce a PSF that resembles a 2D Gaussian, or more precisely, a Moffat profile \citep{moffat1969}. Figure \ref{fi_02:atmospheric_psf_exposures} shows examples of atmospheric PSFs with different exposure times. The atmosphere's effect on the PSF for a long exposure can be approximated by the effect of a spatially varying low-pass filter, thereby broadening the PSF and limiting the telescope's resolution. Astronomers usually use the term \textit{seeing} to refer to the atmospheric conditions of the telescope, and it is measured as the FWHM of the PSF. The loss of resolution due to the atmosphere is one of the main motivations for building space telescopes like \textit{Euclid} and \textit{Roman}, where the PSF is close to the diffraction limit and very stable.

The atmosphere is a heterogeneous medium whose composition changes with the three spatial dimensions and time. The inhomogeneity of the atmosphere affects the propagation of light waves that arrive at the telescope. Instead of supposing that the incoming light waves are plane, as emitted by the faraway source under study, these waves already have some phase lags or leads with respect to an ideal plane wave. The atmosphere introduces a WFE contribution to the optical system. These effects can be resumed as an effective phase-shifting plate, $\Phi_{\text{eff}}(x,y,t)$. However, calculating this effective plate is cumbersome as it involves having a model of the atmosphere and integrating the altitude, $z$, so that we have the spatial distribution, $(x,y)$, of the effective WFEs. The model of the atmosphere is represented by the continuous $C_{n}^{2}(z)$ \citep{roddier1981} profile, which represents the variations of the refractive index due to the atmospheric turbulence as a function of height. However, the $C_{n}^{2}(z)$ is challenging to model and measure, and even if it is possible, it is computationally expensive to exploit. 

\begin{figure}
    \centering
    \includegraphics[width=0.75\textwidth,trim={1cm 4.5cm 0 5cm},clip,]{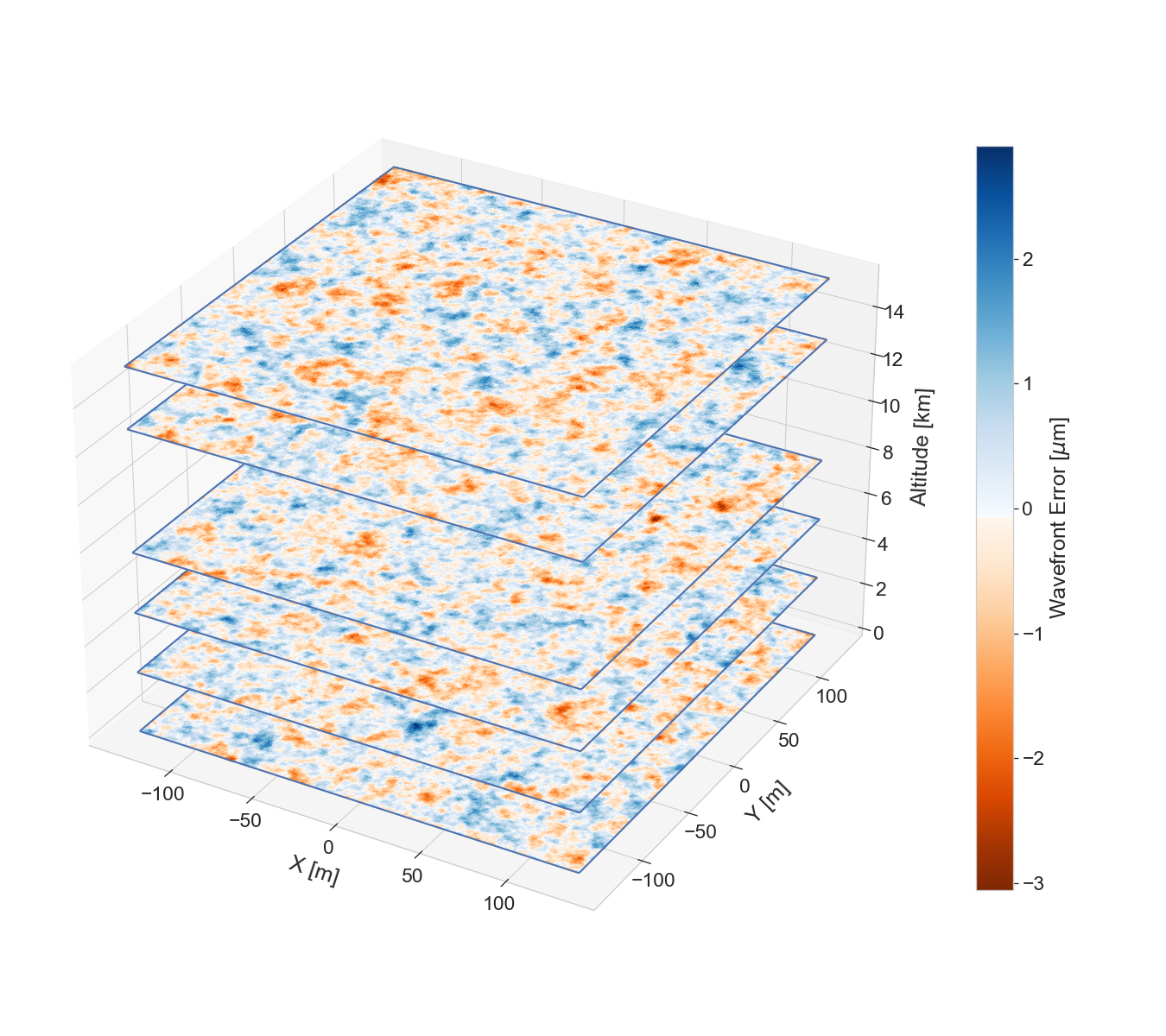}
    \caption{Illustration of six von Kármán phase screen layers at different altitudes simulated for LSST. The simulations were produced with the \textsc{GalSim} package \citep{rowe2015} using the parameters from \citet{jee2011}.}
    \label{fi_02:atmospherical_layers}
\end{figure}

We can discretise the integral over the altitude into $M$ thin-phase screens of variable strength at different altitudes to simulate the effect of the atmosphere. Each phase screen will have specific properties and move at different speeds in different directions. These assumptions are known as the frozen flow hypothesis. Each phase screen will be characterised by its power spectrum that can be modelled by a von Kármán model of turbulence \citep{karman1930}. The power spectrum of the atmosphere's WFE contribution writes
\begin{equation}
    \Psi(\nu) = 0.023 \, r_{0}^{- 5 / 3} \left( \nu^{2} + \frac{1}{L_{0}^{2}} \right)^{- 11 / 6} \,,
    \label{eq_02:karman_turbulence}
\end{equation}
where $\nu$ is a spatial frequency, $r_{0}$ is the Fried parameter, and $L_{0}$ is the outer scale. Both parameters, $r_{0}$ and $L_{0}$, are generally expressed in meters. The Fried parameter relates to the turbulence amplitude, and the outer scale relates to the correlation length. See Figure \ref{fi_02:atmospherical_layers} for an example of atmospherical phase screens. For lengths longer than $L_{0}$, the power of the turbulence asymptotically flattens. If we take the limit of $L_{0}$ to infinite, we converge to the Kolmogorov model of turbulence \citep{kolmogorov1991}. See \citet{sasiela1994} for more information on electromagnetic wave propagation in turbulence.

Once the phase screens, $\Phi_m(x,y|u_i,v_i)$, have been simulated following Equation \ref{eq_02:karman_turbulence}, the temporal variation of the screen has to be taken into account. The phase screens contribute to the WFE of the PSF, which is why it depends on the pupil plane variables $(x,y)$. The temporal variation is usually modelled with the wind's properties at the phase screen's reference altitude. We describe the wind with two components, $v_u$ and $v_v$, where we have assumed that $v_z=0$. We then obtain the effective phase screen by a weighted average of the phase screens at the different altitudes as
\begin{equation}
    \Phi_{\text{eff}}(x,y;t|u_i,v_i) = \sum_{m=1}^{M} c_m \Phi_m(x,y;t|u_i,v_i) \,,
    \label{eq_02:effective_phase_screen}
\end{equation}
where $\{c_m\}$ are some weights. The difficulty of modelling the atmosphere is that the time scales are comparable with the exposure time. Therefore, the PSF we estimate for a given time snapshot will change with respect to another PSF at another snapshot within the same camera exposure. This change means that we need to integrate the instantaneous PSF over time to model the PSF physically, which corresponds to
\begin{equation}
    I_{\text{img}}(\bar{u},\bar{v}|u_i,v_i) = \int_{t_{0}}^{t_{0} + T_{\text{exp}}} I_{\text{img}}(\bar{u},\bar{v};t|u_i,v_i)\; dt \,,
    \label{eq_02:atm_time_integral}
\end{equation}
where $I_{\text{img}}(\bar{u},\bar{v},t|u_i,v_i)$ is the instantaneous image of the object affected by the PSF $\mathcal{H}(u,v;\lambda;t|u_i,v_i)$, $t_{0}$ is a random initial time and $T_{\text{exp}}$ is the exposure time. 

Finally, we need to choose the time step size to discretise the integral from Equation \ref{eq_02:atm_time_integral}. Each instantaneous PSF will look like a speckle. Once we add them up in the integral, the PSF starts becoming rounder and smoother. Figure \ref{fi_02:atmospheric_psf_exposures} shows examples of atmospheric PSFs using different exposure times that were simulated using $6$ phase screens using the parameters from \citet{jee2011} that correspond to an LSST-like scenario. It is interesting to see how the short-exposure PSF looks like a speckle, and then the profile becomes more and more smooth as the exposure time increases. As a reference, the exposure time used for the r-band observations in CFIS is $200$s\footnote{\href{https://www.cfht.hawaii.edu/Science/CFIS/}{https://www.cfht.hawaii.edu/Science/CFIS/}}. \citet{vries2007} studied the PSF ellipticity change due to atmospheric turbulences as a function of the exposure time. \citeauthor{vries2007} observed that the ellipticity of the PSF decreases its amplitude as the exposure time increases.

\begin{figure}
    \centering
    \includegraphics[width=0.24\textwidth]{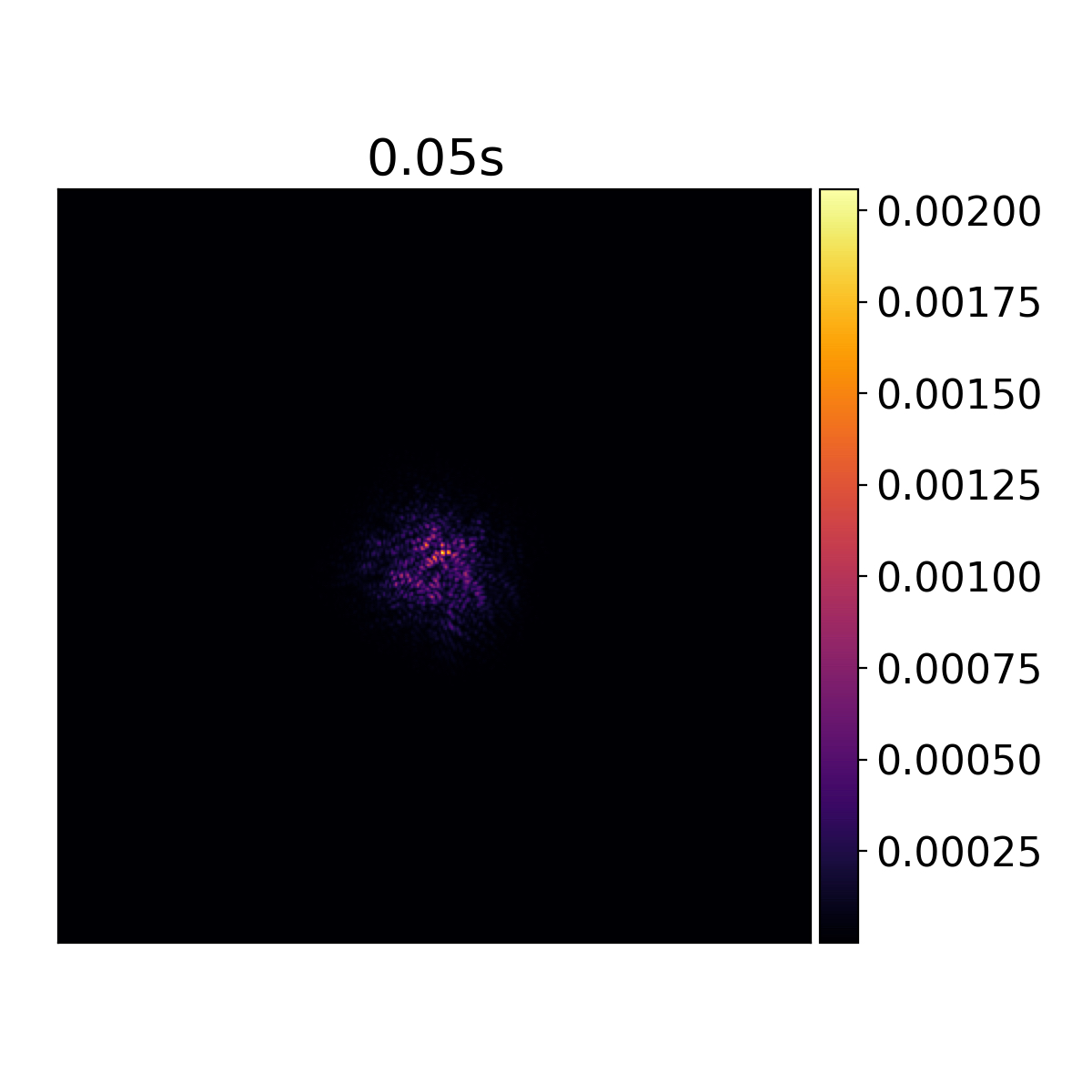}
    \includegraphics[width=0.24\textwidth]{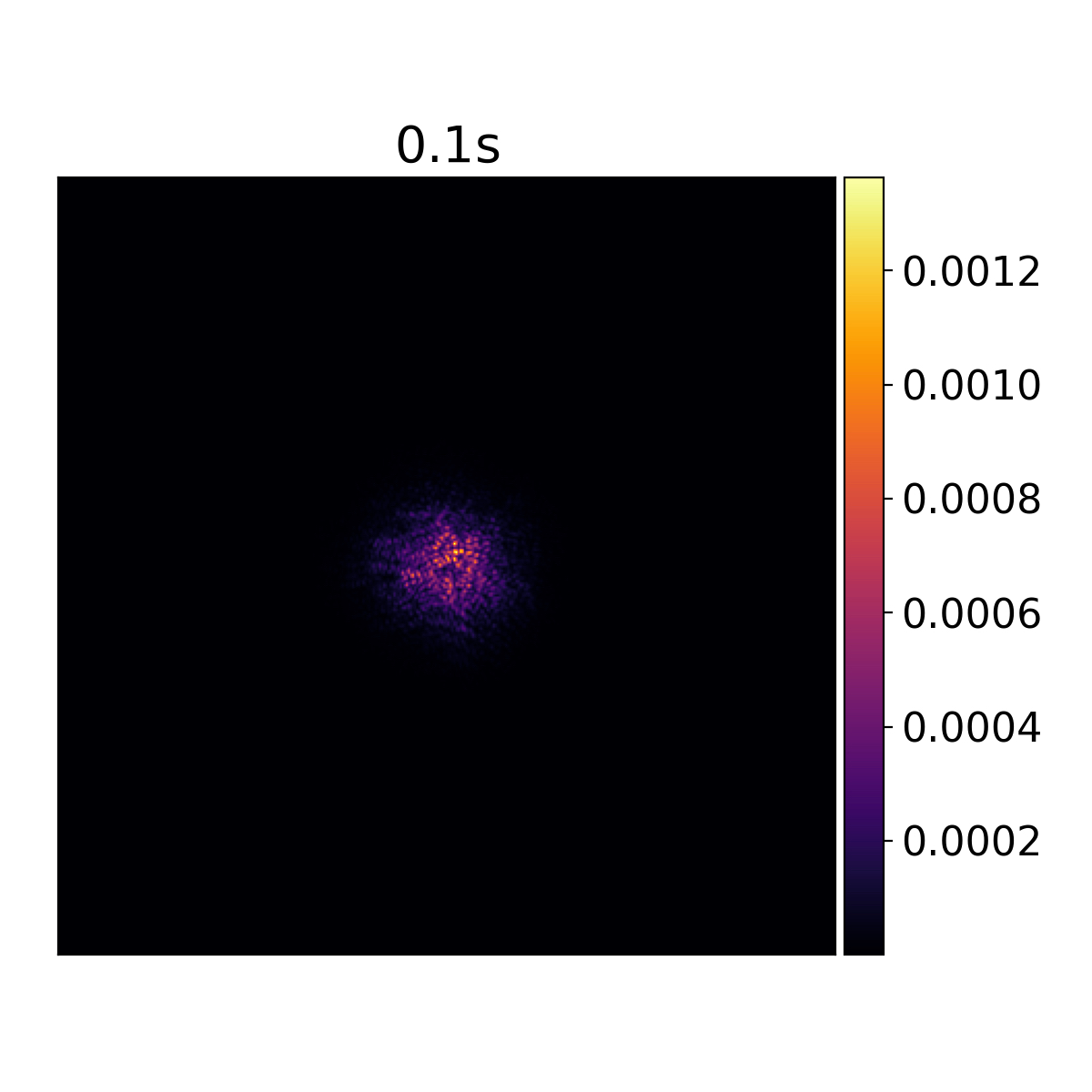}
    \includegraphics[width=0.24\textwidth]{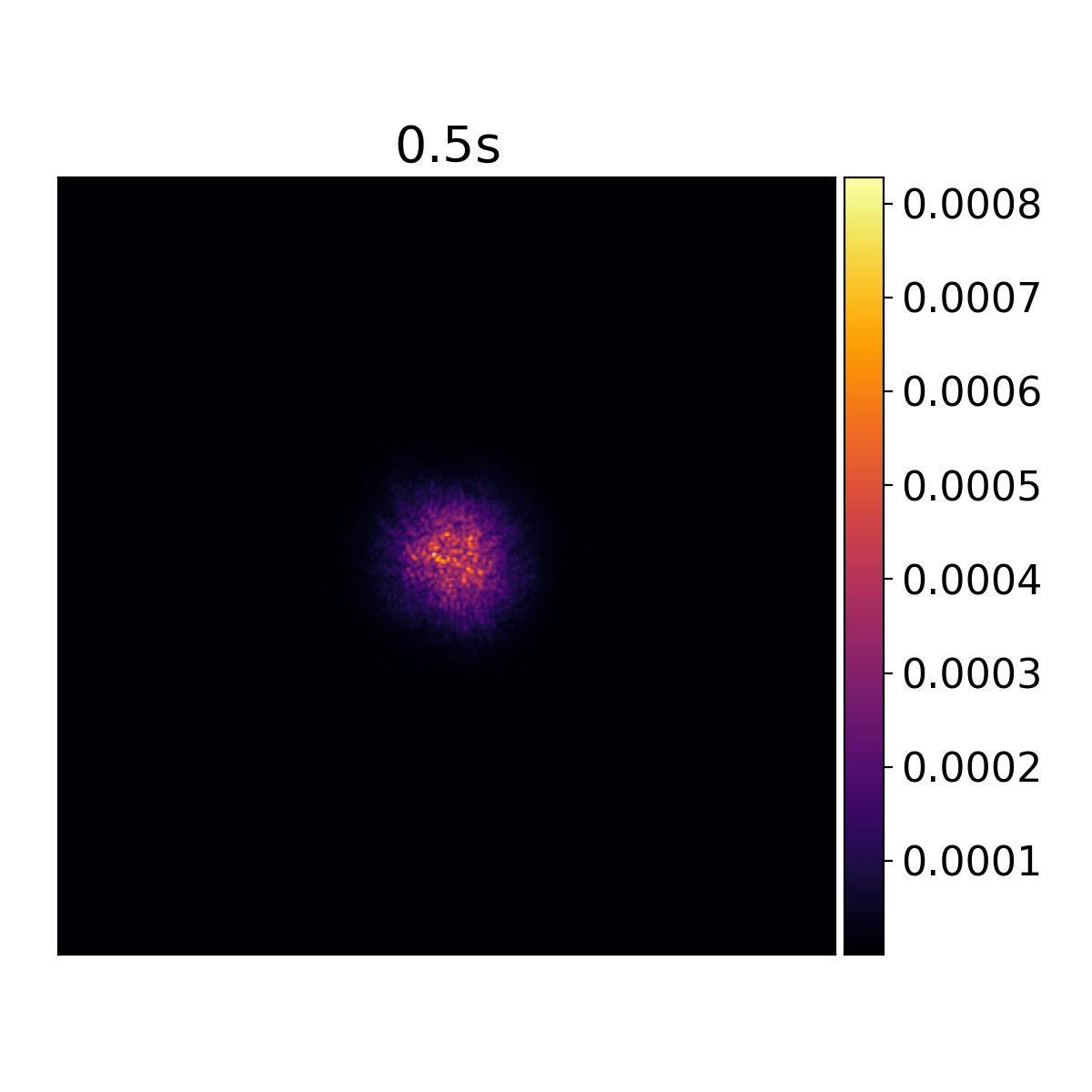}
    \includegraphics[width=0.24\textwidth]{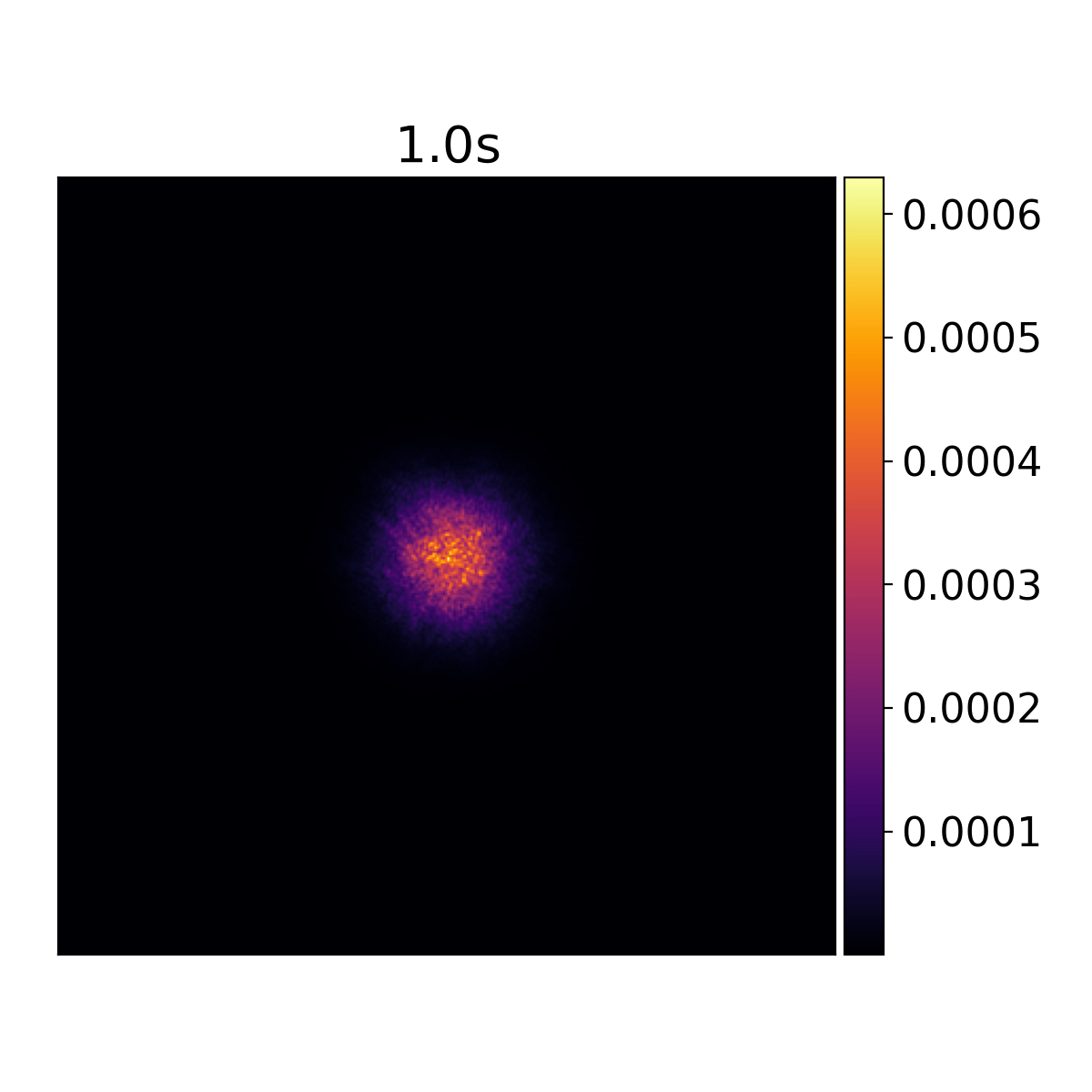}\\
    \includegraphics[width=0.24\textwidth]{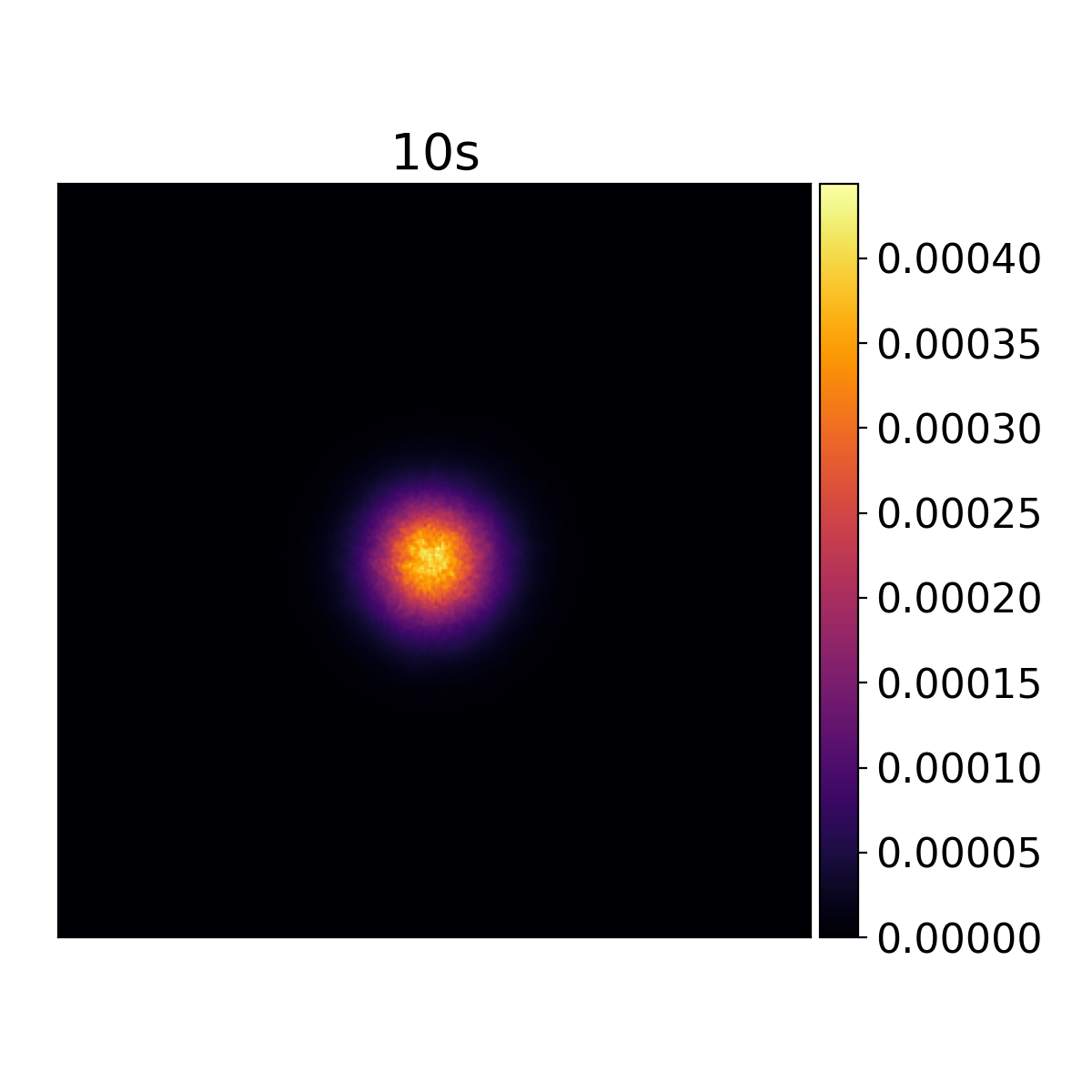}
    \includegraphics[width=0.24\textwidth]{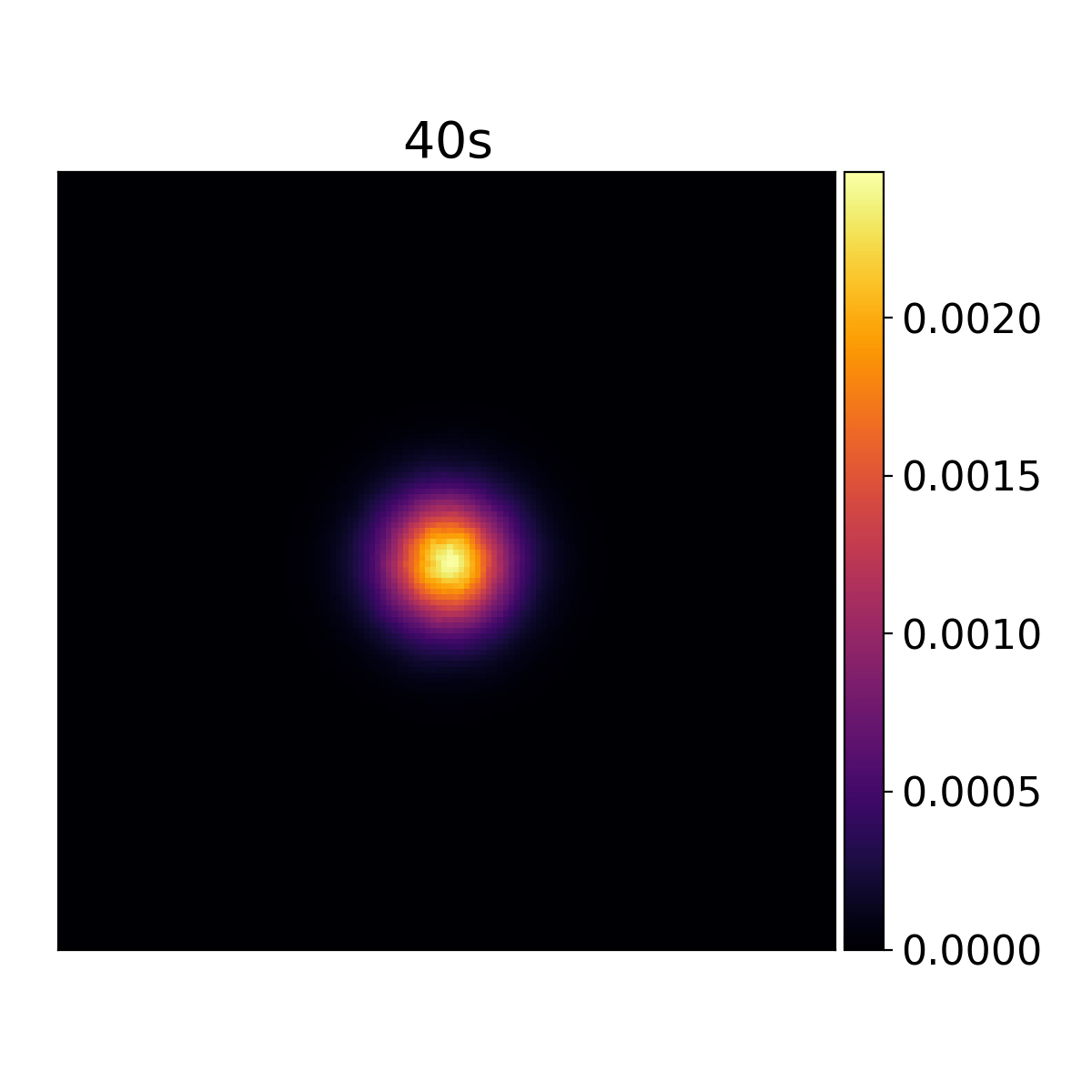}
    \includegraphics[width=0.24\textwidth]{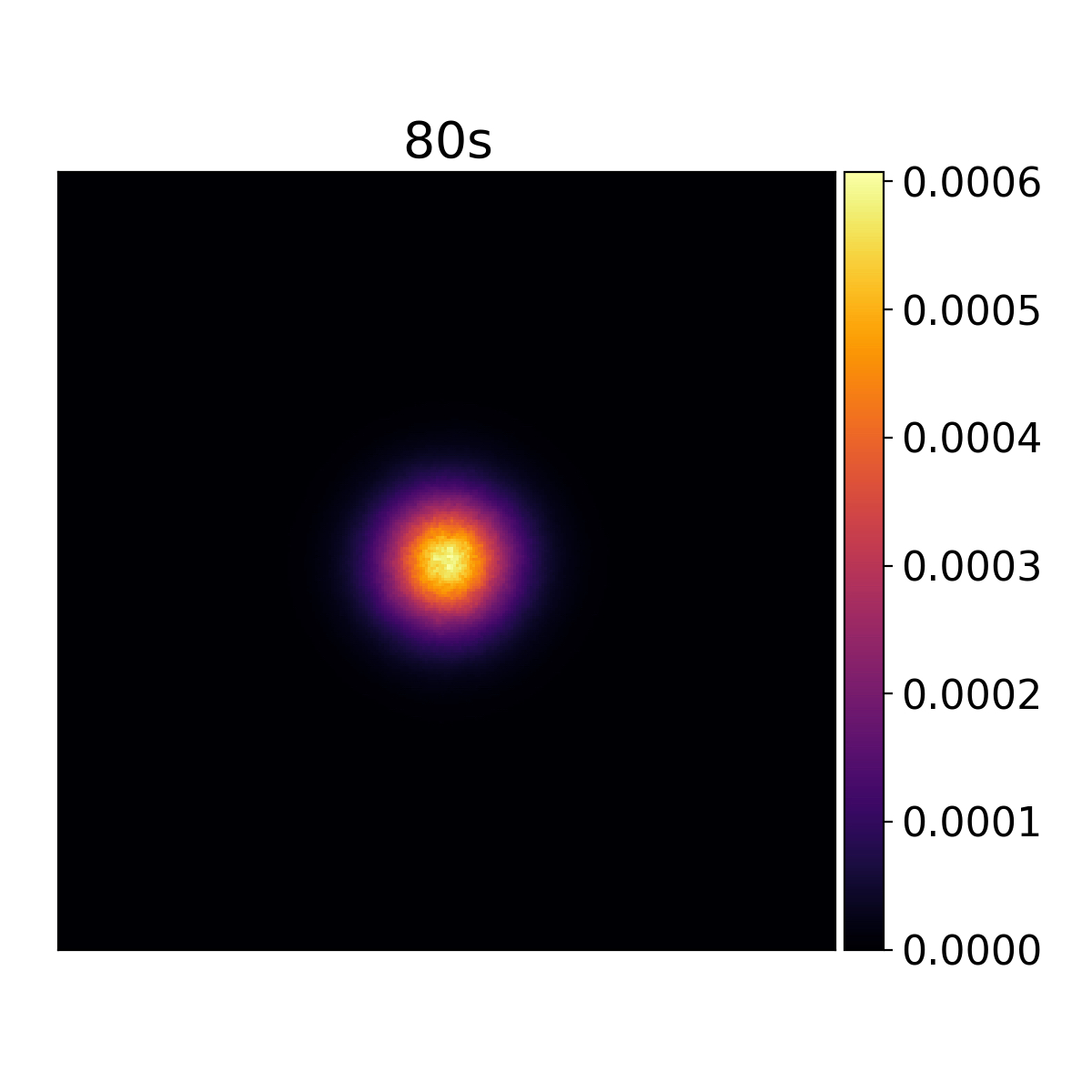}
    \includegraphics[width=0.24\textwidth]{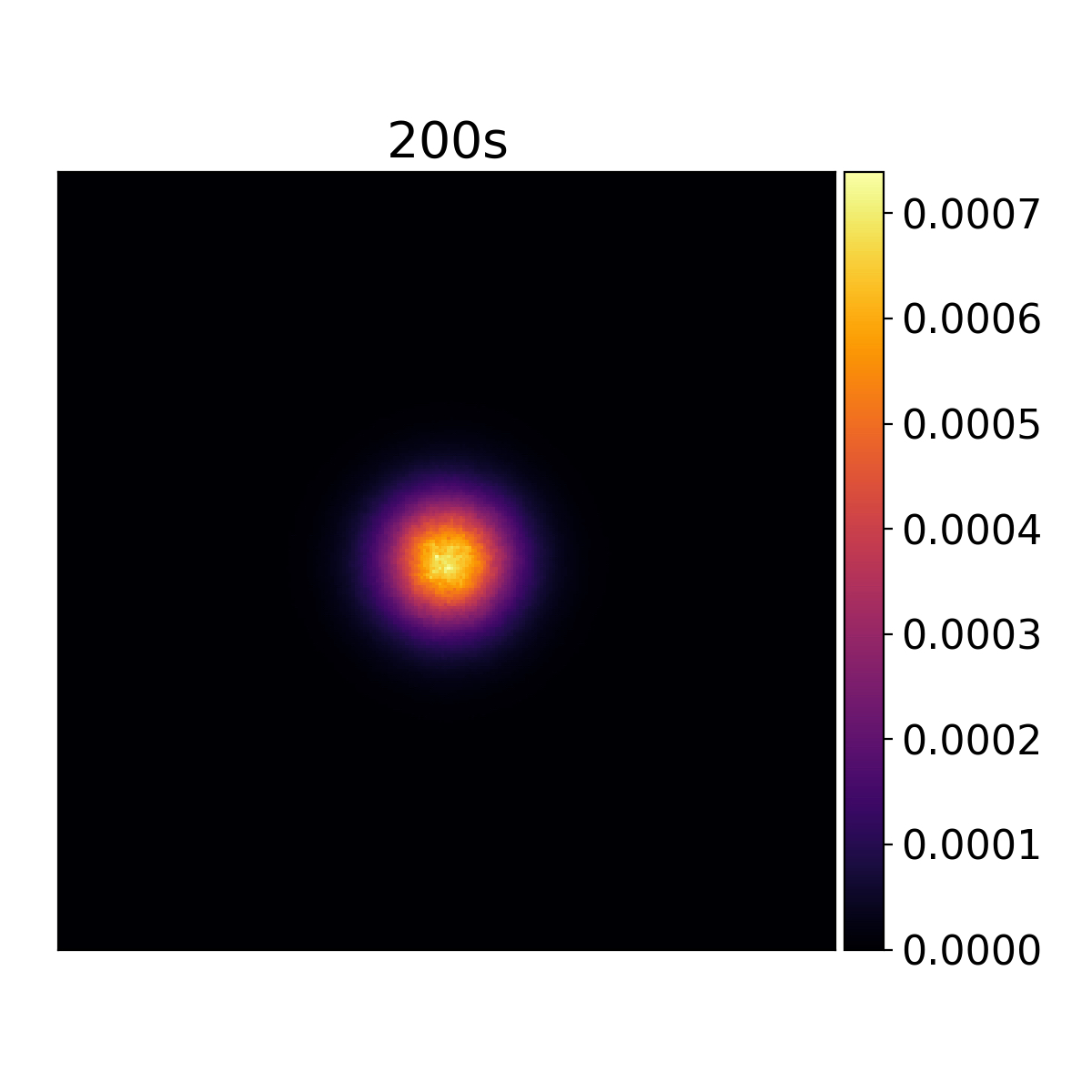}
    \caption{Example of atmospheric PSFs with different exposure times. The simulation was done using the atmospherical parameters from \citet{jee2011} for an LSST-like scenario.}
    \label{fi_02:atmospheric_psf_exposures}
\end{figure}

Another effect that should be considered is \textit{atmospheric differential chromatic refraction}. This effect represents the refraction due to the change of medium from vacuum to the Earth's atmosphere. The effect varies as a function of the zenith angle and wavelength. \citet{meyers2015} performed a study on the impact of the atmospheric chromatic effect on weak lensing for surveys like LSST and DES. 

\citet{heymans2012} performed a study on the impact of atmospheric distortions on weak-lensing measurement with real data from CFHT. \citeauthor{heymans2012} characterised the ellipticity contribution of the atmosphere to the PSF for different exposure times. To achieve this, they computed the two-point correlation function of the residual PSF ellipticity between the observations and a \texttt{PSFEx}-like PSF model (described in detail in Section \ref{sc_02:psf_modelling}). \citet{salmon2009} studied the image quality and the observing environment at CFHT. \citep{xin2018} carried out a study of the PSF and the variation of its width with time and wavelength for the Sloan Digital Sky Survey (SDSS) \citep{york2000, gunn1998}. \citet{jee2011} carried out a simulation study to evaluate the impact of atmospheric turbulence on weak-lensing measurement in LSST. \citet{jee2011} used the atmospherical parameters from \citep{ellerbroek2002} that were measured in the LSST site in Cerro Pachón, Chile. There is an ongoing project at LSST DESC\footnote{\href{https://github.com/LSSTDESC/psf-weather-station}{github.com/LSSTDESC/psf-weather-station}} to leverage atmospheric and weather information at or near the observation site to produce realistically correlated wind and turbulence parameters for atmospheric PSF simulations.

Another way to simulate the atmosphere and the PSF is to use a photon Monte Carlo approach. This line of work was carried out in \citet{peterson2015, peterson2019, peterson2020} with a simulator available coined \textsc{PhoSim}\footnote{\href{https://bitbucket.org/phosim/phosim_release/wiki/Home}{bitbucket.org/phosim/phosim\_release/wiki/Home}} that is capable of simulating several telescopes including the Vera C. Rubin observatory and JWST. The method consists of sampling photons from astronomical sources and simulating their interactions with their models of the atmosphere, the optics and the detectors. \textsc{PhoSim} is characterised by being a remarkably fast simulator regarding the level of simulation complexity handled. This simulator has proved helpful in several studies \citep{walter2015,xin2015,beamer2015,carlsten2018,xin2018,burke2019,sanchez2020,nie2021_2,merlin2023,bretonniere2023,merz2023}.\textsc{PhoSim} is a powerful simulation tool that can help to study the PSF of systems by following a forward model approach where simulations are compared with observations \citep{chang2012,cheng2017}. The \textsc{PhoSim} parameters can then be fitted or modified to match the simulation output with the observed images. Another known simulation software, \textsc{GalSim}\footnote{\href{https://github.com/GalSim-developers/GalSim}{github.com/GalSim-developers/GalSim}} \citep{rowe2015}, incorporates options to simulate atmospheric PSFs from phase-screen exploiting the photon Monte Carlo approach from \citet{peterson2015}.

To conclude, we have seen that it is possible to develop a physical model of the atmosphere based on the optical understanding we have from Section \ref{sc_02:intro_optics} and the studies of atmospheric turbulence of \citeauthor{karman1930} and \citeauthor{kolmogorov1991}. However, this approach has two inconveniences. First, the approach requires physical measurements of the atmosphere at the telescope's site, which is not always available. Second, it is computationally expensive, as there is an altitude and temporal integration, to handle varying atmospherical properties and reach the exposure time, respectively. In practice, it is required to use long exposure times to obtain deeper observations, meaning observing fainter objects that are important for weak-lensing studies. This fact simplifies the PSF modelling task as the long temporal integration smooths the PSF profile and the PSF spatial variations over the FOV. Therefore, a data-driven approach to modelling the PSF can offer a feasible and effective solution in this scenario.

\subsection{Adaptive optics}
\label{sc_02:adaptive_optics}

An alternative approach to work with ground-based observations affected by the atmosphere is to use adaptive optic (AO) systems \citep{beckers1993}. This technology significantly improves the observation resolution in ground-based telescopes that is severely limited due to the atmosphere, as we have seen in Section \ref{sc_02:atmosphere}. An AO system tries to counteract the effect of the atmosphere on the incoming wavefront by changing the shape of a deformable mirror. The key components of the AO system are wavefront sensors (WFS), wavefront reconstruction techniques and deformable mirrors, which operate together inside a control loop. The WFS provide information about the incoming wavefront and usually incorporate a phase-sensitive device. The wavefront reconstruction has to compute a correction vector for the deformable mirrors by estimating the incoming wavefront from the WFS information. The control loop works in real-time sensing and modifying the deformable mirrors so that the wavefront received by the underlying instrument is free from the optical path differences introduced by the atmosphere. \citet{davies2012} provides a detailed review of AO systems for astronomy. The LSST that will carry out weak lensing studies contains an AO system \citep{neill2014, thomas2016, angeli2014}. Exoplanet imaging studies have greatly benefited from AO systems and impose strict requirements for these systems. Some examples are the SPHERE instrument in the VLT \citep{beuzit2019} and the Gemini Planet Imager \citep{graham2007, macintosh2014, macintosh2018} in the Gemini South Telescope.

%%% 
\section{Current PSF models}
\label{sc_02:psf_modelling}

Let us now discuss some of the most known PSF models, which can be divided into two main families, \textit{parametric} and \textit{non-parametric}, also known as \textit{data-driven}.

\subsection{Parametric PSF models}
This family of PSF models is characterized by trying to build a physical optical system model that aims to be as close as possible to the telescope. Once the physical model is defined, a few parameters are estimated using star observations. Such estimation, also called calibration, is required as some events, like launch vibrations, ice contaminations and thermal variations, introduce significant variations in the model. These events prevent a complete on-ground characterization from being a successful model. Parametric models are capable of handling chromatic variations of the PSF as well as complex detector effects. Nevertheless, parametric models have only been developed for space missions and are custom-made for a specific telescope. The parametric model is compelling if the proposed PSF model matches the underlying PSF field. However, if there are mismatches between both models, significant errors can arise due to the rigidity of the parametric models. The difficulty of building a physical model for the atmosphere, already discussed in Section \ref{sc_02:atmosphere}, makes them impractical for ground-based telescopes. 

The parametric model \texttt{Tiny-Tim}\footnote{\href{https://github.com/spacetelescope/tinytim}{github.com/spacetelescope/tinytim}} \citep{krist1993, krist1995b, krist2011} has been used to model the PSF of the different instruments on board the HST. The Advanced Camera for Surveys (ACS) in HST was used to image the Cosmic Evolution Survey (COSMOS), which covers a $2$ $\text{deg}^{2}$ field that was used to create a widely used space-based weak-lensing catalogue. The first WL shape catalogue used the \texttt{Tiny-Tim} model \citep{leauthaud2007}. \citet{rhodes2007} studied the stability of HST's PSF, noticing a temporal change of focus in the images. Besides the parametric model \texttt{Tiny-Tim} model, \citet{anderson2000} developed the concept of \textit{effective PSF}, which is the continuous PSF arriving to the detectors, i.e., Equation \ref{eq_02:intensity_psf}, convolved with the pixel-response function of the detector. \citet{anderson2000} proposed an algorithm to model the effective PSF iteratively from observed stars. \citet{anderson2006} continued the work on the effective PSF, adding some improvements and detailed a model for the HST instruments ACS and Wide Field Camera (WFC). \citet{hoffmann2017} then carried out a comparison between \texttt{Tiny-Tim} and the effective PSF approach for ACS/WFC. The study shows that the effective PSF approach consistently outperforms the \texttt{Tiny-Tim} PSFs, exposing the limitations of parametric modelling. \citet{anderson2016} describes the adoption of the effective PSF approach applied to Wide Field Camera 3 (WFC3/IR) observations, which are undersampled. The software \textsc{Photutils}\footnote{\href{https://github.com/astropy/photutils}{github.com/astropy/photutils}} \citep{bradley2022} provides an implementation of the effective PSF approach from \citet{anderson2000} with the enhancements from \citet{anderson2016}. \citet{schrabback2010} used a data-driven PSF model based on PCA when studying the COSMOS field. 

The early severe aberrations of HST's optical system were an important driver of phase retrieval algorithms. Several efforts to characterise HST were published in an \textit{Applied Optics} special issue \citep{breckinridge1993}. Solving the phase retrieval problem provides a reliable approach to characterise the optical system accurately. \citet{fienup1993_2} studied new phase retrieval algorithms and \citet{fienup1993} present several results characterising HST's PSF.

Regarding recently launched space telescopes, \textit{Euclid}'s VIS parametric model constitutes the primary approach for \textit{Euclid}'s PSF modelling. The model will soon be published and used to work with observations from \textit{Euclid}. JWST has a Python-based simulating toolkit, \textsc{webbpsf}\footnote{\href{https://github.com/spacetelescope/webbpsf}{github.com/spacetelescope/webbpsf}} \citet{perrin2012, perrin2014}. Recent works developed and compared data-driven PSF models for JWST's NIRCam \citep{nardiello2022, zhuang2023}.

\subsection{Data-driven (or non-parametric) PSF models}
The \textit{data-driven} PSF models, also known as \textit{non-parametric}, only rely on the observed stars to build the model in pixel space. They are blind to most of the physics of the inverse problem. These models assume regularity in the spatial variation of the PSF field across the FOV and usually differ in how they exploit this regularity. Data-driven models can easily adapt to the current state of the optical system. However, they have difficulties modelling complex PSF shapes occurring in diffraction-limited settings. One limitation shared by all the data-driven models is their sensitivity to the available number of stars to constrain their estimation. A low star number implies that there might be not enough stars to sample the spatial variation of the PSF. When the number of stars in a FOV falls below some threshold, the model built is usually considered unusable for WL purposes. This family of models has been widely used for modelling ground-based telescope PSFs. Nevertheless, they are not yet capable of successfully modelling the chromatic variations in addition to the spatial variations and the super-resolution. 

We proceed by describing several PSF models in chronological order. The first models, described in more detail, were used to process real data from different surveys, except for Resolved Component Analysis \citep{ngole2016}. The latter models are worth mentioning but have yet to be used to produce a WL shape catalogue with all the validation and testing it implies.

\subsubsection{Shape interpolation}
The first approach for PSF modelling consisted of estimating some parameters of the PSF at the positions of interest. It was done for early studies in WL and is closely related to the widely employed galaxy shape measurement method KSB \citep{kaiser1995}. This precursor method only required the PSF's ellipticity and size at the positions of the galaxies. Therefore, a full-pixel image of the PSF was unnecessary. Then, the KSB method can correct the galaxy shape measurement for the effects of the PSF. The method to interpolate the shape parameters to other positions is usually a polynomial interpolation. For example, this was the case for the early WL study of the Canada-France-Hawaii Telescope Legacy Survey (CFHTLS) \citep{fu2008}. \citet{gentile2013} reviewed the different interpolation methods and studied their performance for WL studies. \citet{viola2011} performed a study showcasing different biases present in the KSB \citep{kaiser1995} method. These biases are a consequence of problematic KSB assumptions: (i) KSB gives a shear estimate per individual image and then takes an average, while WL shear should be estimated from averaged galaxy images (ii) KSB works under the assumption that galaxy ellipticities are small, but in the context of weak lensing, what is considered ``small" pertains to the alteration in ellipticity caused by the shear, and (iii) KSB gives an approximate PSF correction that only holds in the limit of circular sources and does not invert the convolution with the PSF. Recent WL studies no longer use this approach. The WL studies have evolved to more sophisticated galaxy shape measurement techniques that require a full pixel image of the PSF at the position of galaxies.

\subsubsection{Principal component analysis (PCA)}
Principal component analysis is a widely known method for multivariate data analysis and dimensionality reduction. Let us start with a set of star observations in $\mathbb{R}^{p^{2}}$ that we concatenate in a matrix $\bar{\mathbf{I}} = [\bar{I}_{1}, \ldots, \bar{I}_{n}]$. We have flattened the 2D images into an array to simplify expressions. One would like to represent the observations with $r$ components $\{X_{i} \}_{i=1}^{r}$ in $\mathbb{R}^{p^{2}}$, where $r \geq n$, assuming that $p^{2} > n$. The PCA method gives $r$ orthonormal components in $\mathbb{R}^{p^{2}}$ which define directions in the $\mathbb{R}^{p^{2}}$ space where the variance of the dataset $\bar{\mathbf{I}}$ is maximized. 

If $n$ components are used to represent the observations, then the learned components in the PCA procedure represent a basis of the subspace spanned by the observations or the columns of $\bar{\mathbf{I}}$. The method can be interpreted as a linear transformation to a new representation with orthogonal components. As it is usual to observe regularity in the spatial variations of the PSF, most of the dataset variability can be described with a few components. Then, one can only use the first $r$ principal components and achieve a dimensionality reduction of the observations. The dimensionality reduction technique allows denoising the model as the observational noise cannot be represented with $r$ components and only the PSF trends are well described.

The PCA method was used to model the PSF for the SDSS \citep{lupton2001}, although it was referenced as the Karhunen-Loève transform. \citet{jarvis2004} then proposed its use in a WL context. \citet{jee2007} used PCA to model the spatial and temporal variations of the HST PSF. \citet{jee2011} also used PCA to model the PSF in LSST simulations. HST's COSMOS catalogue \citep{schrabback2010} used PCA to model the PSF. PCA showcased the utility and robustness of data-driven methods and the importance of using a pixel representation of the PSF and is the precursor of several of the following models.

\subsubsection{\texttt{PSFEx}}
\texttt{PSFEx}\footnote{\href{https://github.com/astromatic/psfex}{github.com/astromatic/psfex}} \citep{bertin2011} has been widely used in astronomy for weak-lensing surveys, for example, DES year $1$ \citep{zuntz2018}, HSC \citep{mandelbaum2017}, and CFIS \citep{guinot2022}. It was designed to work together with the \texttt{SExtractor} \citep{bertin1996} software which builds catalogues from astronomical images and measures several properties of the observed stars. \texttt{PSFEx} models the variability of the PSF in the FOV as a function of these measured properties. It builds independent models for each CCD in the focal plane and works with polychromatic observations. It cannot model the chromatic variations of the PSF field. The model is based on a matrix factorisation scheme, where one matrix represents PSF features and the other matrix the feature weights. Each observed PSF is represented as a linear combination of PSF features. The feature weights are defined as a polynomial law of the selected measured properties. This choice allows having an easy interpolation framework for target positions. In practice, the properties that are generally used are both components of the PSF's FOV position. The PSF features are shared by all the observed PSFs and are learned in an optimisation problem. The PSF reconstruction at a FOV position $(u_i, v_i)$ can be written as
\begin{equation}
    \bar{I}^{\text{PSFEx}}_{\text{star}}(\bar{u}, \bar{v}|u_i,v_i) = F^{\text{PSFEx}} \underbrace{\left\{ \sum_{\substack{p,q \geq 0 \\ p+q \leq d}} u_{i}^{p} \, v_{i}^{q} \, S_{p,q}(\bar{u}, \bar{v}) \,+\, S_{0}(\bar{u}, \bar{v}) \right\}}_{\bar{H}^{\text{PSFEx}}(\bar{u}, \bar{v}|u_i,v_i)} \,,
    \label{eq_02:psfex_model}
\end{equation}
where $\bar{I}^{\text{PSFEx}}_{\text{star}, (\cdot|u_i, v_i)} \in \mathbb{R}^{p \times p}$ is the \texttt{PSFEx} reconstruction of the observed star $\bar{I}_{(\cdot|u_i, v_i)}$, $S_{p, q} \in \mathbb{R}^{P \times P}$ represents the learned PSF features or \textit{eigenPSFs}, $S_{0} \in \mathbb{R}^{P \times P}$ represents a first guess of the PSF, the polynomial law is defined to be of degree $d$, and $F^{\text{PSFEx}}$ represents the degradations required to match the model with the observations. The model's PSF reconstruction is represented by $\bar{H}^{\text{PSFEx}}$. The first guess can be computed as a function of the median of all the observations. The dimensions $p$ and $P$ will be the same if no downsampling operation is included in $F^{\text{PSFEx}}$.

The PSF features are learned in an optimisation problem that aims to minimise the reconstruction error between the \texttt{PSFEx} model and the observations, which reads
\begin{equation}
    \min_{\substack{S_{p,q} \\ \forall p,q \geq 0 \,,\, p+q \leq d}} \left\{ \sum_{i=1}^{n_{\text{obs}}} \left\| \frac{\bar{I}_{(\cdot|u_i,v_i)} - \bar{I}^{\text{PSFEx}}_{\text{star}, (\cdot|u_i,v_i)}}{\hat{\sigma}_{i}} \right\|_{F}^{2} + \sum_{\substack{p,q \geq 0 \\ p+q \leq d}} \left\| T_{p,q} \, S_{p,q} \right\|_{F}^{2} \right\} \,,
    \label{eq_02:psfex_optim}
\end{equation}
where $\hat{\sigma}_{i}^{2}$ represent the estimated per-pixel variances, $\bar{I}$ represents the noisy observations, and $\| \cdot \|_{F}$ the Frobenius norm of a matrix. The second term in Equation \ref{eq_02:psfex_optim} corresponds to a Tikhonov regularisation, where $T_{p,q}$ represents some regularisation weights to favour smoother PSF models. The PSF recovery at target positions is straightforward. One needs to introduce new positions in the Equation \ref{eq_02:psfex_model} after learning the PSF features $S_{p,q}$. The recovery at a new FOV position $(u_j, v_j)$ simply writes 
\begin{equation}
    \bar{H}^{\text{PSFEx}}(\bar{u}, \bar{v}|u_j,v_j) = \sum_{\substack{p,q \geq 0 \\ p+q \leq d}} u_{j}^{p} \, v_{j}^{q} \, S_{p,q}(\bar{u}, \bar{v}) \,+\, S_{0}(\bar{u}, \bar{v})\,,
    \label{eq_02:psfex_recovery}
\end{equation}
where $\bar{H}^{\text{PSFEx}}$ is the model's PSF reconstruction, and $S_{0}$ and $S_{p, q}$ were learned during the training procedure.

\subsubsection{Resolved component analysis (RCA)}
RCA\footnote{\href{https://github.com/CosmoStat/rca}{github.com/CosmoStat/rca}} \citep{ngole2016} is a state-of-the-art data-driven method designed for the space-based \textit{Euclid} mission \citep{schmitz2020}. The model builds an independent model for each CCD, can handle super-resolution, and, like \texttt{PSFEx}, is based on a matrix factorisation scheme. However, there are three fundamental changes with respect to \texttt{PSFEx}. The first difference is that, in RCA, the feature weights are defined as a further matrix factorisation, and are also learned from the data. The feature weights are constrained to be part of a dictionary\footnote{In the signal processing community, a dictionary is a collection of templates, or basic elements, used to decompose a signal.} built with different spatial variations based on the harmonics of a fully connected undirected weighted graph. The graph is built using the star positions as the nodes and a function of the inverse distance between the stars to define the edge weights. The rationale of the graph-harmonics dictionary is to capture localised spatial variations of the PSF that occur in space-based missions exploiting the irregular structure of the star positions.
The RCA reconstruction of an observed star is then
\begin{align}
    \label{eq:RCA_obs_star}
    &\bar{I}^{\text{RCA}}_{\text{star}}(\bar{u}, \bar{v}|u_i,v_i) = F^{\text{RCA}} \left\{ \bar{H}^{\text{RCA}}(\bar{u}, \bar{v}|u_i,v_i) \right\} , \text{ where}\quad \\
    & \bar{H}^{\text{RCA}}(\bar{u}, \bar{v}|u_i,v_i) = \sum_{r=1}^{N_{\text{comp}}} S_{r}(\bar{u}, \bar{v}) \, A[r,i] = \sum_{r=1}^{N_{\text{comp}}} S_{r}(\bar{u}, \bar{v}) \, (\alpha V^{\top})[r,i] \,, \nonumber
\end{align}
where $\bar{I}^{\text{RCA}}_{\text{star},(\cdot|u_i,v_i)} \in \mathbb{R}^{p \times p}$, $F^{\text{RCA}}$ corresponds to the degradation model of RCA, including downsampling, intra-pixel shifts among others, $S_{r} \in \mathbb{R}^{D \,p \times D \,p}$ corresponds to the data-driven feature, i.e., eigenPSF, $r$ from a total of $N_{\text{comp}}$ features, $D$ is the upsampling factor in case a super-resolution step is required, and $A[r,i]$ is the $r$-th feature weight of the $i$-th star. The feature weight matrix $A$ is decomposed into a sparse matrix $\alpha$ and a dictionary matrix $V^{\top}$ of graph-based spatial variations; see \citet{ngole2016} for more information. 

To regularise the inverse problem, RCA enforces a low-rank solution by fixing $N_{\text{comp}}$, a positivity constraint on the modelled PSF, a denoising strategy based on a sparsity constraint in the starlet \citep{starck2015} domain, which is a wavelet representation basis, and a constraint to learn the useful spatial variations from the graph-harmonics-based dictionary. The optimisation problem that the RCA method targets is
\begin{align}
    \label{eq_03:optim_rca}
    \min_{S_{k}, \alpha_{k}} &\Bigg\{ \frac{1}{2} \sum_{i=1}^{n_{\text{obs}}} \left\lVert \bar{I}_{(\cdot|u_i,v_i)} - F^{\text{RCA}}\left\{ \bar{H}^{\text{RCA}}_{(\cdot|u_i,v_i)} \right\}\right\rVert_{F}^{2} + \sum_{r = 1}^{N_{\text{comp}}} \| w_{r} \odot \Phi S_{r} \|_1 \\
    &+ \iota_+\left(\bar{H}^{\text{RCA}}_{(\cdot|u_i,v_i)} \right) + \iota_\Omega(\alpha) \Bigg\}\, \quad \text{s.t. } \quad \bar{H}_{i}^{\text{RCA}} = \sum_{r=1}^{N_{\text{comp}}} S_{r} \, (\alpha V^{\top})[r,i], \nonumber
\end{align}
where $w_{r}$ are weights, $\Phi$ represents a transformation allowing the eigenPSFs to have a sparse representation, e.g., a wavelet transformation, $\odot$ denotes the Hadamard product, $\iota_+$ is the indicator function of the positive orthant, and $\iota_\Omega$ is the indicator function over a set $\Omega$, which is defined as a set of sparse vectors and is used to enforce sparsity on $\alpha$. 

The second difference with respect to \texttt{PSFEx} corresponds to the regularisations used in the objective function from Equation \ref{eq_03:optim_rca}, which ends up as being non-convex due to the matrix factorisation and non-smooth due to the $\| \cdot\|_{1}$ constraint. The optimisation is solved through a block coordinate descent, as it is a multi-convex problem, and exploiting proximal optimisation algorithms that tackle the non-smooth subproblems \citep{beck2009, condat2013}.

The third difference is handling the PSF recovery at a new position $(u_j,v_j)$. The recovery is carried out by a radial basis function (RBF) interpolation of the learned columns of the $A$ matrix, issuing a vector, $\hat{\mathbf{a}}_{j} \in \mathbb{R}^{N_{\text{comp}}}$, see \citet{schmitz2020} for more details. This way, the spatial constraints encoded in the $A$ matrix are preserved when estimating the PSF at galaxy positions. The interpolated feature weights $\hat{\mathbf{a}}_{j}$ can be introduced in the Equation \ref{eq:RCA_obs_star} formulation to generate the PSF at the new $j$ position.

The RCA model has yet to be used to generate a WL shape catalogue from real observations. \citet{liaudat2020} showed that RCA is not robust enough to handle real ground-based observations from CFIS as some CCDs exhibited significant errors of the PSF shape.

\subsubsection{The Multi-CCD PSF model (MCCD)}

MCCD\footnote{\href{https://github.com/CosmoStat/mccd}{github.com/CosmoStat/mccd}} \citep{liaudat2020} is a state-of-the-art data-driven method originally designed for the ground-based CFIS from CFHT. MCCD can model the full focal plane at once by incorporating the CCD mosaic geometry into the PSF model. The MCCD model rationale is explained by the limitations of other PSF models that build independent PSF models for every CCD, e.g., RCA and \texttt{PSFEx}, are: (i) fundamentally limited in the possible model complexity due to the lack of constraining power of a reduced number of star observations, and (ii) the difficulty of modelling PSF spatial variations spanning the entire focal plane, i.e., several CCDs, from independently modelled CCDs. MCCD overcomes these issues by building a PSF model containing two types of variations, global and CCD-specific. Both variations are modelled by a matrix factorisation approach, building over the success of \texttt{PSFEx} and RCA. The global features are shared between all CCDs, and the local CCD-specific features aim to provide corrections for the global features. The feature weights are defined as a combination of the polynomial variations from \texttt{PSFEx} and the graph-based variations from RCA. The model's optimisation is more challenging than the previous models and is based on a novel optimisation procedure based on iterative schemes involving proximal algorithms \citep{parikh2014}. 

The MCCD model has proven robust enough to handle real observations from CFIS \citep{liaudat2020}, giving state-of-the-art results. MCCD has been incorporated into the recent \textsc{ShapePipe} shape measurement pipeline \citep{farrens2022} originally designed to process the CFIS survey and generate a WL shape catalogue. The first version of the shape catalogue \citep{guinot2022}, spanning $1700~\text{deg}^{2}$, from \textsc{ShapePipe} used \texttt{PSFEx}. However, the next version, spanning $\sim 3500~\text{deg}^{2}$, used the MCCD PSF model and will be released soon.

\subsubsection{\textit{lens}fit}
\textit{lens}fit \citep{miller2007, kitching2008, miller2013} refers to a Bayesian galaxy shape measurement method for WL surveys. It also includes a data-driven PSF model that will also be referenced as \textit{lens}fit and is sparsely described throughout the different publications involving the shape measurement \textit{lens}fit \citep{miller2013, kuijken2015, giblin2020}. This method has been used with real data to produce the WL shape catalogues of CFHTLenS \citep{erben2013, miller2013}, KiDS-+VIKING-450 \citep{wright2019}, KiDS-450 \citep{hildebrandt2016, fenechconti2017_b}, KiDS-1000 \citep{giblin2020}, and VOICE \citep{fu2018}. However, the code is not publicly available.

This PSF model differs from the previous ones. \texttt{PSFEx} and RCA learn some features or \textit{eigenPSFs} that all the PSFs share. The \textit{lens}fit model is fitted on a pixel-by-pixel basis. Each pixel is represented as a polynomial model of degree $d$ of the FOV positions. The \textit{lens}fit model can use all the observations in one exposure, meaning that it uses several CCDs at once. The model uses the low-order polynomials, up to degree $n_{\text{c}} < d$, to be fitted independently for each CCD and the rest of the monomials are fitted using the observations from all the CCDs. This multi-CCD modelling is a significant change with respect to previous methods that built independent models for each CCD. The total number of coefficients \textit{per pixel} is
\begin{equation}
    N_{\text{coeff}} = \frac{1}{2} \left((d + 1)(d + 2) + (N_{\text{CCD}} - 1)(n_{\text{c}} + 1)(n_{\text{c}} + 2) \right) \,,
\end{equation}
where $N_{\text{CCD}}$ is the total number of CCDs in the camera, $d$ represents the degree of the polynomial varying in the full FOV, and $n_{c}$ the polynomial that is CCD-dependent. We can write the description of the pixel $(\bar{u},\bar{v})$ of the PSF model for a FOV position $(u_j,v_j)$ in CCD $k$ as follows
\begin{equation}
    \bar{H}^{\text{lensfit}}(\bar{u},\bar{v}|u_j,v_j) = \sum_{\substack{p,q \geq 0 \\ p+q \leq n_{\text{c}}}} u_{j}^{p} \, v_{j}^{q} \, a_{(p,q),(\bar{u},\bar{v})}^{k} + \sum_{\substack{p+q > n_{\text{c}} \\ p+q \leq d}} u_{j}^{p} \, v_{j}^{q} \, b_{(p,q),(\bar{u},\bar{v})} \,,
\end{equation}
where $a_{(p,q),(\bar{u},\bar{v})}^{k}$ is the coefficient specific of CCD $k$, pixel $(\bar{u},\bar{v})$, and polynomial $(p,q)$ to be fitted to the observations. The coefficient $b_{(p,q),(\bar{u},\bar{v})}$ is shared by all the CCDs.

One thing to notice in this approach is that as the modelling of the PSF is done pixel-by-pixel, and consequently every observation should share the same pixel grid of the PSF. There is no guarantee that an observation will have its centroid aligned with the chosen pixel grid. Therefore, the PSF model has to be aligned with the observations. Other methods, like \texttt{PSFEx} and RCA, interpolate the model to the observation's centroids. However, \textit{lens}fit interpolates all the observations to the model's pixel grid with a sinc function interpolation which implies interpolating noisy images. This procedure is described in \citet{kuijken2015}.

For the KiDS DR2 \citep{kuijken2015}, the hyperparameters used by \textit{lens}fit are $n_c=1$, $d=3$, and $N_{\text{CCD}}=32$ (from CFHT's OmegaCAM instrument), where the images used belong to a $32 \times 32$ pixel grid. When fitting the model's parameters, each star is given a weight that is a function of its SNR with the following empirical formula
\begin{equation}
    w_i = \frac{s_{i}^{2}}{s_{i}^{2} + 50^{2}} \,,
\end{equation}
where $s_{i}$ is the measured SNR of the star $i$. 

\subsubsection{PSFs In the Full Field-of-View (PIFF)}
PIFF\footnote{\href{https://github.com/rmjarvis/Piff}{github.com/rmjarvis/Piff}} \citep{jarvis2020} is a recently developed PSF model that was used for the DES year $3$ WL shape catalogue \citep{gatti2021} replacing \texttt{PSFEx} that was used for the DES year $1$ release. The PIFF model targets the LSST survey. Some improvements of PIFF with respect to \texttt{PSFEx} are the ability to use the full focal plane to build the model and modelling the PSF in sky coordinates rather than pixel coordinates. PIFF offers a modular and user-friendly design that will enable further improvements. The change of modelling coordinates was motivated by the strong tree ring detector effect observed in the DES instrument, Dark Energy Camera, which introduces astrometric distortions that are easier to correct in sky coordinates. Pixel coordinates refer to the coordinates defined on the pixel grid of the instrument. In contrast, sky coordinates refer to the angles in the celestial sphere, known as right ascension (RA) and declination (DEC). The geometric transformations that allow going back and forth between the pixel and sky coordinates are known as World Coordinate System (WCS) transformations.

Being a modular PSF modelling code, PIFF allows choosing between different options for the PSF model and the interpolation method. For example, the model can be an analytical profile like a Gaussian, a Moffat or a Kolmogorov profile, or a more general non-parametric profile called \texttt{PixelGrid}. The interpolation method can be a simple polynomial interpolation, K-nearest neighbours method, a Gaussian process (also known as Kriging), or a \textit{Basis-function polynomial interpolation}. Let us clarify the difference between the first and last mentioned interpolations. The simple polynomial interpolation first fits the PSF model's parameters $\mathbf{p}$ for each observed star. Then, it fits the coefficients of a polynomial of the 2D star positions that will later be used to interpolate. In the \textit{Basis-function polynomial interpolation}, the position polynomial's interpolation coefficients are fitted simultaneously with the model's parameters using all the available stars (from a single CCD or the entire exposure). If this last option is used with the \texttt{PixelGrid} model, it comes closer to the approaches of \texttt{PSFEx} and RCA without the specific characteristics of each model. We have only mentioned position polynomials, but, as in \texttt{PSFEx}, the interpolation polynomial can be built on any parameter of the PSF, as, for example, a colour parameter.

The current PIFF PSF model includes an outlier check after the algorithm has converged. The outlier check is based on the chi-squared, $\chi^{2}$, pixel residual error between the model and the observations. The model implements an iterative refining approach which means that after the model has converged, one (or more) outlier star(s) is(are) removed from the observations. A new iteration then starts with the model being recomputed. Although this approach effectively removes outlier stars not representative of the PSF (because they are binary stars or have some contamination), it can be very computationally demanding. The computing time increases linearly with the number of iterations used, which might be problematic depending on the total area to analyse or the available computing resources. We refer the reader to \citet{jarvis2020} for more details.

The DES year $3$ shape catalogue \citep{gatti2021} used the \textsc{PIFF} model. The model was a \texttt{PixelGrid} with Basis-function polynomial interpolation using a $3$rd order polynomial. Even if PIFF has the \textit{potential} to build a model across several CCD chips, in practice, each model was built independently for each CCD.

\subsubsection{WaveDiff and differentiable optics approaches}

\begin{figure}
    \centering
    \includegraphics[width=0.95\textwidth]{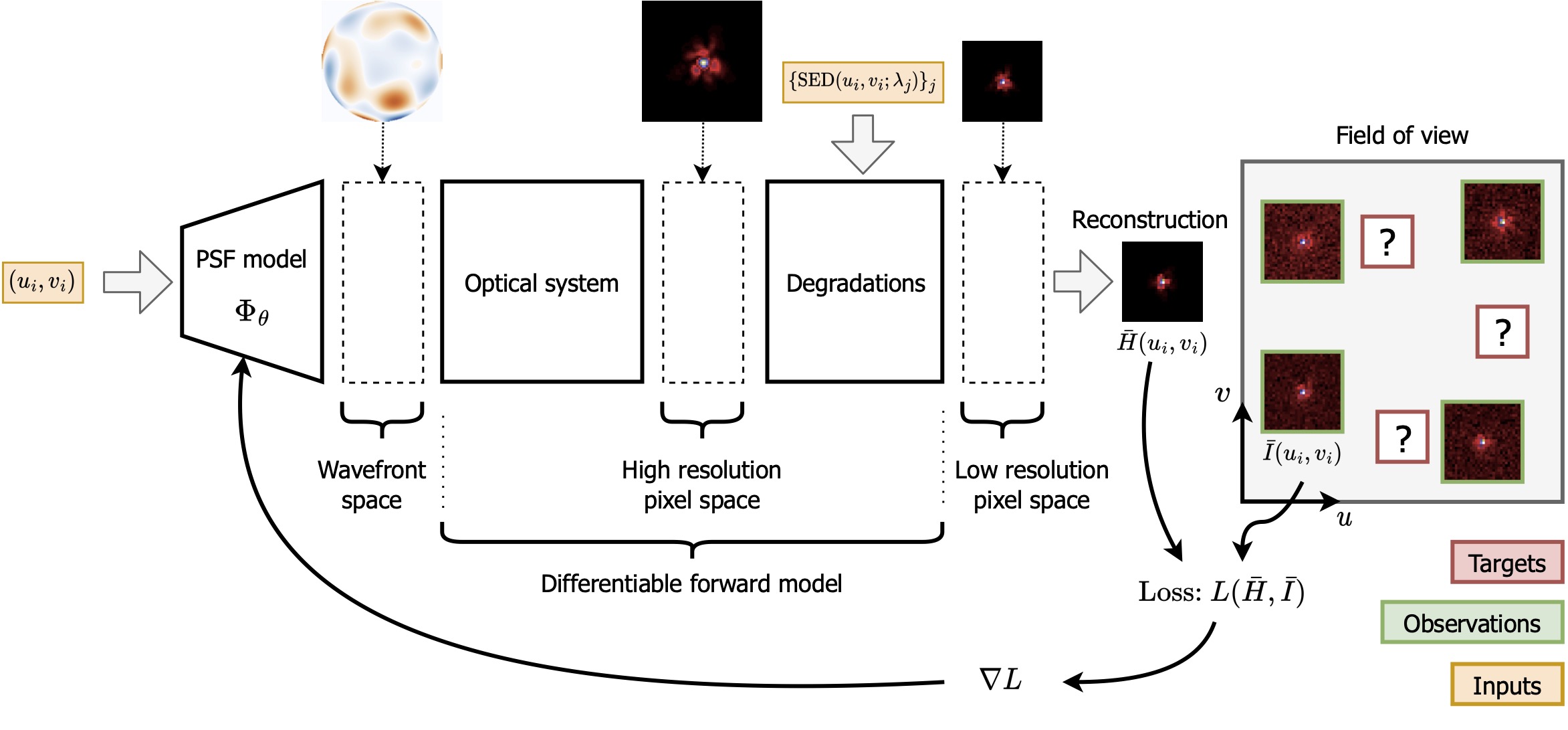}
    \caption{Overview of the WaveDiff approach to model the PSF with a differentiable optical forward model. Figure adapted from \citet{liaudat2023}.}
    \label{fi_02:wavedif_illustration}
\end{figure}

The WaveDiff\footnote{\href{https://github.com/CosmoStat/wf-psf}{github.com/CosmoStat/wf-psf}} \citep{liaudat2023,liaudat2021} PSF model was recently developed targeting space telescopes, in particular the \textit{Euclid} mission \citep{laureijs2011}. WaveDiff proposes a paradigm shift for \textit{data-driven} PSF modelling. Instead of building a data-driven model of the PSF in the pixel space as the previous models, WaveDiff builds its model in the wavefront error (WFE) space. This change relies on a differentiable optical forward model that allows propagating the wavefront from the pupil plane to the focal plane and then computing the pixel PSF. The model is based on two components, a parametric WFE and a data-driven WFE. The parametric WFE can be based on optical simulations, characterisations of the optical system, or complementary measurements such as phase diversity observations. The parametric part should aim to model very complex dependencies that cannot be inferred from the degraded star observations. \citet{liaudat2023} proposes a parametric model built using fixed features, namely the Zernike polynomials \citep{noll1976}. For several reasons, the parametric WFE model obtained often cannot accurately represent the observed PSF, e.g., the telescope changes over time, there are errors in the parametric model built, and relevant effects were not considered or neglected. Consequently, the data-driven WFE should be able to correct the aforementioned mismatches. This data-driven part is based on a matrix factorisation with spatial variations inspired from \texttt{PSFEx} and RCA. It is crucial to model smooth variations that have a reliable generalisation of the PSF to different positions in the FOV. An overview of the model is presented in Figure \ref{fi_02:wavedif_illustration}.

Estimating the model's parameters in the WFE space is a challenging, ill-posed inverse problem known as phase retrieval \citep{fienup1993,fienup1993_2,schechtman2015}, i.e., estimating a complex signal from intensity-only observations. The optimisation problem is non-convex and non-smooth, the star observations are not very informative, and there is no guarantee that the WFE model's structure can represent the underlying ground truth WFE. \citet{liaudat2023} shows that under the aforementioned conditions, targeting the estimation of the ground truth WFE is not the best option. The PSF model's objective is to have a good pixel representation of the PSF. It is, therefore, possible to estimate a WFE manifold far away from the underlying WFE but very close in the pixel space. The data-driven features, the basis of the WFE manifold, are estimated with a stochastic gradient descent method widely used for estimating neural network parameters.

The WaveDiff model can handle spatial variations, super-resolve the PSF, and model chromatic variations thanks to the WFE formulation and the optical forward model, which also considers more general degradations as in Equation \ref{eq_02:approx_star_obs_model}. To the best of our knowledge and at the time of writing, this is the \textit{only} data-driven PSF model able to cope with the spectral variations of the PSF. WaveDiff shows a breakthrough in performance for data-driven PSF models in a simplified \textit{Euclid}-like setting \citep{liaudat2023}. WaveDiff is flexible enough to be adapted to different space telescopes. The framework proposed shows an exciting research direction for future data-driven PSF modelling. The WaveDiff model has yet to be tested with real space-telescope observations and needs to incorporate detector-level effects, which are more naturally modelled in pixel space. We refer the reader to \citet{liaudat2023} and \citet{liaudat2022_thesis} for more details. 

More general approaches based on differentiable optics have been recently emerging, e.g., $\partial$Lux\footnote{\href{https://github.com/LouisDesdoigts/dLux}{github.com/LouisDesdoigts/dLux}} \citet{desdoigts2023}. Approaches based on automatic differentiation can be useful not only to model the PSF but have been used to design apodizing phase-plates coronagraphs \citet{Wong2021} and detector calibration \citet{desdoigts2023}.

\subsubsection{Other PSF models}
\begin{itemize}
    \item \textit{Shapelets}: \citet{refregier2003a} proposed a framework to analyse images based on a series of localised basis functions of different shapes named \textit{shapelets}. Images can then be decomposed using these basis functions. \citet{refregier2003b} continued the work proposing the \textit{shapelet} framework to be used for building shear estimates and modelling the PSF. The PSF modelling consists of decomposing the star images in the \textit{shapelet} basis and then performing an interpolation of the coefficients to positions of interest. Essentially, it is an extension of the approach seen in shape interpolation. Expressing the image in \textit{shapelet} coefficients allows denoising the star images and provides an easier framework for the galaxy-PSF deconvolution. However, capturing all the PSF structures in a finite expansion over analytical functions is not always possible, leading to lost information. \citet{massey2005} extended the framework from Cartesian to polar \textit{shapelets}.
    \item \textit{Moffatlets and Gaussianlets}: \citet{li2016} proposed two other series of basis functions to decompose the PSF named \textit{Moffatlets} and \textit{Gaussianlets}. \citeauthor{li2016} compared the PSF reconstruction using the aforementioned basis with a PCA-based method on LSST-like images. Using analytical basis functions leads to more denoised models, as expected. \citet{nie2021} continued the approach and forced the principal components being learned in the PCA-like algorithm to be built using the \textit{Moffatlets} basis. This choice avoids the principal components of learning noise as the \textit{Moffatlets} basis avoids it. Both analyses lack a performance benchmark with a reference PSF model like \texttt{PSFEx}. In addition, the models' performance is computed at the same position as the observed stars, so the models' generalisation to other positions, a fundamental task of the PSF model, still needs to be studied.
    \item \textit{Fourier-based methods}: \citet{zhang2007} proposed a Fourier-based method for directly measuring the cosmic shear taking into account the PSF, which was further developed in several publications \citep{zhang2011, zhang2015, lu2017, zhang2019}. The method is based on the quadrupole moments of the galaxy images (described in detail in Section \ref{sc_02:moment_based_metrics}) but is measured in Fourier space. The handling of the PSF is also done in Fourier space. \citet{lu2017} explores different interpolation approaches for the PSF in the aforementioned Fourier framework. The 2D power spectrum of the observed PSFs is interpolated to target positions. The interpolation is done pixel-by-pixel, and the best-performing method is a well-parametrised polynomial interpolation. An advantage of the Fourier interpolation is that the 2D power spectrum is automatically centred in Fourier space, simplifying the handling of images with intra-pixel shifts or, what is the same, different centroids. Another Fourier-based shear measurement method is the Bayesian Fourier Domain (BFD) \citep{bernstein2014_2,bernstein2016} built on the Bayesian formalism. However, it does not include a specific PSF model.
    \item \textit{Optimal transport (OT)-based methods}: There exist two approaches involving OT \citep{peyre2019} to tackle the PSF modelling problem. \citet{ngole2017} proposes to use Wasserstein barycenters as a non-linear geometric-aware interpolation procedure of a low-dimensional embedding representation of the PSFs. Although elegant, the performance of the approach does not seem to justify its computational burden. In the comparison method, an RBF interpolation of the principal components obtained through PCA achieves a similar performance. The performance of the PCA method is better in terms of ellipticity but slightly worse in terms of pixel error and PSF FWHM. \citet{schmitz2019} worked on a data-driven PSF model based on RCA that would be able to model the chromatic variations of the PSF through the use of Wasserstein barycenters that were previously developed in \citet{schmitz2018}. The OT-based PSF model coined $\lambda$RCA was compared to RCA. The comparison showed a lower pixel and size error for $\lambda$RCA, although the ellipticity error was similar or better for RCA. This method assumes that the PSF's chromatic variation is smooth over all the passband. This assumption holds if the only chromatic effect of the PSF is due to the diffraction phenomena, which exhibits a smooth variation with the $1/\lambda$ dependence in the WFE that was already presented in Equation \ref{eq_02:intensity_psf}. However, if this is not the case and another non-smooth chromatic variation is present, currently occurring in \textit{Euclid} \citep{venancio2016, baron2022}, there is no straightforward way to adapt the $\lambda$RCA model to account for it.
    \item \textit{Wavefront approach}: \citet{soulez2016} proposed to model the propagation of light through the mirrors of the optical system. The PSF modelling problem is recast into a phase retrieval problem. The article is a proof-of-concept as there are only qualitative results, and many PSF-modelling difficulties remain unaddressed.
    % %
    \item \textit{Exploit out-of-focus images}: Some instruments, like the Dark Energy Camera (DECam) \citep{flaugher2015}, are equipped with wavefront sensors that are helpful for focus, alignment and adaptive optic system (AOS) \citep{roddier93, roodman2010, roodman2012}. The LSST camera \citep{LSST2009} will also be equipped with wavefront sensors \citep{manuel2010, xin2015, xin2016, claver2012}. \citet{roodman2014} proposed to use the data from the wavefront sensors to constrain the optical contribution of the PSF. The work was continued by \citet{davis2016} that proposed a wavefront-based PSF model for the DECam instrument using out-of-focus observations. The PSF model is based on Zernike polynomials fitted to out-of-focus stars, also called doughnuts, that contain considerably more wavefront information than in-focus stars. Then, a new fit is done for each exposure based on the measured quadrupole moments of the in-focus star images. It is not easy to understand at which point the quadrupole moments constrain the proposed PSF model and at which point it is the base physical wavefront measured from the out-of-focus images that is the only part driving the performance of the model. \citet{snyder2016} proposed using the AOS measurements to characterise the atmospheric turbulence in terms of a Zernike decomposition.
    \item \textit{Deep learning approaches}: A model coined PSF-NET was proposed by \citet{jia2020c} and is based on two convolutional neural networks (CNNs) trained jointly. One network has to transform high-resolution images into low-resolution images, while the other has to do the opposite. The CNNs are trained in a supervised way expecting that the first network will learn a PSF manifold. However, it is unclear how the approach handles the spatial variation of the PSF, and it has not been tested for WL purposes. \citet{jia2020b} proposed another approach for PSF modelling based on denoising auto-encoders, but the spatial variation of the PSF remains untackled. Another approach is followed by \citet{herbel2018}, where the PSF profile is modelled by a parametric function consisting of a base profile of two Moffat profiles and several parametrised distortions to increase the expressivity. A CNN is trained in a supervised manner to predict the parameters of the parametric profile from a noisy star observation. The neural network provides a good estimation of the parameters, but the spatial variation of the PSF is, again, not addressed. Having a PSF model that can model the observations is important. However, in PSF modelling for WL analysis, a crucial part is to capture the spatial variations of the PSF and that the model outputs the PSF at different positions in the FOV. 
\end{itemize}

\section{Desirable properties of PSF models}
\label{sc_02:PSF_modelling_comments}
In the previous section, we reviewed some of the most relevant PSF models developed so far. After studying many models, we can conclude on desirable PSF model properties. The PSF model should:
\begin{itemize}
    \item[(a)] \textit{Have an accurate modelling of the PSF light profile.} This modelling is essential for any target task, as the light profile is the convolutional kernel for a given position. The smoothness and structure in the PSF profile are a consequence of the PSF being the Fourier transform, in Fraunhofer's approximation, of a particular finite-length aperture that limits the frequency content of the PSF. This frequency limitation prevents us from having a Dirac distribution as a PSF, as it would require an infinite frequency content to build it. One difficulty is accurately modelling the PSF's wings, or outer region, which is often below the noise level. In ground-based telescopes, the effect of the atmosphere can be interpreted as a low-pass filter for the PSF, smoothing the PSF light profile.
    \item[(b)] \textit{Produce noiseless estimations of the PSF.} 
    The presence of noise in the PSF estimations is an issue for purely data-driven models, which sometimes tend to overfit noisy observations. Some regularisations have to be introduced in the PSF model parameter optimisation to avoid producing noisy PSFs. 
    A seemingly straightforward solution to this problem is to rely on PSF models based on fixed basis functions like Shapelets or Moffatlets. These models will always output denoised PSFs as their basis functions cannot reproduce the noise. However, they will introduce modelling errors if they cannot accurately model the observed PSF light profiles.
    \item[(c)] \textit{Capture the PSF field's variations.} 
    It is often the case that a good quality PSF is required at the position of a certain object where no direct information about the PSF is available. The PSF model first needs to capture most of the relevant information from the observations at other positions and wavelengths. Then, the model exploits this information and predicts the PSF at the required position and wavelength. The PSF model relies upon its generalisation power as it is required to exploit the PSF field information from other positions.
    \item[(d)] \textit{Be able to exploit the structure of the PSF field variations.} 
    This desired property is related to the previous point (c). An exciting approach to obtaining good generalisation power is to learn the structure of the PSF field variations. This structure is a consequence of the physical properties of the telescope's optical system. The subsection below provides a physical understanding of the PSF field structure, which imposes a certain smoothness to the variations. The spatial variations are also structured due to the atmosphere if we study the PSF field of a ground-based telescope. A data-driven PSF model should use a low-complexity representation of the PSF field, which would be able to learn its structure and spatial variations. 
    \item[(e)] \textit{Handle discontinuities of the PSF field.} 
    The CCD misalignments are a source of discontinuity of the PSF's spatial variations. The PSF field is piece-wise continuous, and the borders delimiting the discontinuity are well known as the geometry of the focal plane is accurately known. A straightforward way to handle the discontinuities is to model the PSF for each CCD independently, e.g., \texttt{PSFEx}. Although this is simple to implement, it limits the number of stars available to constrain the PSF model. Another more difficult but potentially more powerful approach is to build a PSF model for the entire focal plane, accounting for the focal plane discontinuities, e.g., MCCD. Another source of discontinuity is segmented mirrors, e.g. JWST's hexagonal mirrors seen in Figure \ref{fi:real_JWST_OPD_a}.
    \item[(f)] \textit{Be robust to outliers and contaminations of the star sample.} 
    Contaminations can come from the selection of stars, the fact that the objects classified as stars are good representations of the PSF and are not small galaxies or binary stars \citep{kuntzer2017}. Outliers can come from imperfect image preprocessing where detector effects, like CTI, have not been adequately removed. In addition, the model should be robust to different observation conditions such as spatial distributions of the observed stars, SNRs, and the number of observed stars.
    \item[(g)] \textit{Work appropiately with the target task.} 
    The PSF model can be exploited differently according to the target task's objective, e.g., estimating a deconvolved object or estimating some summary statistic of the deconvolved image. The model should be developed with the task in mind, as each task might be more or less susceptible to different kinds of errors. 
    \item[(h)] \textit{Be fast.}
    Upcoming surveys will process a vast amount of observations. Consequently, they put significant pressure on the computing time of PSF models as they need to cope with the data intake. The requirements in terms of computing time can drive many design choices in a PSF model, preventing the use of costly physical simulations.
\end{itemize}

Once the PSF model has been developed with all the aforementioned properties in mind, it is essential to validate the model's performance. The validation should ensure that the expected performance of the model is achieved, and it should help to identify sources of problems and provide directions for the improvement of the model. In the next section, we give an overview of validation methods for PSF models.

%%% 
\section{Validation of PSF models}
\label{sc_02:validation_psf}

The validation of PSF models is a challenging problem. To derive a validation method, it would be necessary to quantify the impact of PSF modelling errors on the final objective of our analysis. We consider, as an example, a weak-lensing-based cosmological analysis, where the objective is to derive constraints on the parameter of the cosmological model under analysis. This exercise is challenging, given the analysis' complexity and the large data volume. Nevertheless, with some simplifying assumptions, it was carried out to set the PSF modelling requirements for the \textit{Euclid} mission as shown in Section \ref{sc_02:psf_error_propagation}. In this analysis, some assumptions on the PSF shape used do not always hold for the high complexity of the PSF in a space-based mission like \textit{Euclid}. Even though it is essential to derive requirements for the PSF model, these do not give much information on the nature of the errors and possible problems the PSF model has. Therefore, it is necessary to derive different diagnoses or null tests. \citet{jarvis2016} proposed a set of null tests for the DES WL shear catalogues science verification, including the PSF model validation. 

The most basic rule for any validation of PSF models is to separate the observations in the FOV into two datasets for estimation and testing, i.e., validation, which could be $80\%$ and $20\%$, respectively. The first one should be used to estimate the PSF model. The second one should help validate its performance and not be used in the PSF model estimation. This rule tests the PSF model's generalisation power to unseen positions in the FOV. Next, we will describe the most used PSF diagnosis that will help us to validate the performance of the PSF models.

\subsection{Pixel-based metrics}
The most straightforward diagnostic we can think of is to compute the pixel residual of our PSF model. Once trained, the model is used to recover the PSF at test positions. We can then compute the RMSE of the pixel reconstruction residuals. The PSF model can ideally predict the observed test stars without error, and the reconstruction residual would only contain the observational noise. If we work with simulations, we can produce noiseless stars for our testing set, and the RMSE will directly indicate the pixel reconstruction error. Even though pixel-based metrics can give insight into the PSF model performance, they are not easily interpretable regarding scientific impact. Errors in the PSF core or the PSF wings can impact the observed galaxy's estimated shape differently. With the existing methodology, it is difficult to translate a pixel-based metric into a scientifically meaningful quantity in terms of error propagation.

When working with real data, the PSF model's validation with pixel-based metrics becomes more complicated. The different noise levels in the data can hide the pixel reconstruction error, making it difficult to compare different PSF models or even assess the performance of a single one. \citet{liaudat2020} proposed pixel-based reconstruction metrics for real observations. Let us denote with $I_{\text{star}}(\bar{u},\bar{v}|u_i,v_i), I_{\text{model}}(\bar{u},\bar{v}|u_i,v_i) \in \mathbb{R}^{p \times p}$ a test star and the predicted PSFs, respectively, at the FOV position $(u_i,v_i)$, where $p^{2}$ is the total number of pixels in the image. To simplify notation we write $I(\bar{u},\bar{v}|u_i,v_i) = I_{i}$. In star images, most of the PSF flux is concentrated in the centre of the square image, and the noise level can be considered constant in the image. Therefore, we can mask the image to only consider the central pixels within a given radius of $R$ pixels and compute the pixel RMSE of the masked images. We note $\tilde{I} = I \odot M_{R}$ the masked image, where $M_{R} \in \{0,1\}^{p \times p}$ is a binary mask, and $\odot$ is the Hadamard or element-wise product. Let us define the $\sigma$ value as
\begin{equation}
    \sigma( \tilde{I}_{i} ) = \sigma\left( M_{R} \odot I_{i} \right) = \left( \frac{1}{\tilde{p}^{2}} \sum_{\bar{u},\bar{v} = 1}^{p} \left(\tilde{I}_{i}(\bar{u},\bar{v}) \right)^{2} \right)^{1/2} \,,
\end{equation}
where $\tilde{p}^{2}$ is the number of unmasked pixels, and the sum is done on the unmasked pixel values. The first pixel metric is $Q_{p_1}$ and is defined as
\begin{equation}
    Q_{p_1} = \left( \text{Err}^{2} - \sigma_{\text{noise}}^{2} \right)^{1/2} \,,
\end{equation}
where
\begin{equation}
    \text{Err}^{2} = \frac{1}{N_{s}}\sum_{i=1}^{N_{s}} \sigma\left(\tilde{I}_{\text{star}, i} - \tilde{I}_{\text{model}, i} \right)^{2} ,\, \text{and} \; \sigma_{\text{noise}}^{2} = \frac{1}{N_{s}}\sum_{i=1}^{N_{s}} \sigma\left( \tilde{I}^{*}_{\text{star}, i} \right)^{2}\,,
\end{equation}
where the general noise standard deviation, $\sigma_{\text{noise}}$, is computed from the pixels on the outer region of the test stars, i.e., $\tilde{I^{*}}$, which we define as $\tilde{I^{*}} = I_{\text{star}} \odot M_{R}^{*}$, where $M_{R}^{*}$ is such that $ M_{R}^{*} + M_{R} = 1_{p \times p}$. Subtracting our estimated model, $I_{\text{model}}$, from an observed star, $I_{\text{star}}$, should lead to a residual map containing only noise if the model is perfect. The probability of having our model correlated with the noise is minimal. Therefore, the method with the smallest $Q_{p_1}$ can be considered the best from the $Q_{p_1}$ point of view.

The next two metrics, $Q_{p_2}$ and $Q_{p_3}$, help quantify the model noise. Let us define $\sigma^{2}_{\text{model}, i} = [ \sigma(\tilde{I}_{\text{star}, i} - \tilde{I}_{\text{model}, i})^{2} - \sigma(\tilde{I}^{*}_{\text{star}, i})^{2} ]_{+}$, where the operator $[\cdot]_+$ sets to zero negative values. Then, both metrics are defined as follows
\begin{equation}
    Q_{p_2} = \left( \frac{1}{N_{s}} \sum_{i=1}^{N_{s}} \sigma^{2}_{\text{model}, i} \right)^{1/2}\,,\; Q_{p_3} = \left( \frac{1}{N_{s}} \frac{1}{N_{s}} \sum_{i=1}^{N_{s}} ( \sigma^{2}_{\text{model}, i} - Q_{p_2}^{2})^{2} \right)^{1/4} \,.
\end{equation} 
The $Q_{p_2}$ metric represents the modelling error expectation for a given star, and the $Q_{p_3}$ metric indicates the fluctuation of the modelling error. A perfect PSF model would give values close to zero for the three metrics. We have assumed that there is no background contamination in the observed test stars or that it has been removed.

\subsubsection{Chromatic PSF models}

Some applications or analyses require a chromatic PSF model, and it is essential to validate the chromaticity of the PSF model. This monochromatic validation means validating the PSF at every single wavelength or validating the monochromatic PSF before it is integrated into the instrument's passband. A PSF model with a good performance in reproducing the polychromatic stars does not necessarily have a good monochromatic performance. Supposing that is the case and even if the spectra of the different objects are known in advance, the PSF errors will be more significant when used with objects with considerably different spectra, e.g., galaxies. Chromatic PSF models will generally be required if the observing instrument has a wide passband, e.g. the \textit{Euclid}'s VIS instrument passband goes from $550$nm to $900$nm. The pixel RMSE can be computed for monochromatic PSFs in the passband as done in \citet{liaudat2023}. However, it is cumbersome to validate with real data as we usually do not have access to the monochromatic PSFs of the instrument under study. Consequently, the monochromatic validation might only be possible with simulations.

\subsection{Moment-based metrics}
\label{sc_02:moment_based_metrics}
Weak gravitational lensing analyses are interested in measuring the shape of galaxies as the measured ellipticity is an estimator of the shear. Cosmologists have developed formulations to relate the PSF errors, expressed in terms of shape and size metrics \citep{massey2012}, to the cosmological parameters of interest \citep{cropper2013}. Therefore, it seems natural to have diagnosis metrics based on the ellipticity and size of the PSF. These metrics are determined using the moments of the polychromatic observation $\bar{I}[u, v]$. Following \citet{hirata2003}, we redefine the image moments that we will use in practice, including a weight function as follows
\begin{align}
    \label{eq_03:psf_moments_1}
    <\mu> & = \frac{\int \mu \; \bar{I}(\bar{u},\bar{v}) \; w(\bar{u},\bar{v}) \; \text{d}\bar{u} \, \text{d}\bar{v} }{\int \bar{I}(\bar{u},\bar{v}) \; w(\bar{u},\bar{v}) \; \text{d}\bar{u} \, \text{d}\bar{v}}, \\
    M_{\mu \nu} &= \frac{\int \bar{I}(\bar{u},\bar{v}) \; (\mu - <\mu>) \; (\nu - <\nu>) \; w(\bar{u},\bar{v}) \; \text{d}\bar{u} \, \text{d}\bar{v}}{\int \bar{I}(\bar{u},\bar{v}) w(\bar{u},\bar{v}) \; \text{d}\bar{u} \, \text{d}\bar{v}},
    \label{eq_03:psf_moments_2}
\end{align}
where $\mu , \nu \in \{\bar{u}, \bar{v}\}$, $<\mu>$ denotes the mean of $\mu$, and $w(\bar{u}, \bar{v})$ is a weight function that helps in noisy settings. The weight function is also useful to compute the moments from diffraction-limited PSFs, e.g., an Airy profile, as they prevent the integral from diverging due to the wings of the PSF. Equation \ref{eq_03:psf_moments_1} defines the first-order moments, while Equation \ref{eq_03:psf_moments_2} defines the second-order moments. The ellipticities, or \textit{shape} metrics, are defined as
\begin{equation}
    e = e_1 + {\text i} e_2 = \frac{(M_{\bar{u} \bar{u}} - M_{\bar{v} \bar{v}}) + {\text i} \, 2 M_{\bar{u} \bar{v}}}{M_{\bar{u} \bar{u}} + M_{\bar{v} \bar{v}}} \,,
    \label{eq_03:ellip_definition}
\end{equation}
where ${\text i}$ is the imaginary unit, and the size metric is defined as
\begin{equation}
    R^{2} = T = M_{\bar{u} \bar{u}} + M_{\bar{v} \bar{v}} \,.
    \label{eq_03:size_definition}
\end{equation}
One widely used method to estimate these metrics is the widely-used adaptive moment algorithm from \textsc{GalSim}'s HSM module \citep{hirata2003,mandelbaum2005}. The adaptive moment algorithm measurement provides $\sigma$ as size, which relates to the above-defined size metric as $R^2 = 2\sigma^2$. The measurements are carried out on well-resolved polychromatic images. If the observations are undersampled, as is the case for \textit{Euclid}, a super-resolution step is required for the model. \citet{gillis2020} proposed alternative metrics, based on the image moments, that target the validation of space-based PSFs with emphasis on the HST PSF.

The measurements of the shape parameters based on the image moments are susceptible to image noise. If we are working with real data, we do not have access to ground truth images and are obliged to work with noisy observations. Therefore, we need to average over many objects in order to be able to conclude from the different diagnostics. We continue by presenting different moment-based metrics.

\subsubsection{Shape RMSE}
We start with a set of test stars and their corresponding PSF estimations. Then, the most direct moment-based metric is to compute the RMSE of the ellipticities and size residuals between the observations and the model prediction. However, this metric could be more insightful as it does not provide any information about the composition of the residuals and the estimation biases involved.

\subsubsection{\textit{Meanshapes}}
\setcounter{subfigure}{0}
\begin{subfigure}
\setcounter{subfigure}{0}
    \centering
    \begin{minipage}[b]{0.49\textwidth}
        \centering
        \includegraphics[width=\linewidth,trim={0cm 0cm 0 0},clip,]{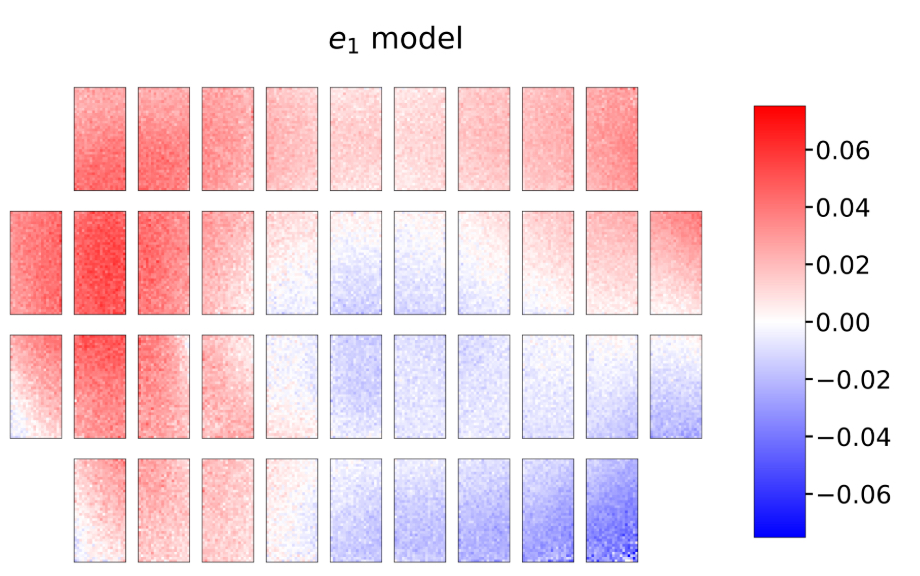}
        \caption{}
        \label{fi_02:meanshapes_e1_model}
    \end{minipage}  
\setcounter{subfigure}{1}
    \begin{minipage}[b]{0.49\textwidth}
        \centering
        \includegraphics[width=\linewidth,trim={0cm 0cm 0 0},clip,]{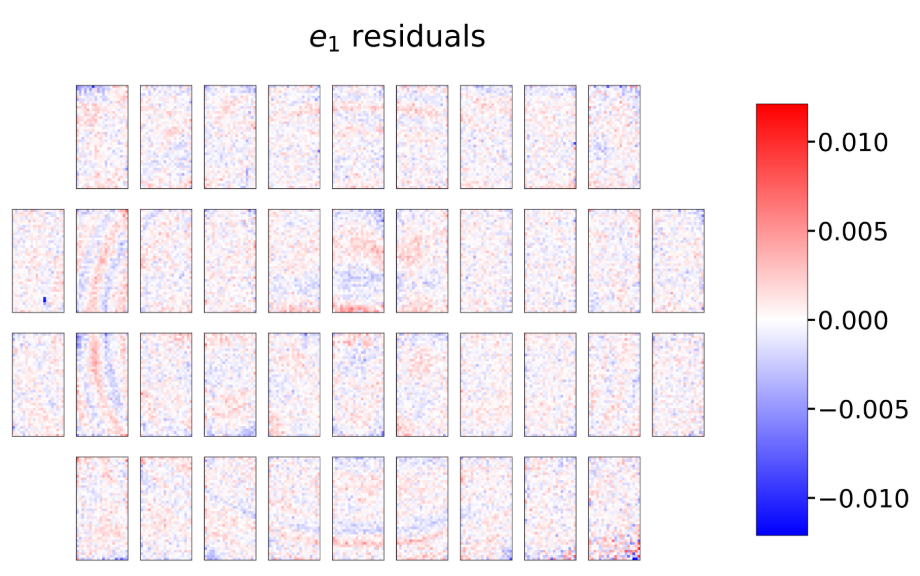}
        \caption{}
        \label{fi_02:meanshapes_e1_residuals}
    \end{minipage}
\setcounter{subfigure}{-1}
    \caption{\textit{Meanshapes} plots showing the first component of the ellipticity of the PSF model in \textbf{(a)} and of its residuals with the observed stars in \textbf{(b)}. The figure shows the $40$-CCD mosaic from the \textsc{MegaCam} instrument's focal plane at the Canada-France-Hawaii Telescope. Figure reproduced from \citet{guinot2022}.} 
    \label{fi:meanshapes_e1}
\end{subfigure}
A useful diagnostic is to compute the spatial distribution of the ellipticities and size residuals, which we coin \textit{meanshapes}. See Figure \ref{fi:meanshapes_e1} for an example. This diagnostic allows us to inspect if there are spatial correlations in the shape and size residuals, indicating that the PSF model is not capturing certain spatial variations from the PSF field. In order to have a finely sampled distribution, we need to average over many exposures, as the available stars from a single exposure are not enough to observe patterns. The shape measurements are also noisy; therefore, averaging over many exposures allows us to smooth out the residuals and observe systematic modelling errors. In practice, we divide the focal plane into several spatial bins, consider several exposures, and then the value of each bin is built by averaging the residuals of all the stars within that bin. A ground-based survey allows us to average the ellipticity contribution of the atmosphere \citep{heymans2012}, as it can be considered a random field with a zero mean. Then, the observed ellipticity distribution over the focal plane is due to the telescope's optical system that is consistent in every exposure. It is also possible to plot the same spatial distribution but observe the positions of the stars. Such a plot will help to observe if there are regions of stellar under-density that could eventually affect the PSF model. 

We have assumed that the ellipticities and size are good descriptors, or summary statistics, of the PSF shape. If this is not the case, the diagnostic could be extended with more accurate descriptors.

\subsubsection{$\rho$-statistics}
\label{sc_02:rho_statistics}
\citet{rowe2010} proposed to compute the auto- and cross-correlations of the ellipticities and their residuals as a diagnostic. The diagnostic was then expanded by \citet{jarvis2016} to a combination of ellipticities, sizes, and residuals. The $\rho$-statistics are helpful to identify PSF modelling biases and detect at which scales it is most affecting the weak lensing analysis. 
Following \citet{jarvis2016}, we define the $\rho$-statistics as follows
\begin{align}
    \rho_{1}(\theta) = & \left \langle \delta e^{*}_{\text{PSF}}(\bm{\theta}') \; \delta e_{\text{PSF}}(\bm{\theta}' + \bm{\theta}) \right \rangle \,, \\
    \rho_{2}(\theta) = & \left \langle e^{*}_{\text{PSF}}(\bm{\theta}') \; \delta e_{\text{PSF}}(\bm{\theta}' + \bm{\theta}) \right \rangle \,, \\
    \rho_{3}(\theta) = & \left \langle \left( e^{*}_{\text{PSF}} \frac{\delta R^{2}_{\text{PSF}}}{R^{2}_{\text{PSF}}} \right) (\bm{\theta}') \; \left( e_{\text{PSF}} \frac{\delta R^{2}_{\text{PSF}}}{R^{2}_{\text{PSF}}} \right) (\bm{\theta}' + \bm{\theta}) \right \rangle \,, \\
    \rho_{4}(\theta) = & \left \langle \delta e^{*}_{\text{PSF}}(\bm{\theta}') \; \left( e_{\text{PSF}} \frac{\delta R^{2}_{\text{PSF}}}{R^{2}_{\text{PSF}}} \right) (\bm{\theta}' + \bm{\theta}) \right \rangle \,, \\
    \rho_{5}(\theta) = & \left \langle e^{*}_{\text{PSF}}(\bm{\theta}') \; \left( e_{\text{PSF}} \frac{\delta R^{2}_{\text{PSF}}}{R^{2}_{\text{PSF}}} \right) (\bm{\theta}' + \bm{\theta}) \right \rangle \,,
    \label{eq_02:rho_stats}
\end{align}
where ${}^{*}$ denotes complex conjugation, $\bm{\theta}$ and $\bm{\theta}'$ denote sky positions, $\theta$ denotes the modulus of $\bm{\theta}$, and $\delta$ denotes the residual error that can be computed as $\delta e_{\text{PSF}} = e_{\text{PSF}} - e_{\text{star}}$ in the case of the PSF ellipticity. Suppose that the ellipticities are random fields that are isotropic and statistically homogenous. In that case, we can compute the correlation $\rho(\bm{\theta}, \bm{\theta}')$ as $\rho(|\bm{\theta} - \bm{\theta}'|) = \rho(\theta)$, using the modulus $\theta$. This choice means we are assuming translational and rotational symmetry, a consequence of the Cosmological Principle. We define several $\theta$ bins in a logarithmic scale, corresponding to $\ln \theta - \Delta \ln \theta / 2 < \ln \theta_{ij} < \ln \theta + \Delta \ln \theta / 2$, where $\theta_{ij} = |\bm{\theta}_{i} - \bm{\theta}_{j}|$ is the distance between two objects at $\bm{\theta}_{i}$ and $\bm{\theta}_{j}$. Consequently, the correlation function at $\theta$ can be computed using the following unbiased estimator of $\rho$ that is
\begin{equation}
    \hat{\rho}(\theta) = \frac{\sum_{i,j} w_i w_j e_{i}^{\text{A}\, *} e_{j}^{\text{B}} }{\sum_{i,j} w_i w_j} \,,
\end{equation}
where we are computing, as an example, the correlation of two ellipticities $e^{\text{A}}$ and $e^{\text{B}}$, and the weights depend on the SNR of the ellipticity measurements. We carry out the weighted sum over the pairs of objects within each bin. 

The $\rho$-statistics are interesting as they can be propagated to the shear two-point correlation function (2PCF) \citep{kilbinger2015}, which allows studying the properties of the weak lensing convergence field. Following \citet{jarvis2020}, we include the PSF errors into the shear 2PCF, making the $\rho$-statistics appear, and then express the systematic error in the shear 2PCF.

\subsubsection{Other shape metrics}
Another shape metric that gives insight into the performance of the PSF model in a weak lensing analysis is the PSF leakage $\alpha$ from \citet{jarvis2016}. It is related to the linear modelling of the shear bias, where it has been decomposed into a multiplicative bias and an additive bias further decomposed into PSF-dependent (leakage) and PSF-independent terms. The PSF leakage helps quantify how the PSF affects the shear estimation through the shape measurement. It measures the leakage of the PSF shape to the galaxy shapes.

\subsection{Weak lensing: PSF error propagation and PSF requirements definition}
\label{sc_02:psf_error_propagation}
The pioneer in the PSF error propagation for WL was \citet{paulin2008}, followed by \citet{massey2012}. The proposed framework is based on the second-order moments of the images, i.e., complex ellipticity $e$ and size $R^{2}$. It expresses how the PSF, or some other effect, affects the observed ellipticity and size. Let us consider the effect of the PSF on the unweighted moments from Equation \ref{eq_03:ellip_definition} and Equation \ref{eq_03:size_definition}, where unweighted represents computing Equation \ref{eq_03:psf_moments_1} and Equation \ref{eq_03:psf_moments_2} without the $w$ weight function. Then, we obtain
\begin{equation}
    e_{\text{obs}} = e_{\text{gal}} + \frac{R^{2}_{\text{PSF}}}{R^{2}_{\text{PSF}} + R^{2}_{\text{gal}}} \left( e_{\text{PSF}} - e_{\text{gal}} \right) \quad \text{and} \quad R^{2}_{\text{obs}} = R^{2}_{\text{gal}} + R^{2}_{\text{PSF}} \,,
    \label{eq_02:psf_error_contribution}
\end{equation}
where the subscript ${}_{\text{obs}}$ refers to the quantity measured to the observed galaxy, the subscript ${}_{\text{gal}}$ refers to the intrinsic quantity of the galaxy, and ${}_{\text{PSF}}$ refers to the quantity measured to the PSF. There are intrinsic assumptions in Equation \ref{eq_02:psf_error_contribution}, that the observational model is $I_{\text{obs}} = I_{\text{gal}} * H_{\text{PSF}}$, and that all the moments are well defined. Then, Equation \ref{eq_02:psf_error_contribution} can be rewritten to express the quantity of interest in weak lensing, the intrinsic galaxy ellipticity, as follows
\begin{equation}
    e_{\text{gal}} = \frac{e_{\text{obs}} \, R^{2}_{\text{obs}} - e_{\text{PSF}} \, R^{2}_{\text{PSF}}}{R^{2}_{\text{obs}} - R^{2}_{\text{PSF}}} \,.
    \label{eq_02:gal_ellip_psf}
\end{equation}
The error propagation consists of expanding the previous equation in a first-order Taylor series with respect to the quantities of interest. In this case, it will be the shape and size of the PSF, and the propagation writes
% \widehat{e_{\text{gal}}}
\begin{equation}
    \hat{e}_{\text{gal}} \approx e_{\text{gal}} + \frac{\partial e_{\text{gal}}}{\partial \left( R^{2}_{\text{PSF}} \right)} \delta \left( R^{2}_{\text{PSF}} \right) + \frac{\partial e_{\text{gal}}}{\partial e_{\text{PSF}}} \delta e_{\text{PSF}} \,,
    \label{eq_02:error_prop_1}
\end{equation}
where $\delta$ refers to errors in the model with respect to the ground truth. It is straightforward to compute the partial derivatives in Equation \ref{eq_02:error_prop_1} from Equation \ref{eq_02:gal_ellip_psf}. We then obtain the following expression
% s
\begin{equation}
    \hat{e}_{\text{gal}} \approx e_{\text{gal}} \left(1 + \frac{\delta \left( R^{2}_{\text{PSF}} \right)}{R^{2}_{\text{gal}} } \right) - \left( \frac{R^{2}_{\text{PSF}}}{R^{2}_{\text{gal}}} \delta e_{\text{PSF}} + \frac{ \delta \left( R^{2}_{\text{PSF}} \right)}{R^{2}_{\text{gal}}} e_{\text{PSF}} \right) \,,
    \label{eq_02:ellip_estimator_psf_error}
\end{equation}
The previous ellipticity estimator can be used to obtain a shear estimator assuming that the intrinsic galaxy distribution has a zero mean. The estimator can then be used in the linear shear bias parametrization from \citet{jarvis2016}. At this point, we can express the additive and multiplicative biases as a function of the elements from Equation \ref{eq_02:ellip_estimator_psf_error}. This analysis shows us that the multiplicative bias is related to the size of the PSF with its estimation error and the size of the galaxy. The result was expected if we paid attention to the first term of Equation \ref{eq_02:ellip_estimator_psf_error}. 

This framework allows us to consider different types of errors. \citet{massey2012} uses it to include errors due to non-convolutional detector effects, imperfect shape measurement, and the fact that the shape measurement method used weighted, i.e., Equation \ref{eq_03:psf_moments_2}, instead of unweighted moments. The procedure consists of adding the desired effect to the galaxy ellipticity expression, Equation \ref{eq_02:gal_ellip_psf}, and then adding their corresponding partial derivatives to the Taylor expansion seen in Equation \ref{eq_02:error_prop_1}. \citet{cropper2013} uses this formalism to derive requirements for a WL mission in space. The aforementioned framework was used to derive the current PSF model requirements for the \textit{Euclid} space mission \citep{laureijs2011}.

The previous formalism is based on \textit{unweighted} moments. However, in practice, the moments are always computed from noisy images using a compact weight function to ensure that the measurement yields significant results. \citep{melchior2011,melchior2012} showed that using a weight function mixes the image's moments. Consequently, the second-order moments are affected by higher-order moments even in the absence of noise, thus exposing the \citet{paulin2008} framework's fundamental limitations. \citet{schmitz2020} then noted and verified empirically in an \textit{Euclid}-like scenario that the propagation is based on second-order moments of the PSF, which do not accurately describe the shape of a space-based PSF. A perfect second-order moment estimation of the PSF would have a zero shear bias contribution in the formalism described. However, the PSF's higher moments error (HME) of the PSF will impact the shear biases and are not considered in the framework proposed by \citet{paulin2008}. The higher the contribution of HME to the PSF, the more significant the deviations will be. A space mission like \textit{Euclid} will have a PSF close to the diffraction limit, meaning that its shape will be complex and not well described by a Gaussian (or by its second-order moments). As a space PSF is not well described by second-order moments, the previous requirements should be used \textit{with caution}. The LSST collaboration, concerned with the previous issue, studied the contribution to systematic biases of the HME of the PSF model on the shear measurement \citep{zhang2021,zhang2022}. \citet{zhang2021} showed that the HME of the PSF model might be a significant source of systematics in upcoming WL analyses. \citet{zhang2022} studied the impact of moments from the $3^{\text{rd}}$ to $6^{\text{th}}$ order to the cosmological parameter inference concluding that the HME of PSF models like \texttt{PSFEx} and PIFF should be reduced for the future surveys like LSST if the WL analysis is to remain unchanged. 

The use and adoption of automatic differentiable \citep{baydin2017} models could make a significant contribution to error propagation. The derivatives of the target estimators with respect to the model's parameters, or intermediate products with physical meaning, would be available. This fact allows us to consider more complex scenarios than the one seen in Equation \ref{eq_02:psf_error_contribution} as we would not require explicitly writing the equations nor their derivatives. A differentiable forward model should be enough to describe how the PSF interacts with the target task.

\section{Conclusions}
\label{sc:conclusions}
This review gives an overview of point spread function (PSF) modelling for astronomical telescopes, emphasising cosmological analyses based on weak gravitational lensing. This application sets the tightest constraints on the PSF models and has driven much of the last progress in PSF models. The development of new instruments and telescopes seeking higher precision and accuracy requires more powerful PSF models to keep up with the reduction of other sources of errors. We differentiate two scenarios that fundamentally change how the PSF is modelled: the ground- and space-based telescopes. The main difference is the atmosphere, how it affects the observations, and how challenging it is to build an atmospherical physical model that can be exploited in a reasonable amount of time. The difficulty of handling the temporal integration of a representative atmospherical model fostered the use of purely data-driven PSF models built in the pixel space for ground-based telescopes. The stability of space-based telescopes allows for exploiting models more physically based on the wavefront.

The optics fundamentals to properly define the PSF and understand its effect on the underlying imaged object are not often introduced in PSF modelling articles. One of our goals was to solve this issue by providing a concise yet comprehensive introduction to optical principles. The provided optical background should cover most of the available PSF models and motivate the general observational model that we have proposed. This observational model can be further simplified and adapted to different use cases, including ground or space telescopes. We described several assumptions that might not always hold. After describing how the PSF affects our observations, we presented an extended list of the leading optical- and detector-level contributors to the PSF field. These contributors are the source of the PSF's spatial, spectral and temporal variations in addition to its morphology. A detailed description of the atmospheric contribution based on phase screens was presented. We then gave a brief description of the most relevant PSF models. 

The \texttt{PSFEx} model has been successful in modelling the PSF for several surveys with its robust and fast implementation. However, the next-generation telescopes set higher PSF requirements, demanding novel models to achieve such performances. Recent models are targetting upcoming telescopes, e.g., the Vera C. Rubin observatory and the \textit{Euclid} and Roman space telescopes. These models are continuously being developed and are pushing forward the capabilities of PSF modelling. A common bottleneck for them is the computing time required to estimate the model from the observations. It is unclear if the solution can be achieved through better software implementations that exploit parallel computing architectures or better-performing programming languages. A refactoring of the methods allowing for simplifications that accelerate calculations might be required, or even both approaches. One big challenge of PSF modelling is to build \textit{fast} still \textit{powerful} models. 

Another challenge is to include complex effects and contributions that cannot be directly constrained from the observations into the PSF model. These contributions can be modelled with simulations and obtained from complementary observations or instrument characterisations. However, this complementary information will not match precisely the state of the telescope during the imaging procedure due to several reasons, e.g., changes in the telescope, measurement errors, and imperfect modelling. The better way to correct this information and adapt it to real observations needs to be further studied.

The validation of PSF models from real observations is a challenging subject that requires further development, as access to the ground truth PSF field is unavailable. Although some validation methods exist, they are generally not very informative or based on second-order moments that are not well suited to describe diffraction-limited PSFs. We have presented the error propagation of a galaxy-shape measurement, where several limitations were exposed. Current error propagation methods have simplifying assumptions, e.g., the PSF is well described by its quadrupole moments that do not hold anymore with recent and upcoming telescopes. Further development of these methods is required to define realistic PSF model requirements and study how PSF modelling errors affect the target task.

\section*{Conflict of Interest Statement}
The authors declare that the research was conducted in the absence of any commercial or financial relationships that could be construed as a potential conflict of interest.

\section*{Author Contributions}
TIL coordinated the entire effort, produced the figures and wrote the text. JLS and MK provided valuable comments and feedback on the different sections of the article.

\section*{Funding}
This work was partially funded by the TITAN ERA Chair project (contract no. 101086741) within the Horizon Europe Framework Program of the European Commission. This work is also partially supported by EPSRC (grant number EP/W007673/1).

\section*{Acknowledgments}
This work was funded by the TITAN ERA Chair project (contract no. 101086741) within the Horizon Europe Framework Program of the European Commission.
Part of this review has been adapted from Tobias Liaudat's PhD thesis \citep{liaudat2022_thesis}. We acknowledge and resume in the following list the open-source simulating and modelling software mentioned in this review:

\begin{itemize}
    \item \textsc{GalSim} \citep{rowe2015}: \href{https://github.com/GalSim-developers/GalSim}{github.com/GalSim-developers/GalSim} ,
    \item \textsc{webbpsf} \citep{perrin2014}: \href{https://github.com/spacetelescope/webbpsf}{github.com/spacetelescope/webbpsf} ,
    \item \textsc{PhoSim} \citep{peterson2015}: \href{https://bitbucket.org/phosim/phosim_release/wiki/Home}{bitbucket.org/phosim/phosim\_release/wiki/Home} ,
    \item \textsc{WaveDiff} \citep{liaudat2023}: \href{https://github.com/CosmoStat/wf-psf}{github.com/CosmoStat/wf-psf} ,
    \item \textsc{ShapePipe} \citep{farrens2022}: \href{https://github.com/CosmoStat/shapepipe}{github.com/CosmoStat/shapepipe} ,
    \item \textsc{PIFF} \citep{jarvis2020}: \href{https://github.com/rmjarvis/Piff}{github.com/rmjarvis/Piff} ,
    \item \textsc{MCCD} \citep{liaudat2020}: \href{https://github.com/CosmoStat/mccd}{github.com/CosmoStat/mccd} ,
    \item \textsc{RCA} \citep{ngole2016,schmitz2020}: \href{https://github.com/CosmoStat/rca}{github.com/CosmoStat/rca} ,
    \item \textsc{PSFEx} \citep{bertin2011}: \href{https://github.com/astromatic/psfex}{github.com/astromatic/psfex} ,
    \item \textsc{Tiny-Tim} \citep{krist2011}: \href{https://github.com/spacetelescope/tinytim}{github.com/spacetelescope/tinytim} ,
    \item \textsc{Photutils} \citep{bradley2022}: \href{https://github.com/astropy/photutils}{github.com/astropy/photutils} ,
    \item \textsc{$\partial$Lux} \citep{desdoigts2023}: \href{https://github.com/LouisDesdoigts/dLux}{github.com/LouisDesdoigts/dLux} ,
    \item \textsc{PSF weather station}: \href{https://github.com/LSSTDESC/psf-weather-station}{github.com/LSSTDESC/psf-weather-station} .
\end{itemize}

\bibliographystyle{Frontiers-Harvard} %  Many Frontiers journals use the Harvard referencing system (Author-date), to find the style and resources for the journal you are submitting to: https://zendesk.frontiersin.org/hc/en-us/articles/360017860337-Frontiers-Reference-Styles-by-Journal. For Humanities and Social Sciences articles please include page numbers in the in-text citations 
\bibliography{PSF_bib}

\end{document}